\documentclass[paper]{JHEP3}
\usepackage{epsfig}
\usepackage{graphicx}
\usepackage{xspace}
\usepackage{amssymb}
\usepackage{subfigure}
\usepackage{multirow}

\newcommand{\be}{\begin{equation}}
\newcommand{\ee}{\end{equation}}

\newcommand{\bea}{\begin{eqnarray}}
\newcommand{\eea}{\end{eqnarray}}
\newcommand{\beanon}{\begin{eqnarray*}}
\newcommand{\eeanon}{\end{eqnarray*}}
\newcommand{\ba}{\begin{array}}
\newcommand{\ea}{\end{array}}
\newcommand{\bd}{\begin{description}}
\newcommand{\ed}{\end{description}}
\newcommand{\bi}{\begin{itemize}}
\newcommand{\ei}{\end{itemize}}
\newcommand{\ben}{\begin{enumerate}}
\newcommand{\een}{\end{enumerate}}
\newcommand{\bc}{\begin{center}}
\newcommand{\ec}{\end{center}}

\newcommand{\toptop}{\mbox{${\mathrm t} \bar {\mathrm t}$}\xspace}

\newcommand{\ordEW}{\mathcal{O}(\alpha_{\scriptscriptstyle EM}^6)\xspace}
\newcommand{\ordQCD}{\mathcal{O}(\alpha_{\scriptscriptstyle EM}^4
  \alpha_{\scriptscriptstyle S}^2)\xspace}
\newcommand{\ordQCDsq}{\mathcal{O}(\alpha_{\scriptscriptstyle EM}^2
  \alpha_{\scriptscriptstyle S}^4)\xspace}

\newcommand{\eqn}[1]{Eq.(\ref{#1})}

\newcommand{\tbn}[1]{Tab.~\ref{#1}}
\newcommand{\tbns}[2]{Tabs.~\ref{#1}--\ref{#2}}
\newcommand{\tbnsc}[2]{Tabs.~\ref{#1},~\ref{#2}}

\newcommand{\sect}[1]{Sect.~\ref{#1}}

\newcommand{\Phantom}{{\tt PHANTOM}\xspace}

\newcommand{\MadEvent}{{\tt MADEVENT}\xspace}

\def\pl #1 #2 #3 {{\it Phys.~Lett.} {\bf#1} (#2) #3}   
\def\np #1 #2 #3 {{\it Nucl.~Phys.} {\bf#1} (#2) #3}
\def\zp #1 #2 #3 {{\it Z.~Phys.} {\bf#1} (#2) #3}
\def\pr #1 #2 #3 {{\it Phys.~Rev.} {\bf#1} (#2) #3}
\def\prep #1 #2 #3 {{\it Phys.~Rep.} {\bf#1} (#2) #3}
\def\prl #1 #2 #3 {{\it Phys.~Rev.~Lett.} {\bf#1} (#2) #3}
\def\intj #1 #2 #3 {{\it Int. J. Mod. Phys.} {\bf#1} (#2) #3}
\def\mpl #1 #2 #3 {{\it Mod.~Phys.~Lett.} {\bf#1} (#2) #3}
\def\rmp #1 #2 #3 {{\it Rev. Mod. Phys.} {\bf#1} (#2) #3}
\def\cpc #1 #2 #3 {{\it Comp. Phys. Commun.} {\bf#1} (#2) #3}
\def\epj #1 #2 #3 {{\it Eur. Phys. J.} {\bf#1} (#2) #3}
\def\jhep #1 #2 #3 {{\it JHEP} {\bf#1} (#2) #3}

\title{A complete parton level analysis of boson-boson scattering and
ElectroWeak Symmetry Breaking in  \boldmath$\ell\nu$ + four jets
production at the LHC.}

\author{
Alessandro Ballestrero$^a$,
Giuseppe Bevilacqua$^c$ and
Ezio Maina$^{a,b}$\\
$^a$ INFN, Sezione di Torino, Italy,\\
$^b$ Dipartimento di Fisica Teorica, Universit\`a di Torino, Italy\\
$^c$ Institute of Nuclear Physics, NCSR Demokritos, 15310 Athens, Greece.
}

\preprint{DFTT 31/2008}

\abstract{
A complete parton level analysis of l$\nu$ + four jets 
production at the LHC is presented, including all processes at order
$\ordEW$, $\ordQCD$ and $\ordQCDsq$.
The infinite Higgs mass scenario, which is considered as a benchmark for strong
scattering theories and is the
limiting case for composite Higgs models, is confronted with the Standard Model
light Higgs predictions in order to determine whether a composite Higgs signal
can be detected as an excess of events in boson--boson scattering.
}

\begin{document}

\section{Introduction}
\label{sec:intro}
The mechanism of Electro--Weak Symmetry Breaking (EWSB) will be a central issue
in the physics program at the LHC\footnote{Detailed reviews and extensive
bibliographies can be found in 
Refs.\cite{HiggsLHC,djouadi-rev1,ATLAS-TDR,Houches2003,CMS-TDR}}.
The Standard Model (SM) describes this phenomenon
in an extremely simple and economical fashion through the Higgs mechanism.
The fit of EW precision data is in agreement 
with the SM predictions to an unprecedented accuracy and gives an upper limit on
the Higgs mass of about 182 GeV \cite{lepewwg07}, while direct searches imply
$m_H > 114$ GeV \cite{lepewwg}.
Any attempt to go beyond the SM is severely constrained and 
made difficult by the very success the SM has achieved.

The first question to which the LHC must provide an answer is whether or not a
light Higgs exists.
If the Higgs is not found then the SM and its most promising extension,
the MSSM, will be ruled out.

In this case scattering processes between longitudinally polarized vector bosons
will play a prominent role because, without a Higgs, the corresponding
amplitudes grow with energy and violate perturbative unitarity at
about one TeV \cite{reviews}. 
However, since unitarity is essentially the statement of conservation of total probability
it cannot be violated in Nature and some new phenomena 
must intervene at an energy scale within reach for the LHC.

Many alternative mechanisms of EWSB have been explored. For instance EWSB may result
from strong dynamics in a hitherto undiscovered sector, as  postulated by Technicolor
theories.
Alternatively, EWSB may be a consequence of boundary conditions satisfied by gauge fields
in compactified additional space dimensions. In this case, new states, the tower of
Kaluza--Klein modes generated by expanding the fields along the compactified directions,
would tame the growth of Vector--Vector (VV) scattering.
However, typically, Technicolor and Higgsless theories have difficulties with
EW precision data and
with the generation of fermion masses. 

A more pragmatical approach is based on the language of the Effective 
Electro--Weak Lagrangian \cite{EEWL},
in which the SM is interpreted as the first term in an
expansion in $E/\Lambda$ of an unknown high energy theory characterized by a
mass scale $\Lambda$ which acts as a cut--off compared with the energy $E$ at
which the theory is probed,
or equivalently on the anomalous
couplings \cite{AQC} description.
This strategy can be supplemented by the adoption of 
one of the many schemes for turning
perturbative scattering amplitudes into amplitudes which satisfy by construction
the unitarity constraints. This procedure,
in analogy with low energy QCD, which can be expressed by exactly the same
formalism which describes the Higgs sector in the SM, leads to expect the
presence of resonances in 
VV scattering. Unfortunately the mass, spin and even number of these
resonances are not uniquely determined \cite{unitarization,Fabbrichesi,Kilian}
from theory.
Again, a detailed study of the two--boson mass distribution in VV
scattering will be mandatory in order to clarify the details of the
resulting spectrum.

The discovery of one or more Higgs particle(s) will trigger the effort to
measure its properties as accurately as possible, beginning with its mass and
its coupling to the other particles.
It should be noted that the Goldstone theorem and the Higgs mechanism do not
require the existence of elementary scalars. It is conceivable and widely
discussed in the literature that composite states are responsible for EWSB
as nicely recently reviewed in Ref.~\cite{Giudice_review}.
For instance in \cite{H_Goldstone} the Higgs is identified with a
pseudoGoldstone boson, while more recently additional possibilities
have been explored: the Little Higgs \cite{LittleH1}, the gauge--Higgs
unification
\cite{gaugeHiggsU1,gaugeHiggsU2} and the holographic Higgs \cite{HologHiggs}.
These theories are characterized by the presence of new states which could be
produced at the LHC, if light enough. In view of the large number of different
proposals it is useful to determine the model independent features of this class
of theories.
There has been recent progress in this area 
\cite{Contino07,Giudice:2007fh,Barbieri}, using the effective theory language.
The leading effects are described by two parameters (one for a universal
modification 
of all Higgs couplings, and the other one for a universal modification of Higgs
couplings to 
fermions) characterized by the ratio $v^2 /f^2$, where $v$ is the Higgs vacuum
expectation value
and $f$ is the $\sigma$--model scale.
Again, a test of Higgs compositeness at the LHC requires an analysis of
longitudinal gauge--boson scattering.
Indeed, because of the modified Higgs couplings,
longitudinal gauge--boson scattering amplitudes violate unitarity at high
energy, even in the presence of a 
light Higgs \cite{Giudice:2007fh}. The $E^2$--growing amplitude is a factor
$v^2 /f^2$ smaller than in the Higgsless case and the violation
is moved to a larger energy regime.
As a consequence, the scattering cross section for an infinitely
massive Higgs in the SM represents, 
at large energies, an upper limit for VV scattering processes in this class
of theories, and can be taken as
a benchmark for the observability of signals of strong scattering
and of Higgs compositeness in boson--boson reactions.
It should be mentioned that in any case the rise of
the cross section at large invariant VV mass will be masked by the decrease of
the parton luminosities at large
momentum fractions and, as a consequence, will be particularly challenging to
detect.
   
Therefore, even if a light Higgs is discovered, boson--boson
scattering is a crucial process to 
study, which can give us useful information on the nature of the Higgs boson.
It is worth pointing out that, in this framework, since the Higgs can be
viewed as an approximate fourth Goldstone boson, its properties 
are related to those of the exact (eaten) Goldstone bosons. Strong gauge-boson
scattering will 
be accompanied by strong Higgs pair production \cite{Giudice:2007fh}.

Scattering processes among vector bosons have been scrutinized since a long
time \cite{history1,history2}. 
In most cases previous studies of boson--boson scattering at high energy
hadron colliders
have resorted to some approximation, either the Equivalent Vector Boson
Approximation (EVBA) \cite{EVBA}, or a production times decay approach.
In Ref.~\cite{Accomando:2005hz,Accomando:2006vj} an analysis of 
l$\nu$ + four jets and $\mu^+\mu^-$ + four jets production at the LHC
has been presented, with the
limitation of taking into account only purely electroweak processes. 
Preliminary results concerning the inclusion of the $\ordQCD$ background, which
include $VV+2j$ and top--antitop production
have appeared in Ref.~\cite{Ambroglini:2009mg}.
In the last few years QCD corrections to boson--boson 
production via vector boson fusion \cite{JagerOleariZeppenfeld}
at the LHC have been computed and turn out to be below 10\%.
While the present paper was being finalized, {\tt VBFNLO} \cite{Arnold:2008rz}
a Monte Carlo program for vector boson fusion, double and triple vector boson
production at NLO QCD accuracy, limited to the leptonic decays of
vector bosons, has been released

\begin{figure}
\begin{center}
%\mbox{\epsfig{file=diag/VVfusion_new2.pdf,width=12.cm}}
\mbox{\epsfig{file=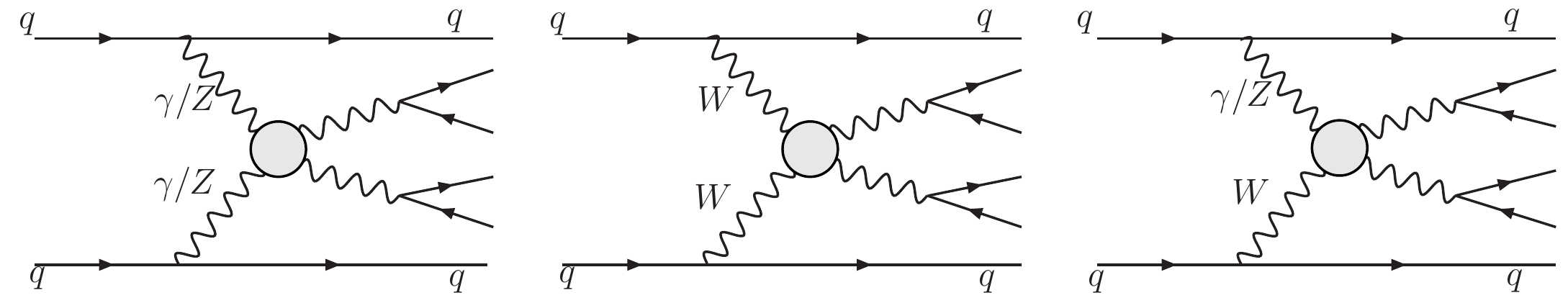,width=12.cm}}
\caption{ Vector boson fusion processes}
\label{VV-diag}
\end{center}
\end{figure}

\begin{figure}
\begin{center}
%\mbox{\epsfig{file=diag/nonreso_new2.pdf,width=12.cm}}
\mbox{\epsfig{file=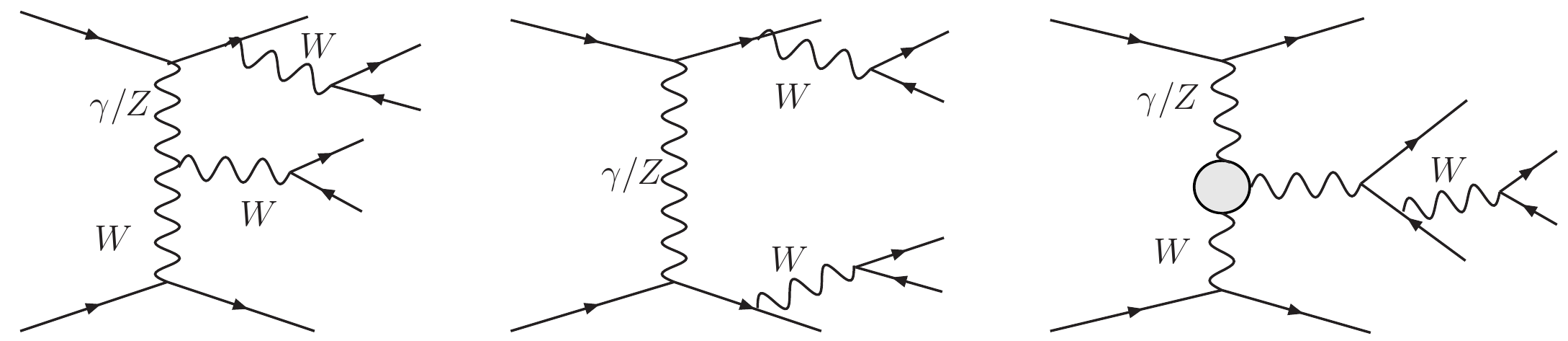,width=12.cm}}
\caption{ Non fusion and non doubly resonant two vector boson production.}
\label{nonreso-diag}
\end{center}
\end{figure}

\begin{figure}
\begin{center}
%\mbox{\epsfig{file=diag/higgs_new3.pdf,width=12.cm}}
\mbox{\epsfig{file=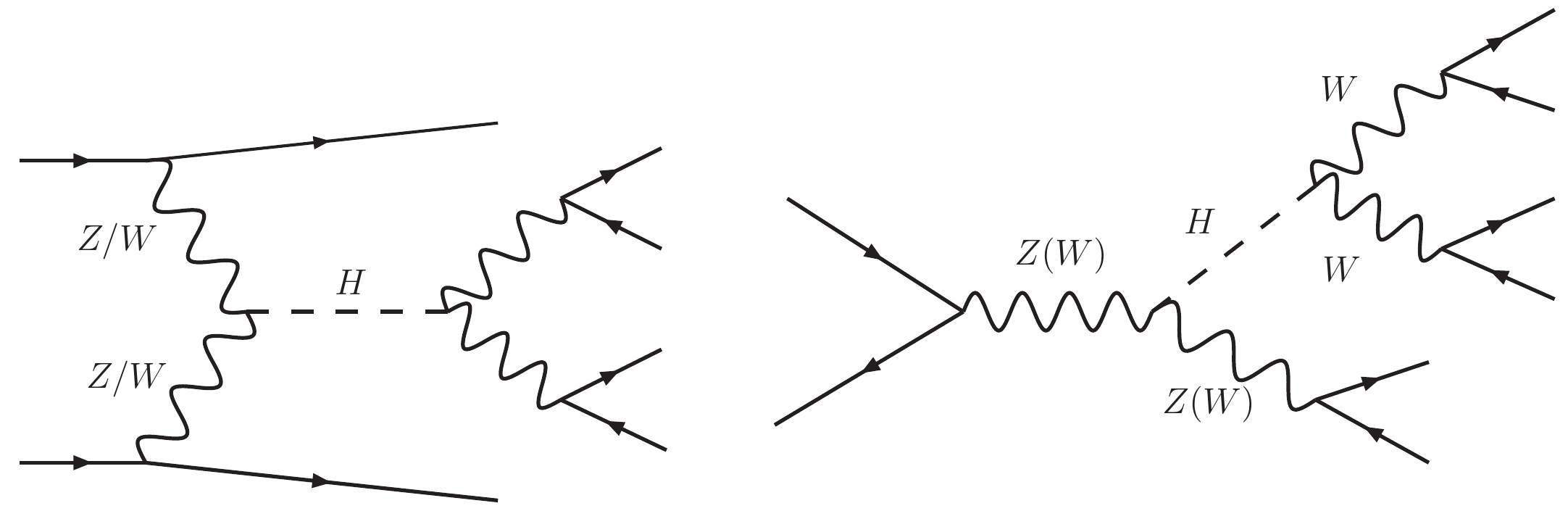,width=12.cm}}
\caption{Higgs boson production via vector boson fusion
  and Higgstrahlung.}
\label{higgs-diag}
\end{center}
\end{figure}

\begin{figure}
\begin{center}
%\mbox{\epsfig{file=diag/toptop_new3.pdf,width=12.cm}}
\mbox{\epsfig{file=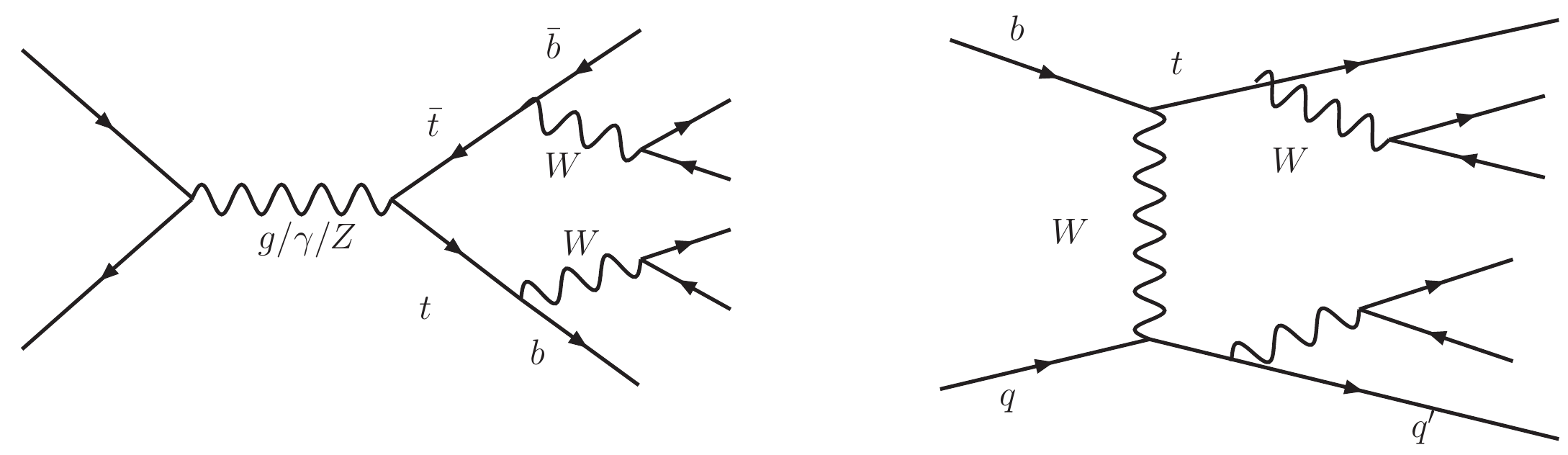,width=12.cm}}
\caption{ Electroweak \toptop and single top production.}
\label{top-diag}
\end{center}
\end{figure}

\begin{figure}
\begin{center}
%\mbox{\epsfig{file=diag/TGC_new3.pdf,width=14.cm}}
\mbox{\epsfig{file=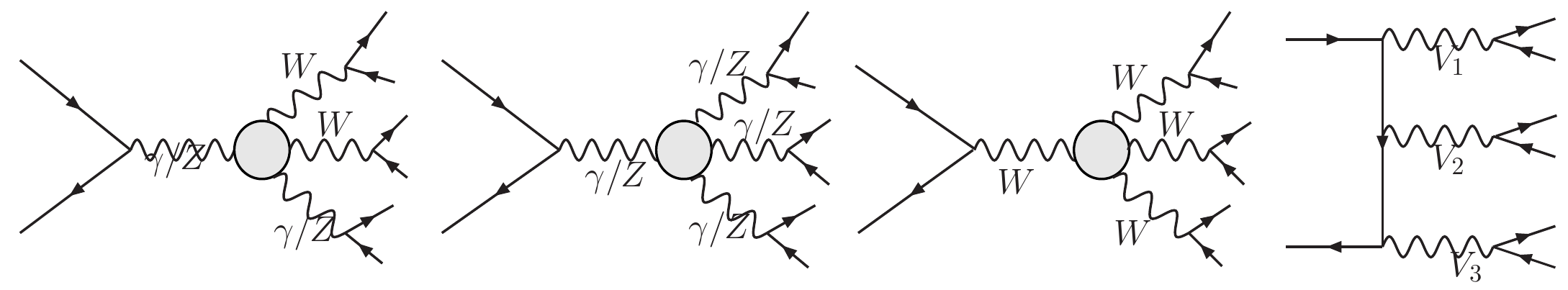,width=14.cm}}
\caption{ Three vector boson production.}
\label{tgc-diag}
\end{center}
\end{figure}

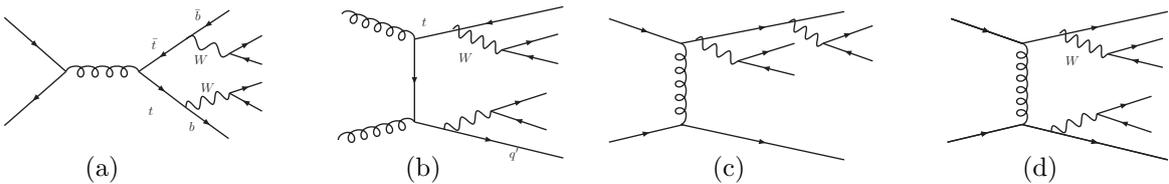
\begin{figure}[h!tb]
\centering
\hspace*{-0.5cm}
\epsfig{file=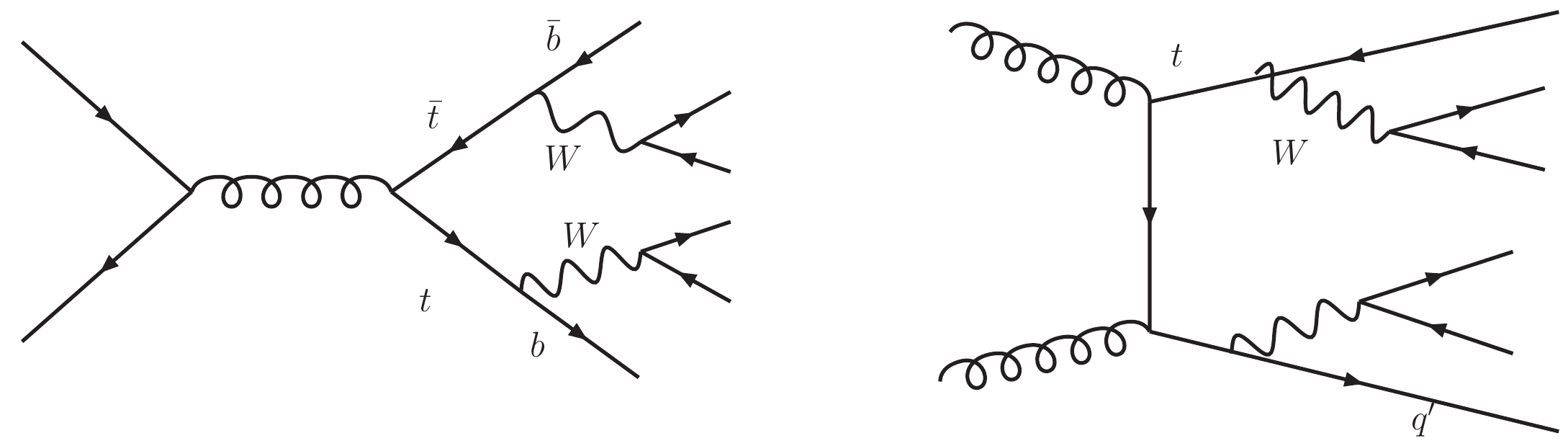,width=7.5cm}
\hspace*{0.2cm}
\epsfig{file=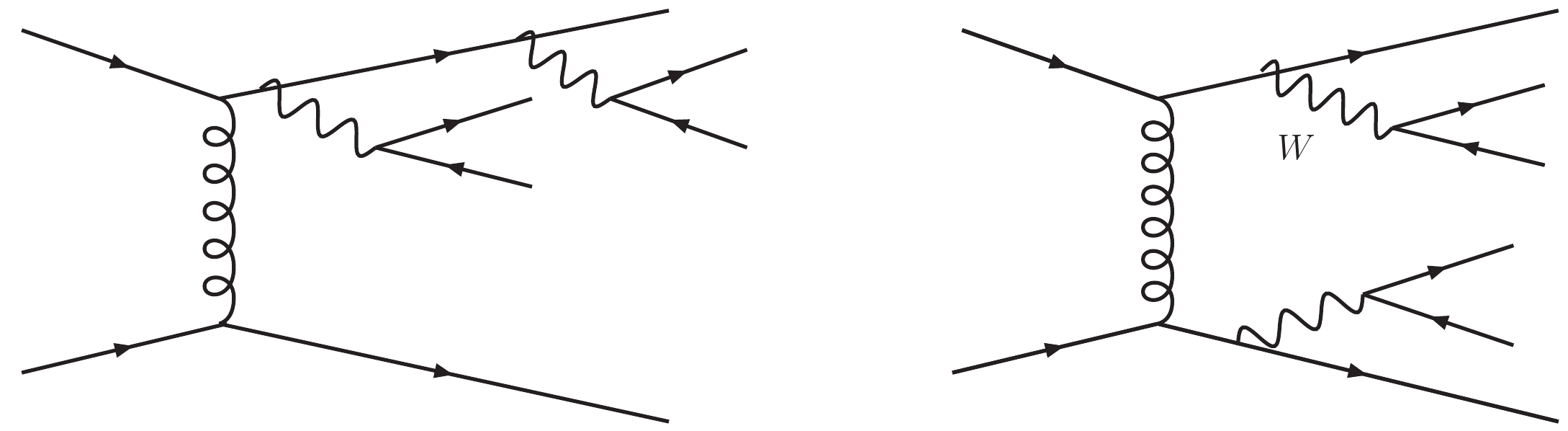,width=7.5cm}
\hspace*{-0.7cm}
\begin{picture}(0,0) (0,0)
  \put(-394,-5) {\small{(a)}}
  \put(-274,-5) {\small{(b)}}
  \put(-159,-5) {\small{(c)}}
  \put(-43,-5) {\small{(d)}}
\end{picture}
\caption{Examples of contributions to the QCD irreducible background: $t\bar{t}$
production (a,b) and $VV+2j$ (c,d)}
\label{fig:VBSbckgr_QCD}
\end{figure}

\begin{figure}[htb]
\centering
\mbox{
\includegraphics*[width=0.29\textwidth]{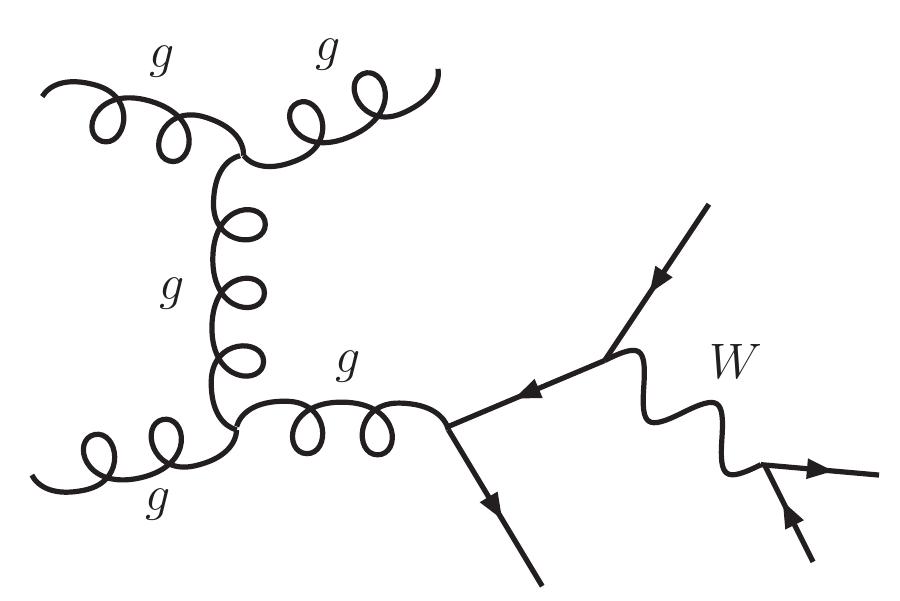}\hspace*{0.5cm}
\includegraphics*[width=0.20\textwidth]{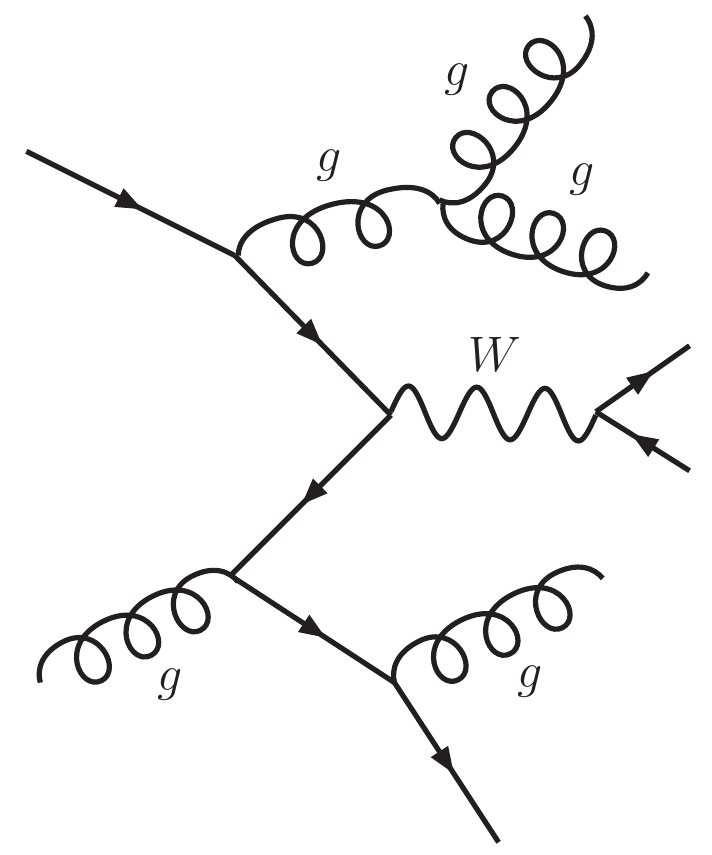}\hspace*{0.6cm}
\includegraphics*[width=0.18\textwidth]{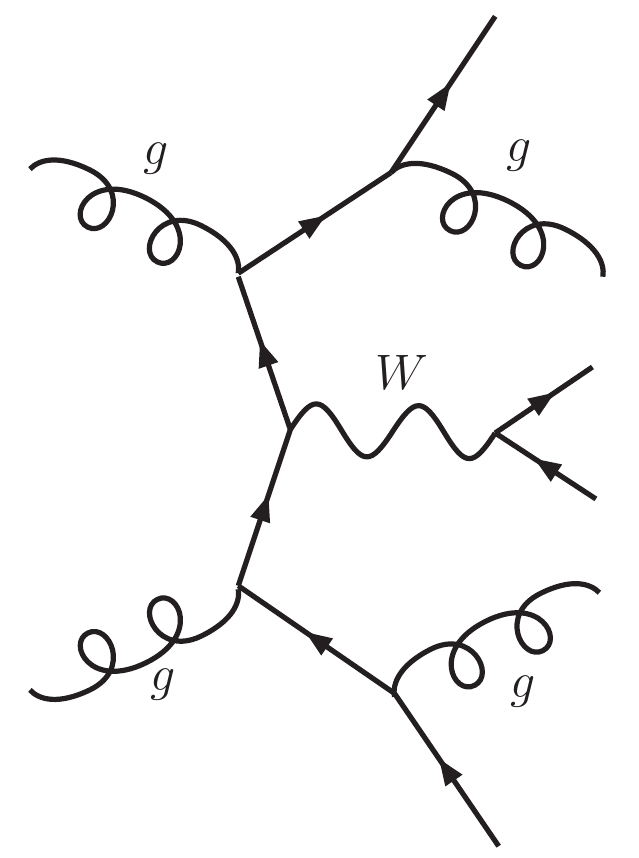}\hspace*{0.6cm}
\includegraphics*[width=0.19\textwidth]{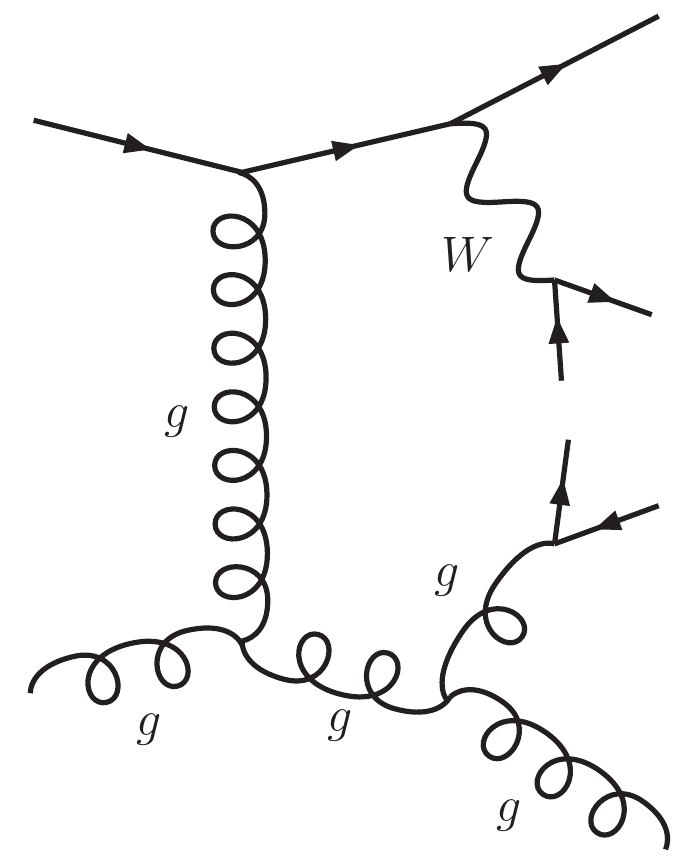}
}\\[0.2cm]
\begin{picture}(0,0) (0,0)
  \put(-180,-5) {\small{(a)}}
  \put(-35,-5) {\small{(b)}}
  \put(65,-5) {\small{(c)}}
  \put(160,-5) {\small{(d)}}
\end{picture}
\caption{}
Representative Feynman diagrams for the $\ordQCDsq$, $W+4j$
production processes at the LHC.
\label{fig:diag_as4}
\end{figure}

In this paper we study at parton level
the process $pp\rightarrow \ell\nu + 4j$, including all backgrounds
contributing to this six parton final state. We use complete tree level
matrix elements.
We consider two scenarios:
a light Higgs SM framework with $M_H = 200$ GeV and an infinite mass Higgs
scenario. The production cross section for $\ell\nu + 4j$ has been shown to be
much larger \cite{Accomando:2005hz,Accomando:2006vj}
than that for $\mu^+\mu^- + 4j$ making it the most promising candidate for
boson--boson scattering studies provided the full QCD background can be kept
under control.
Processes in which both vector bosons decay leptonically suffer from a much
smaller rate despite a reduced QCD background.
  
%%%%%%%%%%%%%%%%%%%%%%%%%%%%%%%%%%%%%%%%%%%%%%%%%%%%%%%%%%%%%%%%%%%%%%%%%
\section{Outline of the analysis} 
\label{sec:Outline_of_the_analysis}

The observation of strong boson--boson
scattering as an excess of events compared to the
SM prediction requires, as an essential condition, that
a signal of VV scattering is extracted from the background.
At the same time the selection strategy must be capable to maximize the
differences between the light Higgs and the no--Higgs cases.
Three perturbative orders contribute to the background.
At $\ordEW$ there are a large number of diagrams which cannot be interpreted as
boson--boson
scattering and which cannot be separated in any sensible way from the
scattering type diagrams due to large cancellations between the two sets
\cite{Accomando:2006vj}.
At $\ordQCD$ we have to deal with the production of two electroweak bosons
plus two jets without any scattering
contribution. At $\ordQCDsq$ only one electroweak boson is effectively produced,
while the additional jets, which do not
peak at any particular mass, populate the full available phase space with a
production rate which is much larger than the signal one.
   
In this section we briefly describe how the analysis on $\ell\nu +4j$ final 
states has been performed. Basically, the whole procedure can be summarized in 
three steps:
\begin{itemize}
\item apply a set of kinematical cuts to isolate a sample of candidate 
      scattering events;
\item define a Vector Boson Scattering (VBS) signal on this enriched sample; 
\item develop a statistical treatment of the signal to estimate the probability
      of observing an excess in terms of confidence levels.
\end{itemize}
The first step is concerned with the identification of a suitable kinematical 
signature which allows to capture the essence of $VV$ scattering. 
The selection of events widely separated in pseudorapidity is a well established
technique for enhancing the scattering contributions at LHC \cite{history1,history2}.
Looking at the topology of the diagrams
embedding the gauge boson scattering as a subprocess in 
Fig.~\ref{VV-diag}, one concludes that it is appropriate to associate the
two most 
forward/backward jets to the tag quarks which radiate the bosons which initiate
VV scattering and to relate
the two most central jets to the hadronic decay of a $W$ or a $Z$ in the final
state.
The main purpose of this kinematical selection is to isolate a sample of genuine
$VV+2j$ events while suppressing the contribution of irreducible backgrounds 
such as three boson production or top quark production, either from single top 
or top--antitop pairs. Additional cuts have been imposed in order to discriminate
more effectively between the light Higgs and the no Higgs scenarios.
Unfortunately, this procedure does not fully screen from the QCD 
background entering $VV+2j$ and additional, \textit{ad hoc}, cuts must be 
applied to this purpose.

Having isolated a sample of candidate scattering events, one needs to
define an observable quantity which is as susceptible as possible to the details
of the mechanism of EWSB in order to maximize the sensitivity to effects of 
alternative models such as strong scattering.
This task is straightforward only apparently. As already 
mentioned, a number of fake hits is expected to come from the QCD 
background, mainly in the form of $W+4j$, as a consequence of the large cross 
section and gluon combinatorics which characterize this kind of contributions.
The classical approach is to focus on the invariant mass distribution of the
final state boson pair. The large QCD background, with its large scale
uncertainty, makes this procedure rather dubious. 
A possible way out for this problem is to look instead at the invariant mass of
the two central jets ($M_{j_cj_c}$) for events with large
vector pair mass.

Provided a convenient set of kinematical cuts has been applied, the 
$\ordEW + \ordQCD$ cross section is dominated by the $W$ and $Z$ peaks, while 
the $\ordQCDsq$ ($W+4j$) contribution is non-resonant in this respect. 
When restricting to the window between 70 and 100 GeV, which covers 
completely the $W$ and $Z$ resonances, we find that the $W+4j$ distribution is 
essentially flat and therefore can reliably be measured from the sidebands
of the physical region of interest.
This procedure has two advantages. On one side,
it drastically reduces the theoretical uncertainty associated to the scale 
dependence of the cross section, which mainly affects the $\ordQCDsq$
contribution. On the 
other side, it allows to subtract the dominant contribution to the irreducible 
QCD background, thus enhancing the visibility of genuine EWSB effects.

Once the 
non-resonant background has been subtracted, one is left with a 
peak whose size is strictly related to the regime of the EWSB dynamics: a 
strongly-coupled scenario would result in a more prominent peak than a 
weakly-coupled one. This feature suggests to take the 
integral of the peak as the discriminator among different 
models. At a given collider luminosity, the number of expected 
events can be derived. With a slight abuse of language, we call this number the
VBS signal.
It is by analyzing the probability density function (p.d.f.) associated with 
this discriminator that we can determine, in the last step, the confidence level
for a given experimental result to be or not to be SM-like, in the same spirit 
of the statistical procedure adopted for the search of the Standard Model Higgs 
boson at LEP \cite{Barate:2003sz}.

%%%%%%%%%%%%%%%%%%%%%%%%%%%%%%%%%%%%%%%%%%%%%%%%%%%%%%%%%%%%%%%%%%%%%%%%%
\section{Calculation}
\label{sec:calc}

As discussed in \sect{sec:Outline_of_the_analysis},
three perturbative orders contribute to $\ell\nu +4j$ at the LHC.

The $\ordEW$ and $\ordQCD$ samples have been generated with \Phantom
\cite{Ballestrero:2007xq,method,phact}, while the $\ordQCDsq$ sample 
has been produced with \MadEvent \cite{MadeventPaper}.
Both programs generate events 
in the Les Houches Accord File Format \cite{LHAFF}.
In all samples full $2 \rightarrow 6$ matrix elements, without any
production times decay approximation, have been used.
The cuts in \tbn{tab:cuts_0} have been applied at generation level.
 
\begin{table}[h!tb]
\begin{center}
\begin{tabular}{|c|}
\hline
\textbf{Generation cuts} \\
\hline
\hspace*{1cm} $p_T(\ell^\pm) > 20 \mbox{ GeV}$ \hspace*{1cm} \\
\hline
$|\eta(\ell^\pm)| < 3.0$ \\
\hline
$p_T(j) > 30 \mbox{ GeV}$ \\
\hline
$|\eta(j)| < 6.5$ \\
\hline
$M(jj) > 60 \mbox{ GeV}$ \\
\hline
\end{tabular}
\caption{Standard acceptance cuts applied in the event generation and present
in all results. Here $j = d,u,s,c,b,g$: any pair of colored partons must have
mass larger than 60 GeV.} 
\label{tab:cuts_0}
\end{center}
\end{table}

For the Standard Model parameters we use the input values:
\begin{equation}
\begin{array}[b]{lcllcllcl}
%\begin{eqnarray}
\label{eq:SMpar}
\mathrm {M_W} & = & 80.40 , \qquad &
\mathrm {M_Z} & = & 91.187~\mathrm{GeV}, \\
G_{\mu} & = & 1.16639~10^{-5}~\mathrm{GeV}^{-2}, \qquad &
\alpha_s(\mathrm {M_Z})  & = & 0.118\\
\mathrm {M_t} & = & 175.0~\mathrm{GeV}, \qquad &
\mathrm {M_b} & = & 4.8~\mathrm{GeV}. 
%\begin{eqnarray}
\end{array}
\end{equation}

The masses of all other partons have been set to zero.
We adopt the standard $G_{\mu}$-scheme to compute the remaining parameters.

All samples have been generated using CTEQ5L \cite{CTEQ5} 
parton distribution functions.
For the $\ordEW$ and $\ordQCD$ samples, generated with \Phantom,
the QCD scale has been taken as:
\be
\label{eq:LargeScale}
Q^2 = M_W^2 + \frac{1}{6}\,\sum_{i=1}^6 p_{Ti}^2
\label{scale}
\ee
while for the $\ordQCDsq$ sample the scale
has been set to $Q^2 = M_Z^2$. This difference in the scales
conservatively leads to a definite relative enhancement of
the  $W\, +  \, 4j$ background. Tests in
comparable reactions at $\ordQCDsq$
have shown an increase of about a factor of 1.5 
for the processes computed at $Q^2 = M_Z^2$ with
respect to the same processes computed with the larger scale
\eqn{eq:LargeScale}. 

We work at parton level with no showering and hadronization.
The two jets with the largest and smallest rapidity are identified as forward and
backward tag jet respectively. The two intermediate jets are considered as
candidate vector boson decay products.

The neutrino momentum is reconstructed according to the usual prescription,
requiring the 
invariant mass of the $\ell \nu$ pair to be equal to the $W$ boson nominal mass,
\begin{equation}
\label{eq:nu_reco_equation}
(p^{\ell}+p^{\nu})^2 = M_W^2 ,
\end{equation}
in order to determine the longitudinal component of the neutrino momentum.
This equation has two solutions,
\begin{equation}
\label{eq:nu_reco}
p_{z}^{\nu} = \frac{\alpha p_z^\ell \pm \sqrt{\alpha^2 p_z^{\ell 2} - 
 (E^{\ell 2} - p_z^{\ell 2})(E^{\ell 2} p_T^{\nu 2} - \alpha^2)}}
  {E^{\ell 2} - p_z^{\ell 2}}  \; ,
\end{equation}
where
\begin{equation}
\alpha = \frac{M_W^2}{2} + p_x^{\ell}p_x^{\nu} + p_y^{\ell}p_y^{\nu}  \; .
\end{equation}

\begin{figure}[htb]
\begin{center}
\hspace*{-2cm}
\includegraphics[width=8.3cm]{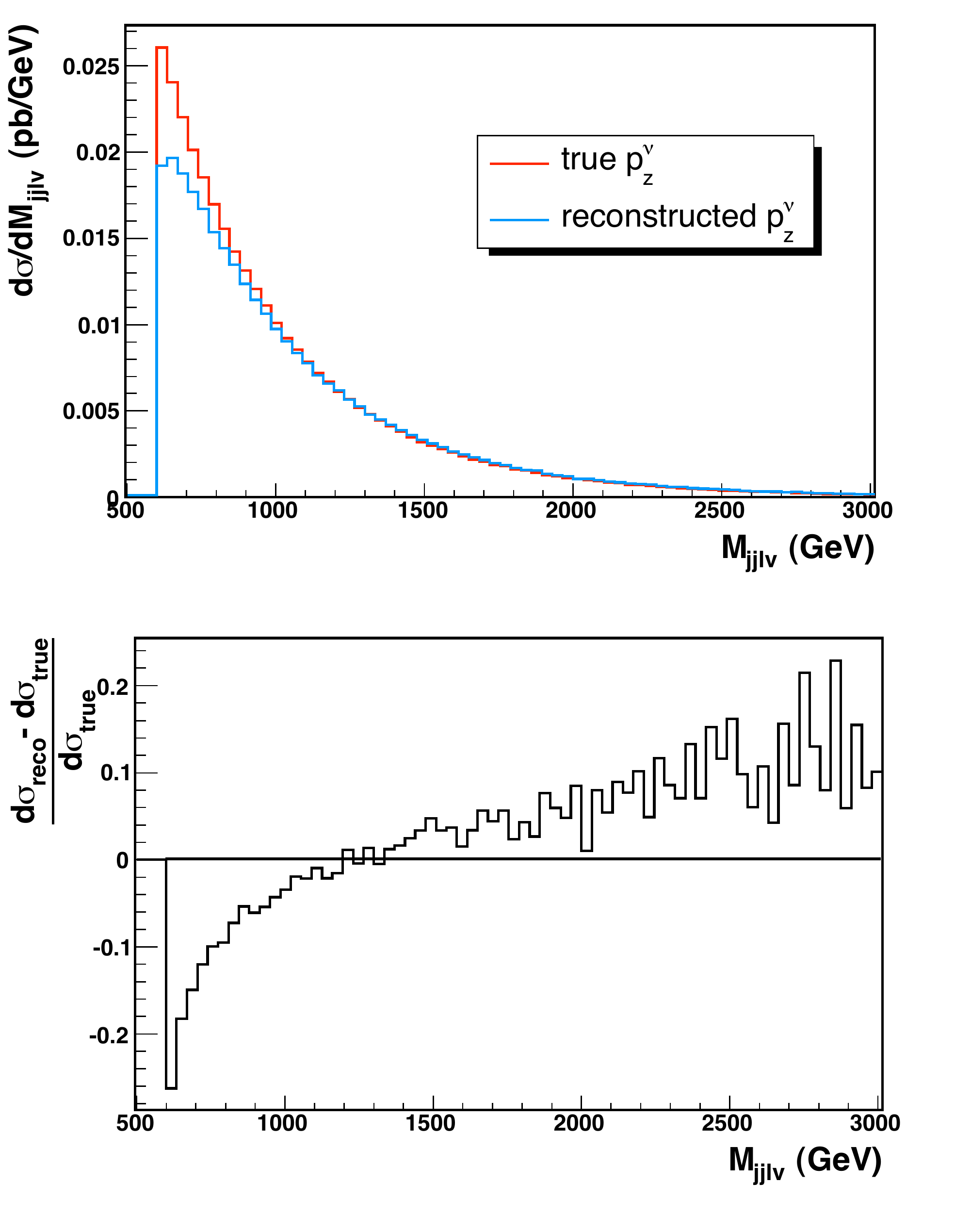}
\hspace*{-0.6cm}
\includegraphics[width=8.3cm]{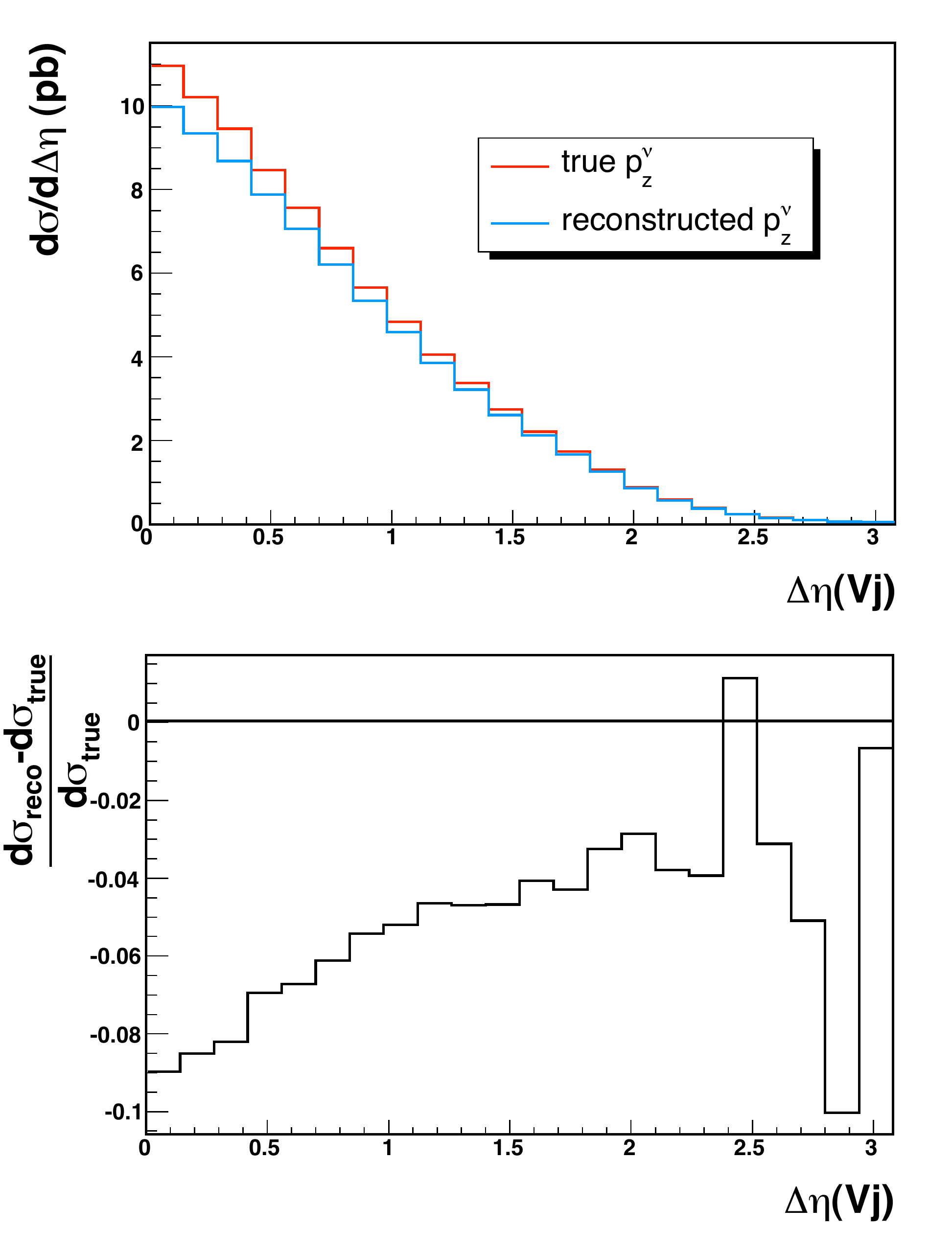}
\hspace*{-2cm}
\caption{Distribution of the invariant mass $M(j_cj_cl\nu)$ (left hand side) and
of the minimum $\Delta \eta$ between any reconstructed heavy boson
and any tag jet (right hand side). The red line is obtained using the actual
longitudinal momentum of the neutrino while the blue line is obtained using the
reconstruucted value. The lower row plots present the relative difference
between the two results.
Cuts as listed in \tbn{tab:cuts_0},with the
addition of $|\Delta\eta(j_fj_b)| > 3.8$ and $
M_{j_cj_c\ell^\pm\nu} > 600 \mbox{ GeV}$.
The numbers refer to the $\mu\nu + 4j$ channel only.}
\label{fig:vrecoVSvtrue}
\end{center}
\end{figure}

If the discriminant of Eq.(\ref{eq:nu_reco}) is negative, which happens only if
the actual momenta satisfy $(p^{\ell}+p^{\nu})^2 > M_W^2$,
it is reset to zero.
The corresponding value of $p_{z}^{\nu}$ is adopted.
This value of $p_{z}^{\nu}$ results in the smallest possible value for
the mass of the $\ell \nu$ pair which is compatible with the measured components
of $p^{\ell}$ and $p^{\nu}$. The corresponding mass is always larger than $M_W$.
If the determinant is positive and the two solutions
for $p_{z}^{\nu}$ have opposite sign we choose the solution whose sign coincides
with that of $p_{z}^{\ell}$. If they have the same sign we choose the
solution with the smallest $\Delta R = \sqrt{\Delta\eta^2 + \Delta \phi^2}$
with the charged lepton.
The reconstructed value is used for computing all physical observables.

The reconstruction procedure detailed above is commonly used by the experimental
collaborations, see for instance Ref.~\cite{CMS_Note}.
While unavoidable, any neutrino reconstruction
necessarily modifies the actual event kinematics and introduces some uncertainty
in the measurement. It should be noticed, however,
that systematic biases introduced
by the reconstruction algorithm, once identified through the analysis of 
Monte Carlo events, can be corrected for when analyzing actual data.
In order to estimate the effect of the neutrino reconstruction on our results,
in Fig.~\ref{fig:vrecoVSvtrue} we show  the distribution of the invariant mass
$M(j_cj_cl\nu)$ (left hand side) and
of the minimum $\Delta \eta$ between any reconstructed heavy boson
and any tag jet (right hand side). These two quantities are the main kinematical
variables which depend on the reconstructed neutrino longitudinal momentum and
which will be used in the analysis. The remaining one is $M(jl\nu)$ which enters
the antitop selection and which normally affects events close to the top pair
production threshold, which is sizably smaller than the typical energy scale at
which $VV$ scattering is studied.
 
The total cross section is about 10\% larger
with actual neutrino momenta than with the reconstructed ones. The difference is
concentrated at small  $M(j_cj_cl\nu)$ masses and small $\Delta \eta$
separations. Since in the following we are going to focus on the region of large
invariant masses and we are going to require a minimum separation 
between any reconstructed heavy boson
and any tag jet, $|\Delta\eta(Vj)| > 0.6$, the reconstruction effects
are modest. We are mainly concerned with the comparison
of the light Higgs and the no--Higgs scenarios which are affected in the same
way by the neutrino reconstruction. Moreover, we are in any case forced
to consider measurable quantities, which can only be defined through
reconstructed variables.
  
\begin{table}[h!tb]
\begin{center}
\begin{tabular}{|c|}
\hline
\textbf{Basic selection cuts} \\
\hline
$70 \mbox{ GeV} < M(j_cj_c) < 100 \mbox{ GeV}$ \\
\hline
$M(j_fj_b)<70 \mbox{ GeV} ; M(j_fj_b)>100 \mbox{ GeV}$ \\
\hline
$|M(jjj;j\ell^\pm\nu_{rec}) - M_{top}| > 15 \mbox{ GeV}$ \\
\hline
$\Delta R(jj) > 0.3$ \\
\hline
\end{tabular}
\caption{These cuts have been adopted in order to separate Boson Fusion
from \toptop and  single--top production and from three--vector--boson production.}
\label{tab:cuts_1}
\end{center}
\end{table}

For very large Higgs masses, all Born diagrams with Higgs propagators become
completely
negligible in the Unitary Gauge we work in
and the expectations for all processes
reduce to those in the $M_H \rightarrow \infty$ limit.

On the generated samples we have applied some basic selection cuts. 
There are a number of backgrounds which can be distinguished from the Boson Fusion
signal because of the mass distribution of their final states.
They are \toptop and  single--top production (see Fig.~\ref{top-diag} and the
left half of Fig.~\ref{fig:VBSbckgr_QCD}) and three--vector--boson production
(see Fig.~\ref{tgc-diag}).
For this purpose,
we have required that no jet triplet satisfies
\be
\label{eq:topcut1}
\vert M_{jjj} - M_t \vert < 15 \,{\mathrm GeV}
\ee
and no jet satisfies 
\be
\label{eq:topcut2}
\vert M_{jl\nu} - M_t \vert < 15 \,{\mathrm GeV}.
\ee

The invariant mass of the two central jets is required to be compatible
with the decay of an electroweak boson, 
\be
\label{eq:vectorcut}
70 \mbox{ GeV} < M(j_cj_c) < 100 \mbox{ GeV}.
\ee
Moreover we have required the mass of the forward and backward jet pair to lie outside
the mass window of the electroweak bosons in order to exclude 
three--vector--boson production. 

At large transverse momentum, jet pairs with mass comparable to the  mass of
electroweak bosons
or even larger can merge into one single jet when an angular measure like $\Delta R(jj)$ is adopted for
reconstructing jets. Therefore we have imposed that all partons satisfy  $\Delta R(jj) > 0.3$.
In \sect{sec:discuss} we discuss in more detail the effect of removing the angular separation constraint
or, on the contrary, of imposing a more stringent requirement $\Delta R(jj) > 0.5$.
For later convenience these further cuts are summarized in \tbn{tab:cuts_1}

%%%%%%%%%%%%%%%%%%%%%%%%%%%%%%%%%%%%%%%%%%%%%%%%%%%%%%%%%%%%%%%%%%%%%%%%%
\section{Pure Electroweak processes}
\label{sec:EW6}

In this section, as a preliminary step,
we compare in some detail the $\ordEW$ sample with a 200 GeV Higgs with the corresponding
sample with the Higgs mass taken to infinity, using a simplified set of cuts
compared to Ref.~\cite{Accomando:2005hz}.
In Ref.~\cite{Accomando:2005hz} the analysis was primarily based on a neural
network approach. Here we prefer to apply a set of physically transparent cuts
which can be generalized to the complete analysis.
The distributions in \cite{Accomando:2005hz} clearly show that, at large invariant masses of the boson--boson pair,
the exact value of the Higgs mass is irrelevant, provided it is
sizably smaller than the boson--boson mass range under investigation.
Therefore the case of a 200
GeV mass discussed in this paper is representative of the full light Higgs
mass range up to a few hundreds GeV.

\begin{table}[hbt]
\begin{center}
\begin{tabular}{|c|c|c|c|c|c|}
\cline{1-5}
\multirow{2}{*}{$M_{cut}$} & \multicolumn{2}{|c|}{no Higgs} & \multicolumn{2}{|c|}{$M_H = 200$ GeV} & \multicolumn{1}{c}{}\\
\cline{2-6}
 & $\sigma$ & events & $\sigma$ & events & ratio  \\
\hline
 400 GeV & 19.875 fb & 1988 & 18.254 fb & 1825 & 1.09 \\
 600 GeV &  9.803 fb &  980 &  7.951 fb &  795 & 1.23 \\
 800 GeV &  4.910 fb &  491 &  3.848 fb &  385 & 1.28 \\
1000 GeV &  2.624 fb &  262 &  2.075 fb &  208 & 1.26 \\
1200 GeV &  1.413 fb &  141 &  1.132 fb &  113 & 1.25 \\
1400 GeV &  0.753 fb &   75 &  0.614 fb &   61 & 1.23 \\
1600 GeV &  0.377 fb &   38 &  0.374 fb &   37 & 1.03 \\
\hline
\end{tabular}
\caption{Integrated $\ordEW$ cross section for $M(j_cj_cl\nu)> M_{cut}$ and number of expected events after one
year at high luminosity ($\mathcal{L} = 100 \mbox{ fb}^{-1}$) with the set of cuts listed in
\tbnsc{tab:cuts_0}{tab:cuts_1}. The numbers refer to the $\mu\nu + 4j$ channel only.}
\label{tab:xsec_cuts1}
\end{center}
\end{table}

We start presenting in \tbn{tab:xsec_cuts1} the integrated cross section for $M(j_cj_cl\nu)> M_{cut}$ and 
the number of expected events after one
year at high luminosity ($\mathcal{L} = 100 \mbox{ fb}^{-1}$) for several $M_{cut}$ values. These results have been
obtained with the set of cuts listed in \tbnsc{tab:cuts_0}{tab:cuts_1} and refer to a single
lepton family.
Taking $M_{cut} = 1000$ GeV as an example we have a SM prediction of about 200
events per year and an excess of
about 50 events in the no Higgs case.

%\begin{table}[h!tb]
\begin{table}[htb]
\begin{center}
\begin{tabular}{|c|}
\hline
\textbf{Selection cuts} \\
\hline
$|\Delta\eta(j_fj_b)| > 4.0$ \\
\hline
$|\eta(\ell\nu)| < 2.0$ \\
\hline
\end{tabular}
\caption{Selection cuts applied in the analysis of the purely
electroweak sample in addition to the basic cuts in \tbnsc{tab:cuts_0}{tab:cuts_1}}
\label{tab:cuts_2}
\end{center}
\end{table}

%\begin{figure}[h!tb]
\begin{figure}[hbt]
\begin{center}
\hspace*{-2cm}
\includegraphics[width=8.3cm,height=6.5cm]{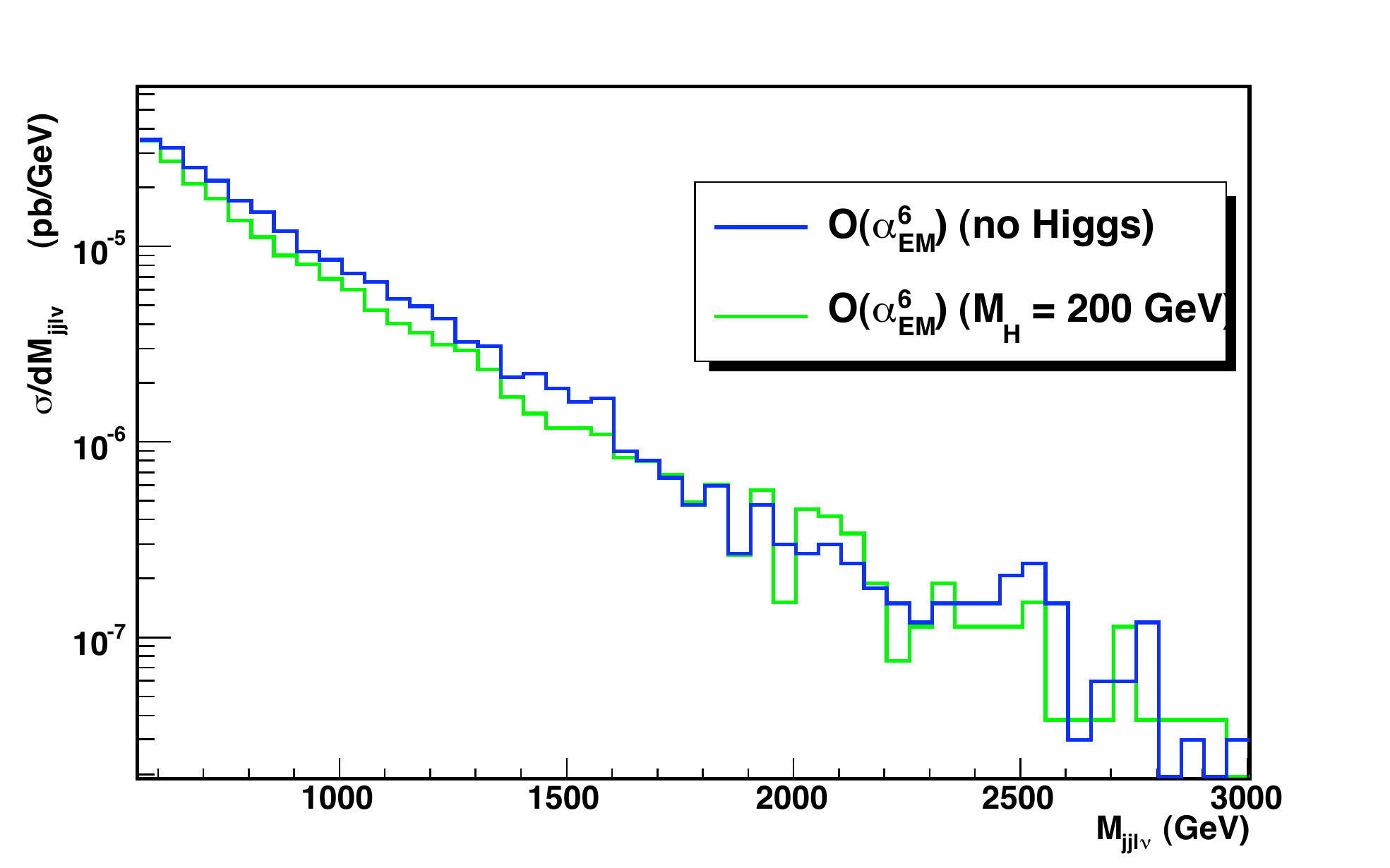}
\hspace*{-0.6cm}
\includegraphics[width=8.3cm,height=6.5cm]{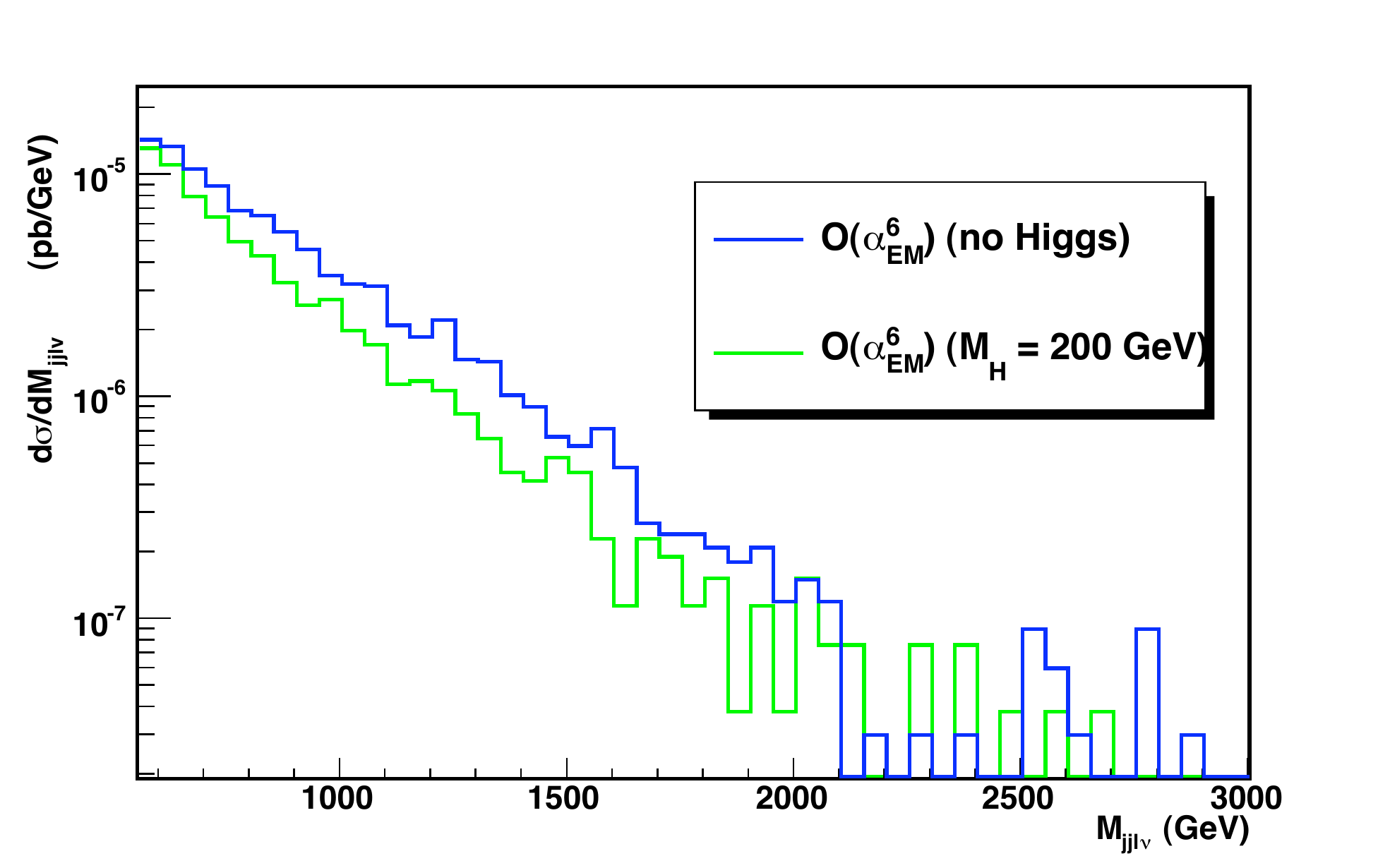}
\hspace*{-2cm}
\caption{Invariant mass $M(j_cj_cl\nu)$ with the set of cuts listed in
\tbns{tab:cuts_0}{tab:cuts_1} on the left hand side and with the addition of the cuts in \tbn{tab:cuts_2}
on the right hand side. The numbers refer to the $\mu\nu + 4j$ channel only.}
\label{fig:Mjjlv2}
\end{center}
\end{figure}

\begin{table}[htb]
\begin{center}
\begin{tabular}{|c|c|c|c|c|c|}
\cline{1-5}
\multirow{2}{*}{$M_{cut}$} & \multicolumn{2}{|c|}{no Higgs} & \multicolumn{2}{|c|}{$M_H = 200$ GeV} 
                 & \multicolumn{1}{c}{}\\
\cline{2-6}
 & $\sigma$ & events & $\sigma$ & events & ratio  \\
\hline
 400 GeV & 8.327 fb & 833 & 6.511 fb & 651 & 1.28 \\
 600 GeV & 4.138 fb & 414 & 2.829 fb & 283 & 1.46 \\
 800 GeV & 2.129 fb & 213 & 1.272 fb & 127 & 1.68 \\
1000 GeV & 1.111 fb & 111 & 0.626 fb &  63 & 1.76 \\
1200 GeV & 0.594 fb &  59 & 0.316 fb &  32 & 1.84 \\
1400 GeV & 0.283 fb &  28 & 0.159 fb &  16 & 1.75 \\
1600 GeV & 0.137 fb &  14 & 0.079 fb &   8 & 1.75 \\
\hline
\end{tabular}
\caption{Integrated $\ordEW$ cross section for $M(j_cj_cl\nu)> M_{cut}$ and number of expected events after
one year at high luminosity ($\mathcal{L} = 100 \mbox{ fb}^{-1}$) with the set of cuts listed in
\tbns{tab:cuts_0}{tab:cuts_1} and \ref{tab:cuts_2}.
The numbers refer to the $\mu\nu + 4j$ channel only.}
\label{tab:xsec_cuts2}
\end{center}
\end{table}

The distribution of the mass of the charged lepton, (reconstructed) neutrino and the two most central jets is shown on the
left hand side of Fig.~\ref{fig:Mjjlv2}.
The separation between the two cases, which we estimate from the ratio of the
expected number of signal events in the two scenarios,
can be increased imposing a large separation between the tag jets, which eliminates
most of the non-scattering background, and requiring the reconstructed $W$ which decays 
leptonically  to be rather central, since the
vector bosons tend to be produced more centrally in the absence of the Higgs boson.
The applied cuts are summarized in \tbn{tab:cuts_2} and the corresponding cross sections as a function of $M_{cut}$ 
are presented in \tbn{tab:xsec_cuts2}. The distribution of the VV mass is shown on the right hand side of
Fig.~\ref{fig:Mjjlv2}. The improvement in the separation of the two curves is
clearly visible.

%\vspace{-2cm}

%\eject
%\newpage

%%%%%%%%%%%%%%%%%%%%%%%%%%%%%%%%%%%%%%%%%%%%%%%%%%%%%%%%%%%%%%%%%%%%%%%%%
\section{The VVjj QCD background}
\label{sec:QCDbkg}

The $\ordQCD$ background is large. With the basic generation cuts in
\tbn{tab:cuts_0} it is
about one hundred times larger than the $\ordEW$ contribution and it is
dominated
by \toptop and single--top production. Once the top--veto in 
\tbn{tab:cuts_1} is imposed this background is severely reduced.

\begin{figure}[htb]
%\begin{center}
%\includegraphics[width=0.48\textwidth,height=5.5cm]{analisi_giuseppe3.2/fig/c_pT_lv_cuts5.pdf}
%\includegraphics[width=0.48\textwidth,height=5.5cm]{analisi_giuseppe3.2/fig/c_pT_lv_cuts5_NORM1.pdf}
%\includegraphics[width=0.48\textwidth,height=5.5cm]{analisi_giuseppe3.2/fig/c_DEta_Vj_cuts5.pdf}
%\includegraphics[width=0.48\textwidth,height=5.5cm]{analisi_giuseppe3.2/fig/c_DEta_Vj_cuts5_NORM1.pdf}
\centering
\subfigure[]{
\hspace*{-2.1cm}
\includegraphics*[width=8.3cm,height=6.5cm]{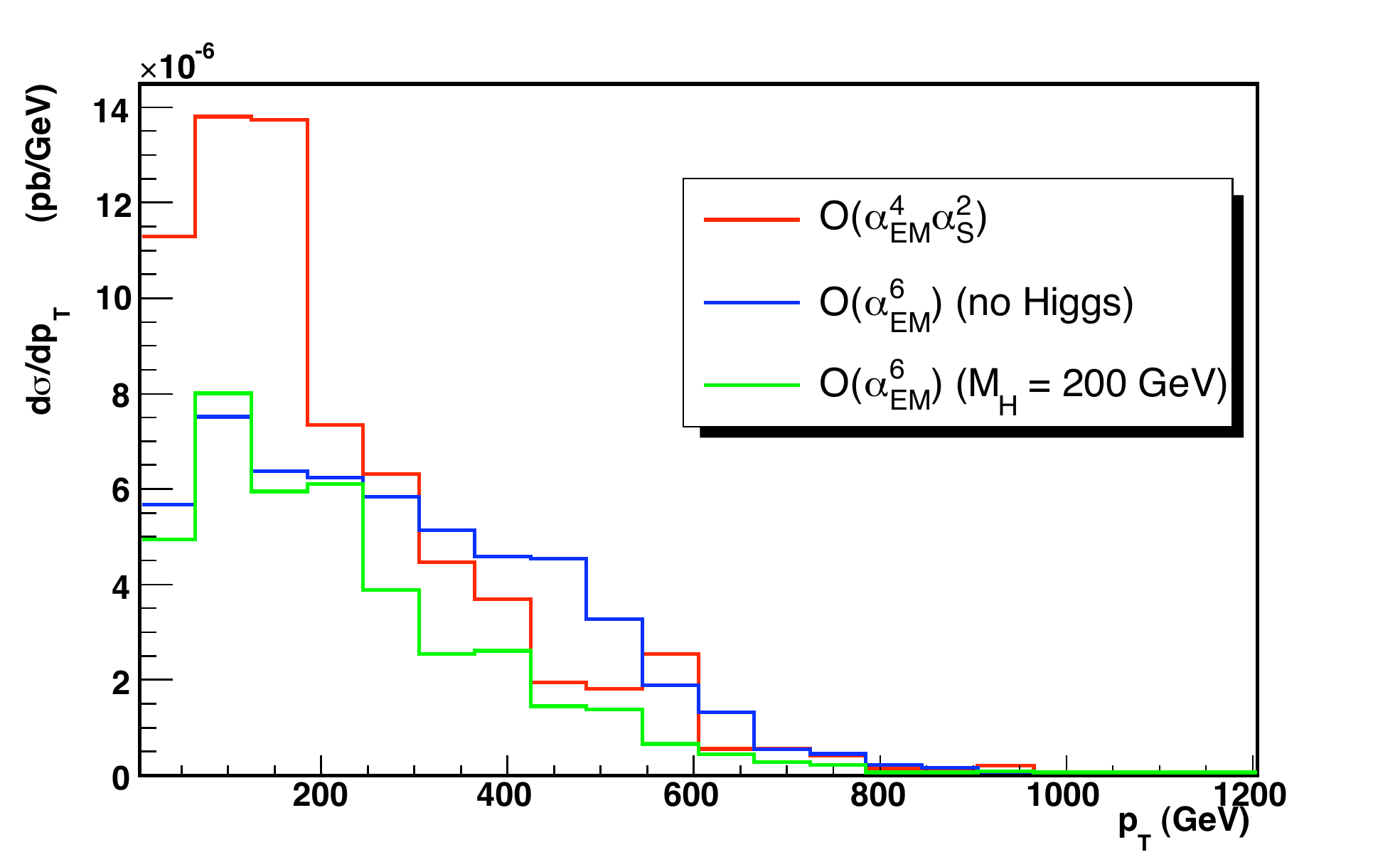}
\hspace*{-0.7cm}
\includegraphics*[width=8.3cm,height=6.5cm]{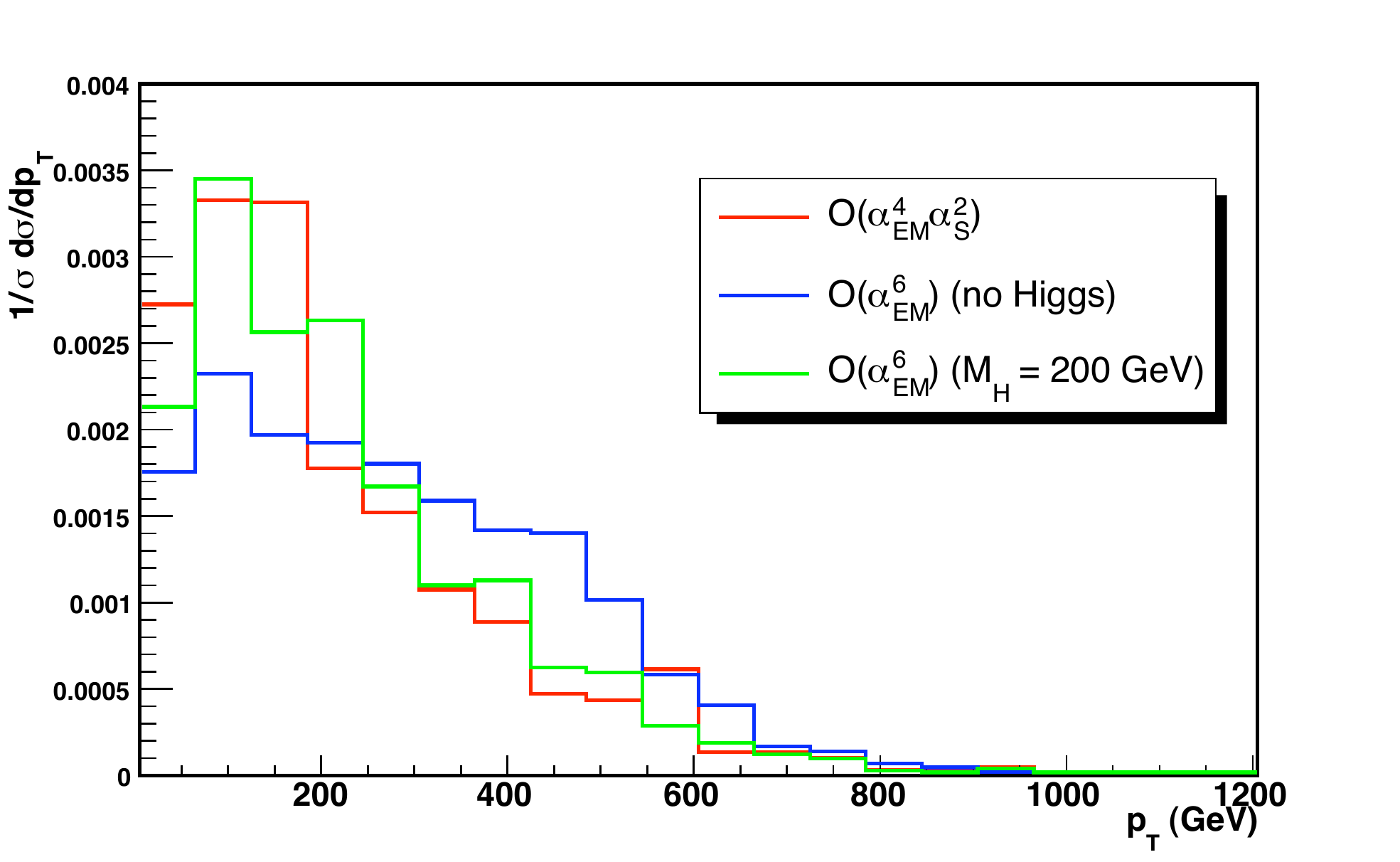}
\hspace*{-3cm}
}
\subfigure[]{
\hspace*{-2.1cm}
\includegraphics*[width=8.3cm,height=6.5cm]{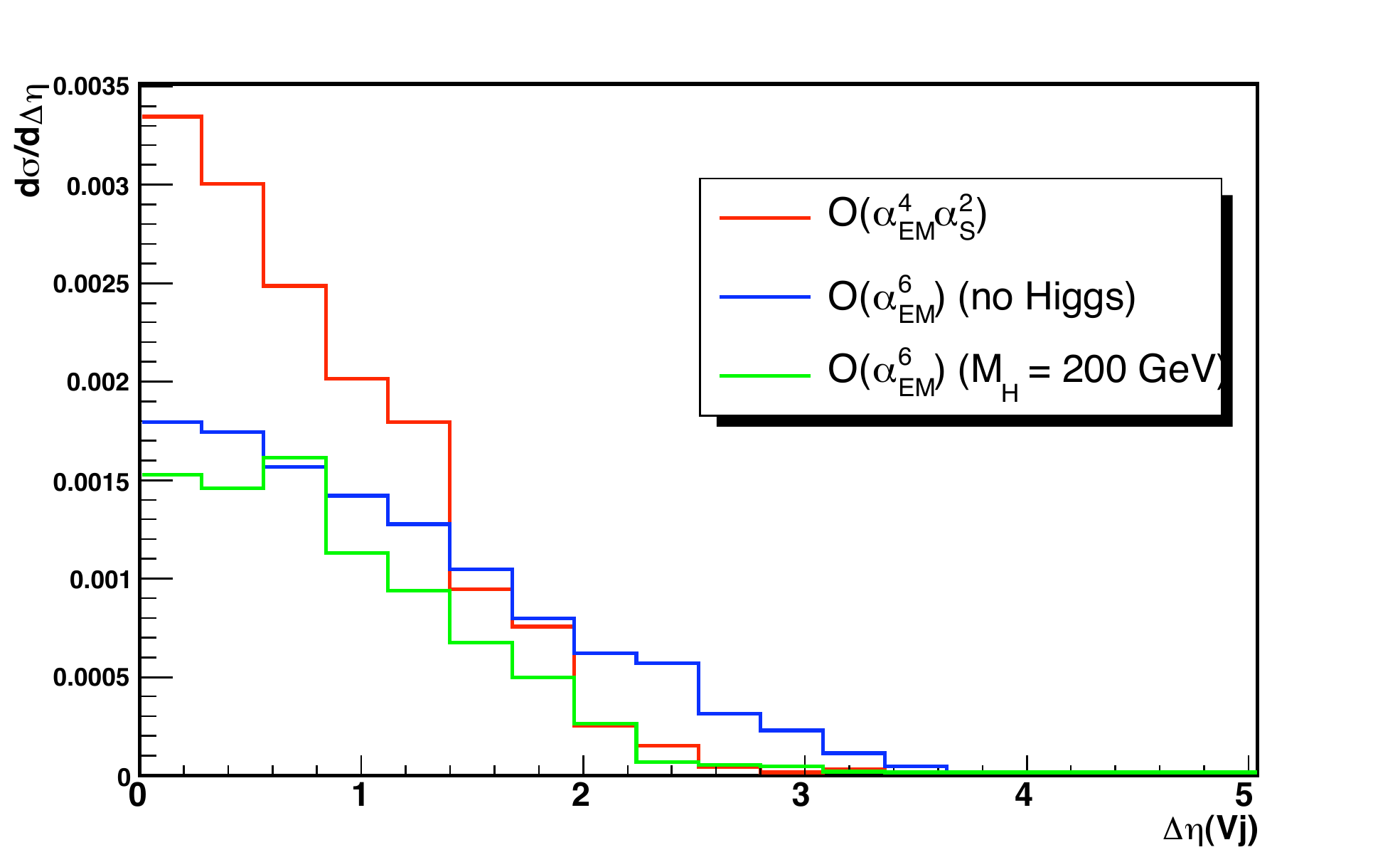}
\hspace*{-0.7cm}
\includegraphics*[width=8.3cm,height=6.5cm]{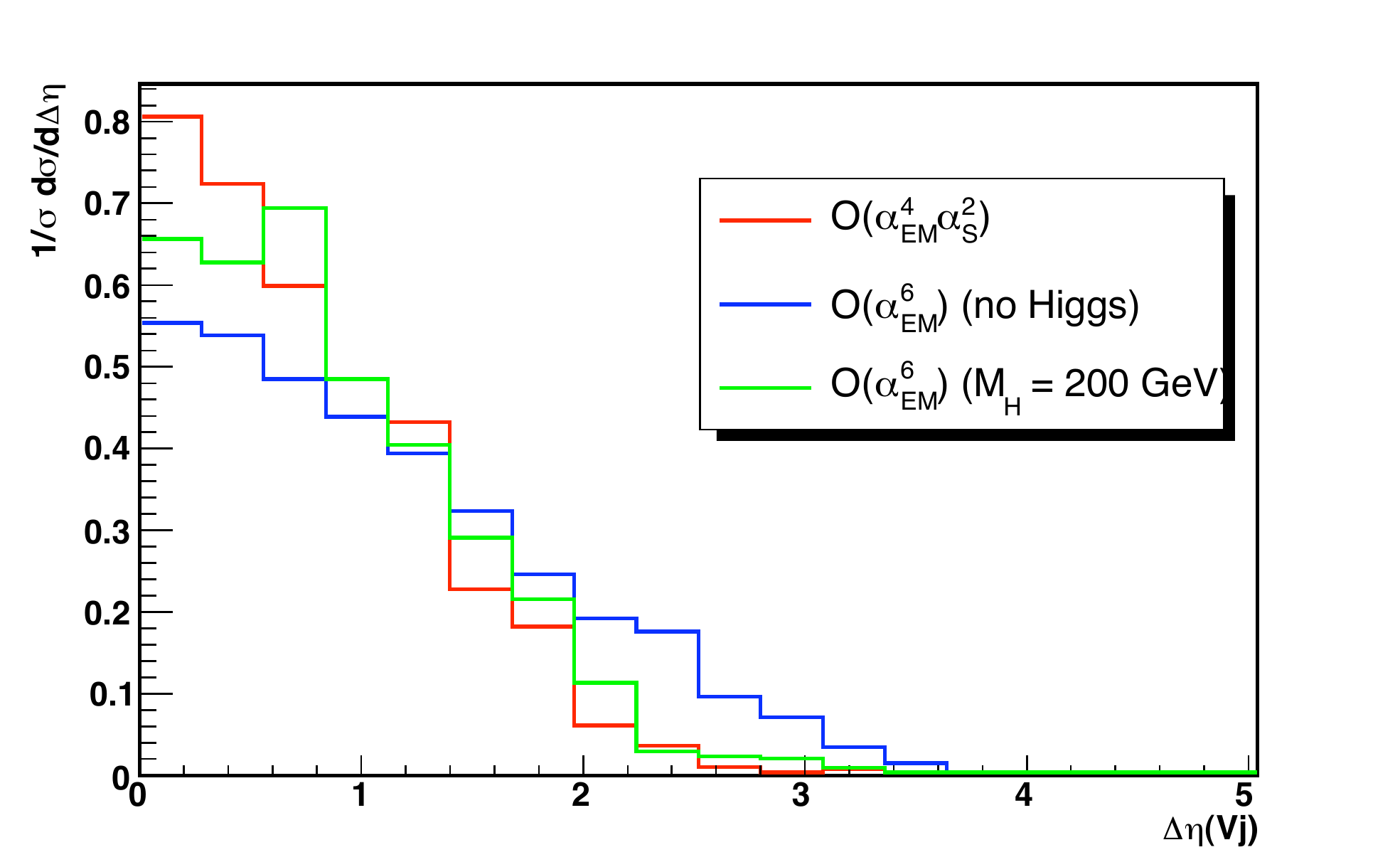}
\hspace*{-3cm}
}
%\end{center}
\caption{\textit{Top:} distribution of the transverse momentum of the $W$
reconstructed from leptons.
\textit{Bottom:} minimum $\Delta \eta$ between any reconstructed heavy boson
and any tag jet. Cuts as listed in \tbns{tab:cuts_0}{tab:cuts_1},with the
addition of $|\Delta\eta(j_fj_b)| > 4.0$
The plots on the right are the analogous of the ones on the left, but
normalized to one.
The numbers refer to the $\mu\nu + 4j$ channel only and to the region
$M(j_cj_cl\nu) > 800 \mbox{GeV}$}
\label{pTW+EtaVj}
\end{figure}

%\begin{table}[h!tb]
\begin{table}[htb]
\begin{center}
\begin{tabular}{|c|}
\hline
\textbf{Selection cuts I} \\
\hline
$|\Delta\eta(j_fj_b)| > 4.0$ \\
\hline
$p_T(\ell\nu) > 200 \mbox{ GeV}$ \\
\hline
$|\Delta\eta(Vj)| > 0.6$ \\
\hline
\end{tabular}
\caption{Selection cuts applied in the analysis of the $\ordEW$ and $\ordQCD$
samples in addition to the cuts in \tbnsc{tab:cuts_0}{tab:cuts_1}.}
\label{tab:cuts_3}
\end{center}
\end{table}

Even in the absence of $W+4j$ contributions, the previous selection procedure 
does not ensure a good separation between typical expectations from the 
Standard Model with a light Higgs and the benchmark no Higgs 
scenario. This is essentially due to the fact that the contribution of the QCD 
diagrams with a gluon exchanged in the $t$ channel,
Fig.\ref{fig:VBSbckgr_QCD}(c,d), is not substantially affected.
However, a preliminary study at $\ordEW+\ordQCD$ has devised a collection of 
observables which appear sensitive to the different kinematics of this kind of
background\cite{Ambroglini:2009mg,Accomando:2006vj}.
More specifically, it has been found that the transverse momentum of 
the reconstructed $\ell\nu$ pair and the angular separation between tag jets
and the final state bosons are suitable signal-to-background 
discriminators.
It was also realized that, in the presence of the
$\ordQCD$ background, requiring a large transverse momentum of the
reconstructed  $\ell\nu$ pair is more effective than the centrality requirement
used previously.
The distributions for $p_T(\ell\nu)$ and $|\Delta\eta(Vj)|$ are shown in 
Fig.~\ref{pTW+EtaVj}.

\begin{table}[htb]
\begin{center}
\begin{tabular}{|c|c|c|c|c|c|}
\cline{1-5}
\multirow{2}{*}{$M_{cut}$} & \multicolumn{2}{|c|}{no Higgs} & \multicolumn{2}{|c|}{$M_H = 200$ GeV} & \multicolumn{1}{c}{}\\
\cline{2-6}
 & $\sigma$ & events & $\sigma$ & events & ratio \\
\hline
 600 GeV & 4.530 (2.640) fb & 453 (264) & 3.500 (1.427) fb & 350 (143) & 1.29 (1.85) \\
 800 GeV & 2.584 (1.553) fb & 258 (155) & 1.763 (0.754) fb & 176 ( 75) & 1.47 (2.07) \\
1000 GeV & 1.354 (0.861) fb & 135 ( 86) & 0.881 (0.404) fb &  88 ( 40) & 1.53 (2.15) \\
1200 GeV & 0.750 (0.478) fb &  75 ( 48) & 0.499 (0.219) fb &  50 ( 22) & 1.50 (2.18) \\
1400 GeV & 0.357 (0.228) fb &  36 ( 23) & 0.270 (0.132) fb &  27 ( 13) & 1.33 (1.77) \\
1600 GeV & 0.203 (0.124) fb &  20 ( 12) & 0.186 (0.077) fb &  19 (  8) & 1.05 (1.50) \\
\hline
\end{tabular}
\caption{Integrated $\ordEW+\ordQCD$ cross section for $M(j_cj_cl\nu)> M_{cut}$ and number of expected
events after one year at high luminosity ($\mathcal{L} = 100 \mbox{ fb}^{-1}$) with the set of cuts listed in
\tbnsc{tab:cuts_0}{tab:cuts_1} and \tbn{tab:cuts_3}.
In parentheses the results for the $\ordEW$
contribution alone is also given.
Interferences between the different perturbative orders are neglected.
The numbers refer to the $\mu\nu + 4j$ channel only.}
\label{tab:xsec_cuts3}
\end{center}
\end{table}

\begin{figure}[htb]
\begin{center}
\hspace*{-2cm}
\includegraphics[width=8.3cm,height=6.5cm]{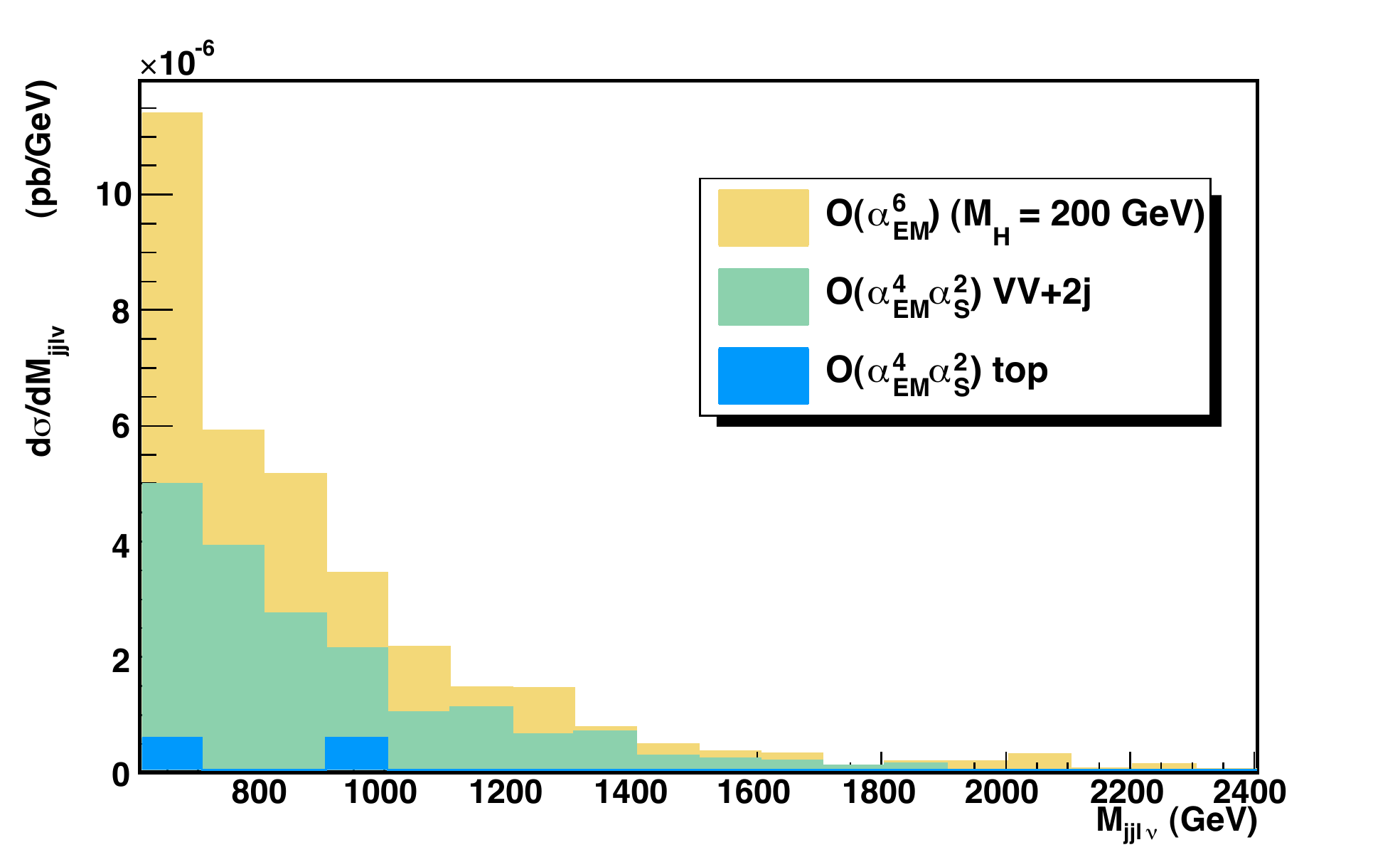}
\hspace*{-1cm}
\includegraphics[width=8.3cm,height=6.5cm]{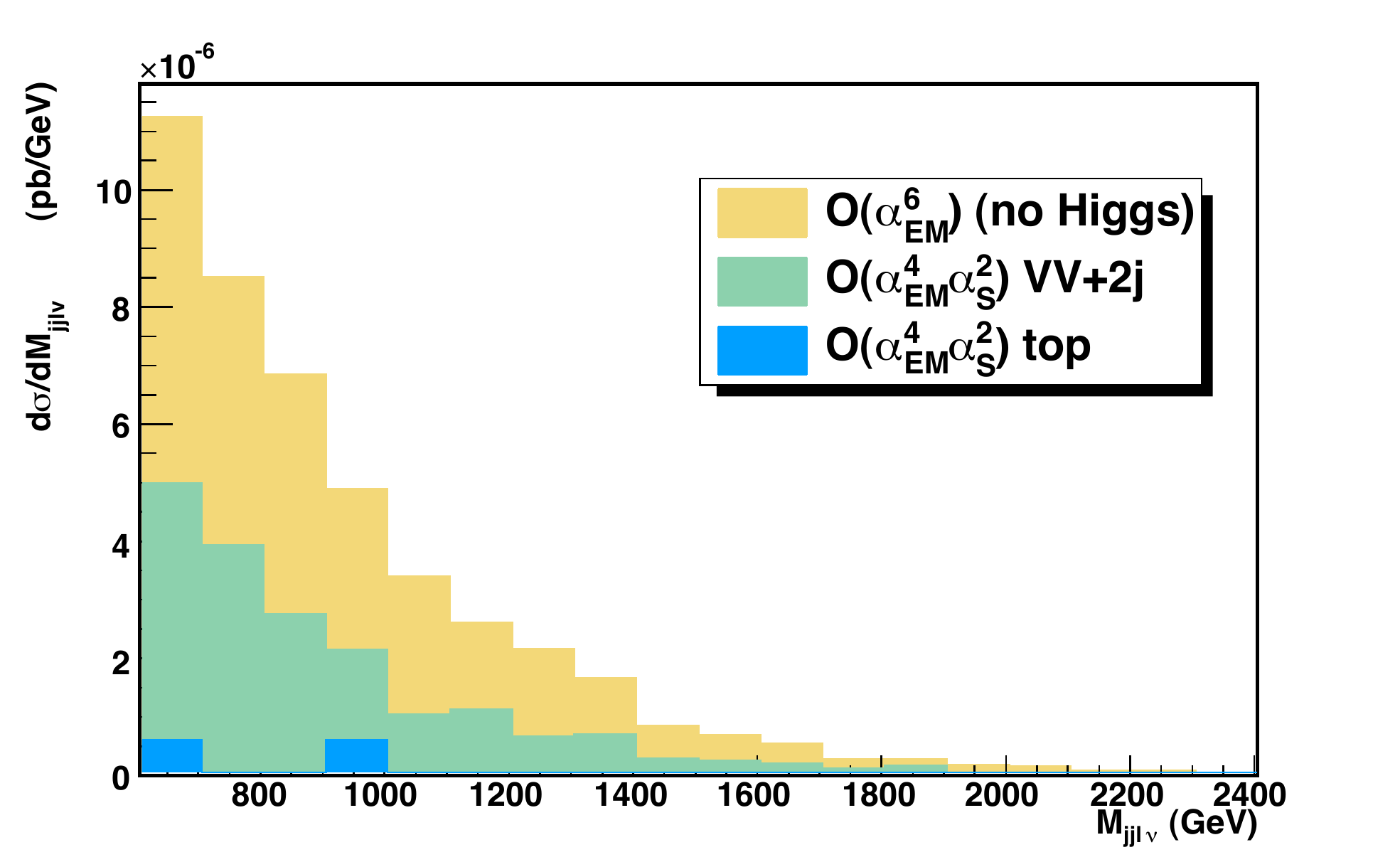}
\hspace*{-2.5cm}
\end{center}
\caption{Invariant mass distribution of the two leptons and the two most 
central jets in the Standard Model with a light Higgs (on the left) and in the 
no--Higgs scenario (on the right). The cuts applied are listed in 
\tbnsc{tab:cuts_0}{tab:cuts_1} and \tbn{tab:cuts_3}.
$\ordEW$ (EW) and $\ordQCD$ (QCD) contributions to
the differential cross section have been isolated and are shown separately.
The $\ordQCD$ contributions are further split into \textit{top background} (in blue)
and $VV+2j$ (in green). The numbers refer to the $\mu\nu + 4j$ channel only.}
\label{fig:Mjjlv4}
\end{figure}

The integrated cross section for $M(j_cj_cl\nu)> M_{cut}$ and number of expected events after one
year at high luminosity ($\mathcal{L} = 100 \mbox{ fb}^{-1}$) for a number of $M_{cut}$ values obtained
with the cuts in \tbnsc{tab:cuts_0}{tab:cuts_1} and \tbn{tab:cuts_3}
are presented in \tbn{tab:xsec_cuts3}.
In brackets the results for the $\ordEW$ contribution alone is also given for the same set of cuts.
The corresponding distribution of the mass of the charged lepton, (reconstructed) neutrino and the
two most central jets is shown on the left hand side of Fig.~\ref{fig:Mjjlv4} for a Higgs mass of 200 GeV
(on the left) and for the no Higgs case (on the right).
Comparing the results in \tbn{tab:xsec_cuts3} with those in \tbn{tab:xsec_cuts2}
shows that, even with the inclusion of the 
$\ordQCD$ background which contributes equally to the two scenarios,
we have been able to maintain a good ratio between the expected number of events
in the two cases. If one takes into account only the EW processes the ratio
has actually improved.
The number of excess
events in the no Higgs case has remained almost constant, particularly at large
$M_{cut}$. The additional cuts in  \tbn{tab:cuts_3} lead to
a reduction of the number of signal events by about 45\% at
$M_{cut} = 600 \mbox{ GeV}$ and by about 15\%  at $M_{cut} = 1000 \mbox{ GeV}$
in the no Higgs case. For a light Higgs the reduction is slightly larger.
Fig.~\ref{fig:Mjjlv4} shows that the top--related background has been reduced to a
negligible level and that the bulk of the background is now 
given by $VV+2j$ production.

%%%%%%%%%%%%%%%%%%%%%%%%%%%%%%%%%%%%%%%%%%%%%%%%%%%%%%%%%%%%%%%%%%%%%%%%%
\begin{figure}[hbt]
%\begin{figure}[h!tb]
\begin{center}
\includegraphics[width=11cm,height=8cm]{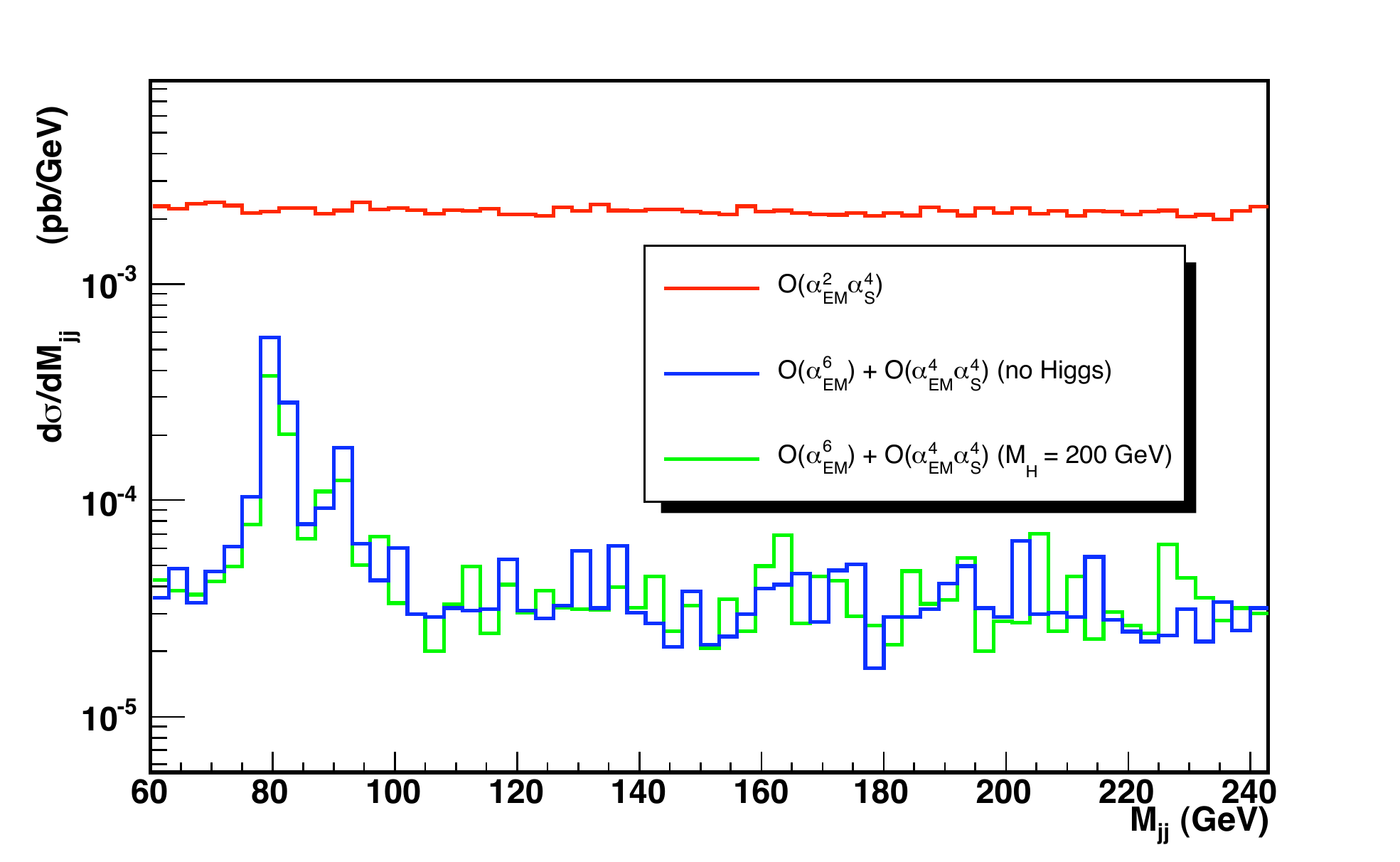}
\end{center}
\caption{Distribution of the invariant mass of the most central jets. Cuts as in 
\tbnsc{tab:cuts_0}{tab:cuts_1} and \tbn{tab:cuts_3},
with the exclusion of the constraint
$70 \mbox{ GeV} < M_{j_cj_c} < 100 \mbox{ GeV}$, in the mass region 
$M_{j_cj_c\ell^\pm\nu} > 600 \mbox{ GeV}$.}
\label{fig:Mjcjc4}
\end{figure}

\section{Full analysis}
\label{sec:full}

In this section we finally analyze all contributions to $\ell\nu + 4j$ simultaneously,
building on the experience gained from the study of partial samples in Sect.~\ref{sec:EW6}
and \ref{sec:QCDbkg}.
In the following we will concentrate on the large invariant mass region
for the VV pair $M_{j_cj_c\ell^\pm\nu} > 600 \mbox{ GeV}$.
In this section and the following one we will consider as signal the sum of the
$\ordEW$+$\ordQCD$ contributions while we refer to the $\ordQCDsq$ processes as
background. While this does modify the standard statistical significance,
$S/\sqrt{B}$, which will turn out to be rather large,
it does not affect the more refined treatment in terms of
confidence levels that we are going to discuss shortly.

\begin{figure}[htb]
\centering
\subfigure[]{
\hspace*{-2.1cm}
\includegraphics*[width=8.3cm,height=6.5cm]{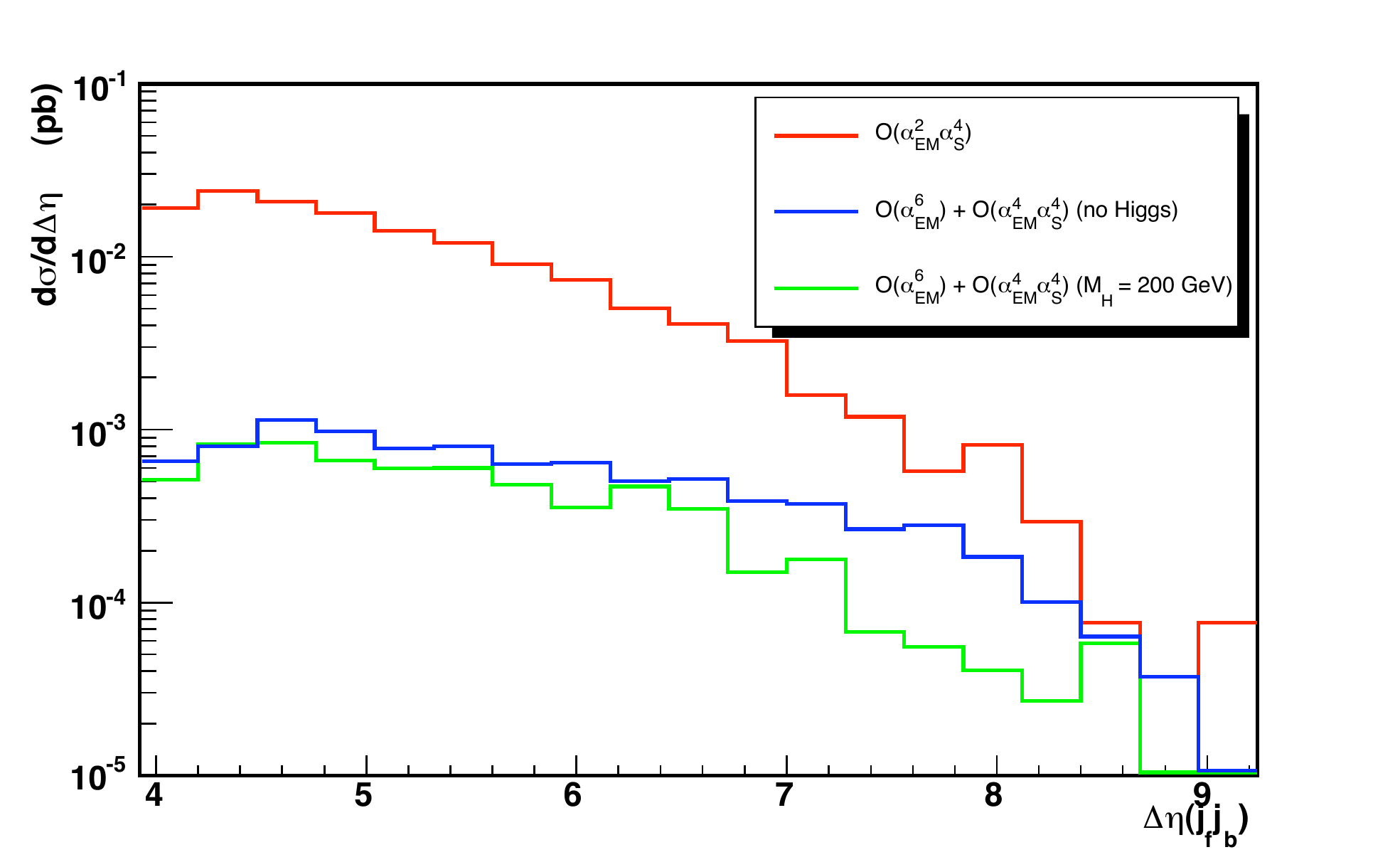}
\hspace*{-0.7cm}
\includegraphics*[width=8.3cm,height=6.5cm]{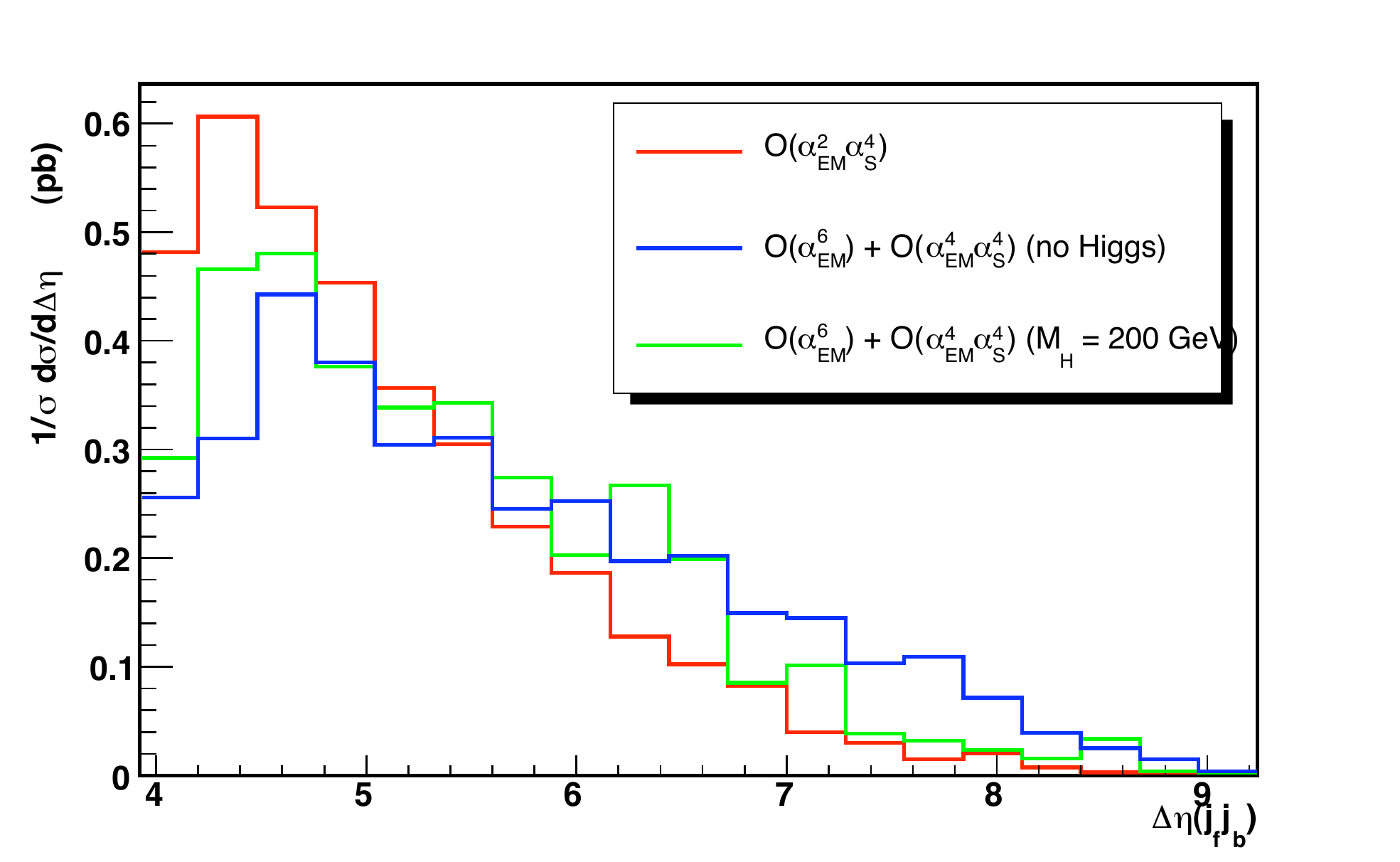}
\hspace*{-3cm}
}
\subfigure[]{
\hspace*{-2.1cm}
\includegraphics*[width=8.3cm,height=6.5cm]{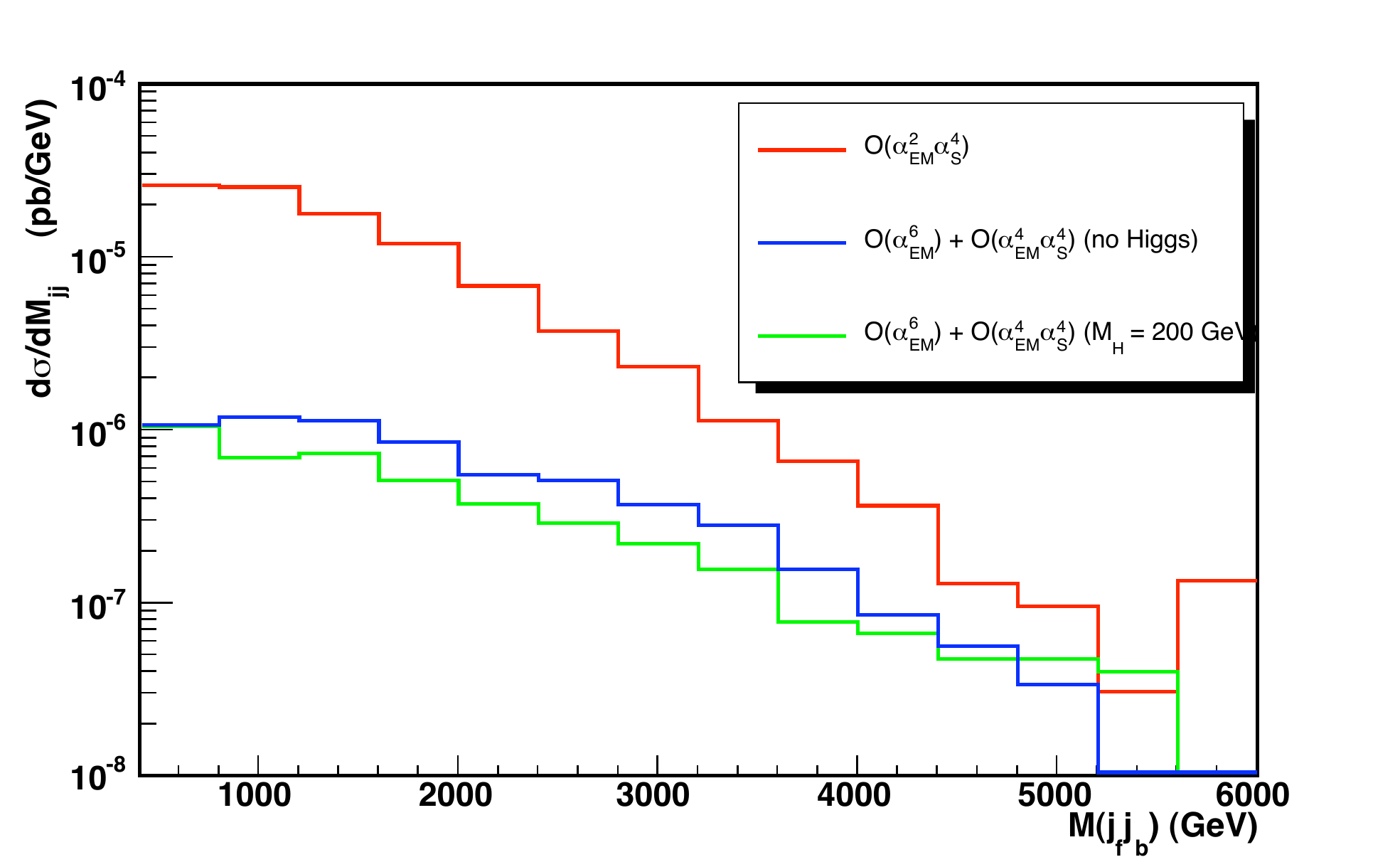}
\hspace*{-0.7cm}
\includegraphics*[width=8.3cm,height=6.5cm]{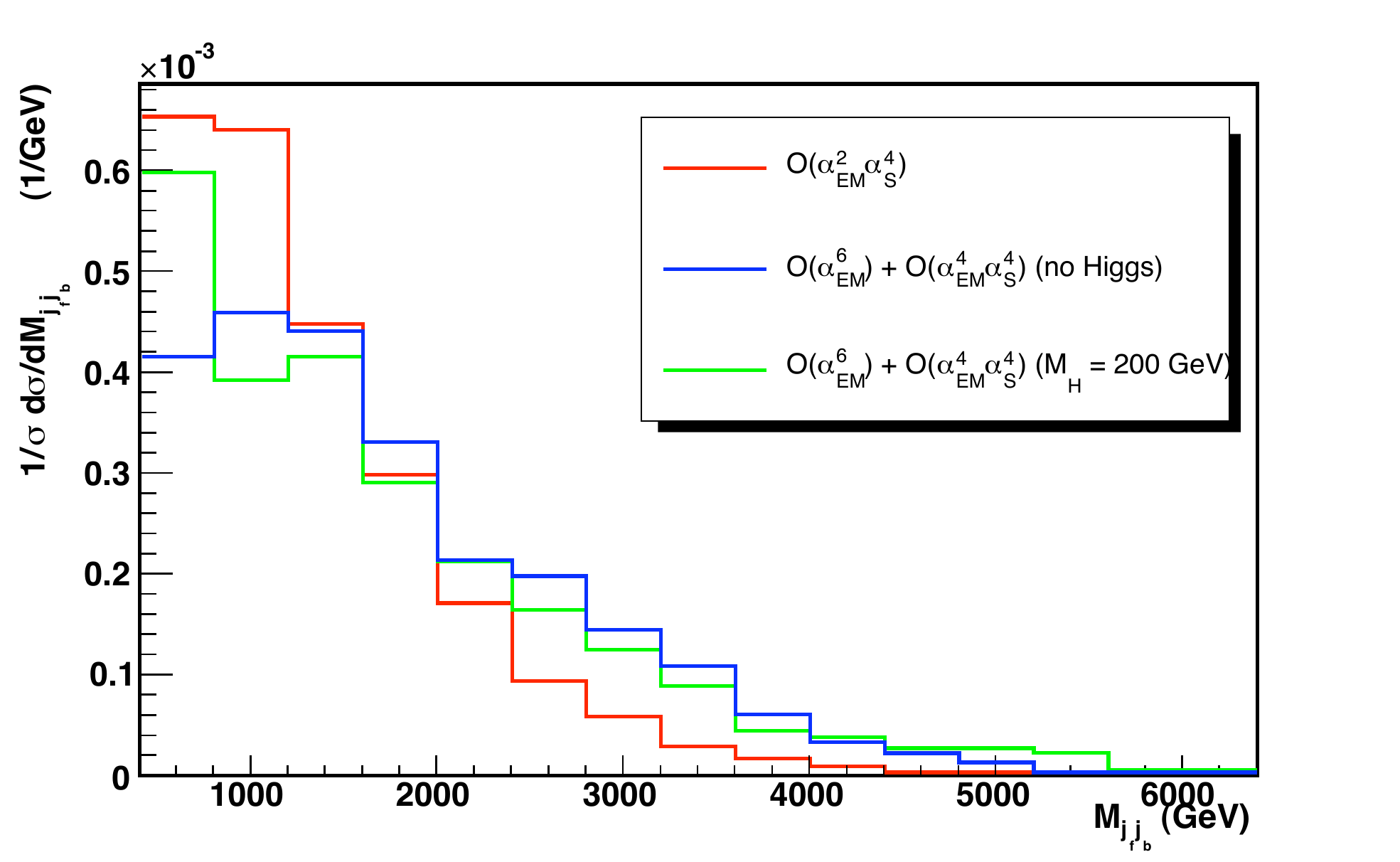}
\hspace*{-3cm}
}
\caption{\textit{From the top:} distribution of the difference in pseudorapidity
between tag jets and of the invariant mass of the two tag jets with the cuts
of \tbnsc{tab:cuts_0}{tab:cuts_1} and of \tbn{tab:cuts_3}.
The plots on the right are
the same as those on the left, but normalized to one. The numbers refer to the
$\mu\nu + 4j$ channel only and to the region
$M(j_cj_cl\nu) > 800 \mbox{GeV}$.
Interferences between the two perturbative orders are neglected.}
\label{fig:plots_step2_1}
\end{figure}

\begin{table}[htb]
\begin{center}
\begin{tabular}{|c|c|c|c|c|c|}
\cline{1-5}
\multirow{2}{*}{$M_{cut}$} & \multicolumn{2}{|c|}{no Higgs} & \multicolumn{2}{|c|}{$M_H = 200$ GeV} & \multicolumn{1}{c}{}\\
\cline{2-6}
 & $\sigma$ & events & $\sigma$ & events & ratio \\
\hline
 600 GeV & 71.581 fb & 7158 & 70.551 fb & 7055 & 1.01 \\
 800 GeV & 42.303 fb & 4230 & 41.482 fb & 4148 & 1.02 \\
1000 GeV & 25.359 fb & 2536 & 24.886 fb & 2489 & 1.02 \\
1200 GeV & 15.817 fb & 1582 & 15.566 fb & 1557 & 1.02 \\
1400 GeV &  9.982 fb &  998 &  9.895 fb &  990 & 1.01 \\
1600 GeV &  6.506 fb &  651 &  6.489 fb &  649 & 1.00 \\
\hline
\end{tabular}
\caption{Integrated $\ordEW+\ordQCD+\ordQCDsq$ cross section for $M(j_cj_cl\nu)> M_{cut}$ and number of
expected events after one year at high luminosity ($\mathcal{L} = 100 \mbox{ fb}^{-1}$) with the set of cuts listed
in \tbnsc{tab:cuts_0}{tab:cuts_1} and \tbn{tab:cuts_3} in the 
mass region $M_{j_cj_c\ell^\pm\nu} > 600 \mbox{ GeV}$ and 
$70 \mbox{ GeV} < M(j_cj_c) < 100 \mbox{ GeV}$.
Interferences between the different perturbative orders are neglected.
The numbers refer to the $\mu\nu + 4j$ channel only.}
\label{tab:xsec_cuts3all}
\end{center}
\end{table}

We disregard in the following all effects of the interference among the
different perturbative orders. The interference  betwen the $\ordEW$ and
$\ordQCD$ contributions has been examined for typical generation cuts and found
to be of the order of 1$\div$2\% compared to the sum of the two contributions.
The interference between these two perturbative orders with the $\ordQCDsq$ term
has not been studied but it is expected to be at most of the order of the
percent with respect to the non double resonant background, since most of the
contributions at $\ordQCDsq$ do not have a corresponding term in the other
orders with the same external particles and color configuration.

In Fig.~\ref{fig:Mjcjc4} the invariant mass distribution of the two central jets is shown for the
sum of the signal and $VV+2j$ background in both Higgs scenarios and for the $\ordQCDsq$ background.
The plot demonstrates that the $\ordEW$+$\ordQCD$ contribution is dominated by the production of two
electroweak bosons while no particular peak appears around 100 GeV for the $W+4j$ distribution,
as expected.

Fig.~\ref{fig:Mjcjc4} clearly shows that the $\ordQCDsq$ background is much larger than the VV scattering signal,
with the present set of selection cuts,
even in the neighborhood of the electroweak boson mass peaks and taking into account only
large invariant masses for the pair of reconstructed bosons.
In order to fully appreciate the degree to which the signal is overwhelmed by the $W+4j$,
the finite resolution in the jet pair invariant mass should be taken into account and signal and background should
be compared after integrating over a reasonable mass window around the vector boson peaks.

The integrated cross section for the sum of all three perturbative orders 
$M(j_cj_cl\nu)> M_{cut}$ and the number of expected events after one
year at high luminosity ($\mathcal{L} = 100 \mbox{ fb}^{-1}$) for a number of $M_{cut}$ values 
are shown in \tbn{tab:xsec_cuts3all}. These results have been obtained
with the cuts in \tbnsc{tab:cuts_0}{tab:cuts_1} and \tbn{tab:cuts_3}. In particular they refer to the 
mass window $70 \mbox{ GeV} < M_{j_cj_c} < 100 \mbox{ GeV}$.
The excess of events in the no Higgs case is within
the statistical uncertainty of the SM prediction.

\begin{figure}[htb]
\centering
\vspace*{-1.5cm}
\subfigure[]{
\hspace*{-2.1cm}
\includegraphics*[width=8.3cm,height=6.2cm]{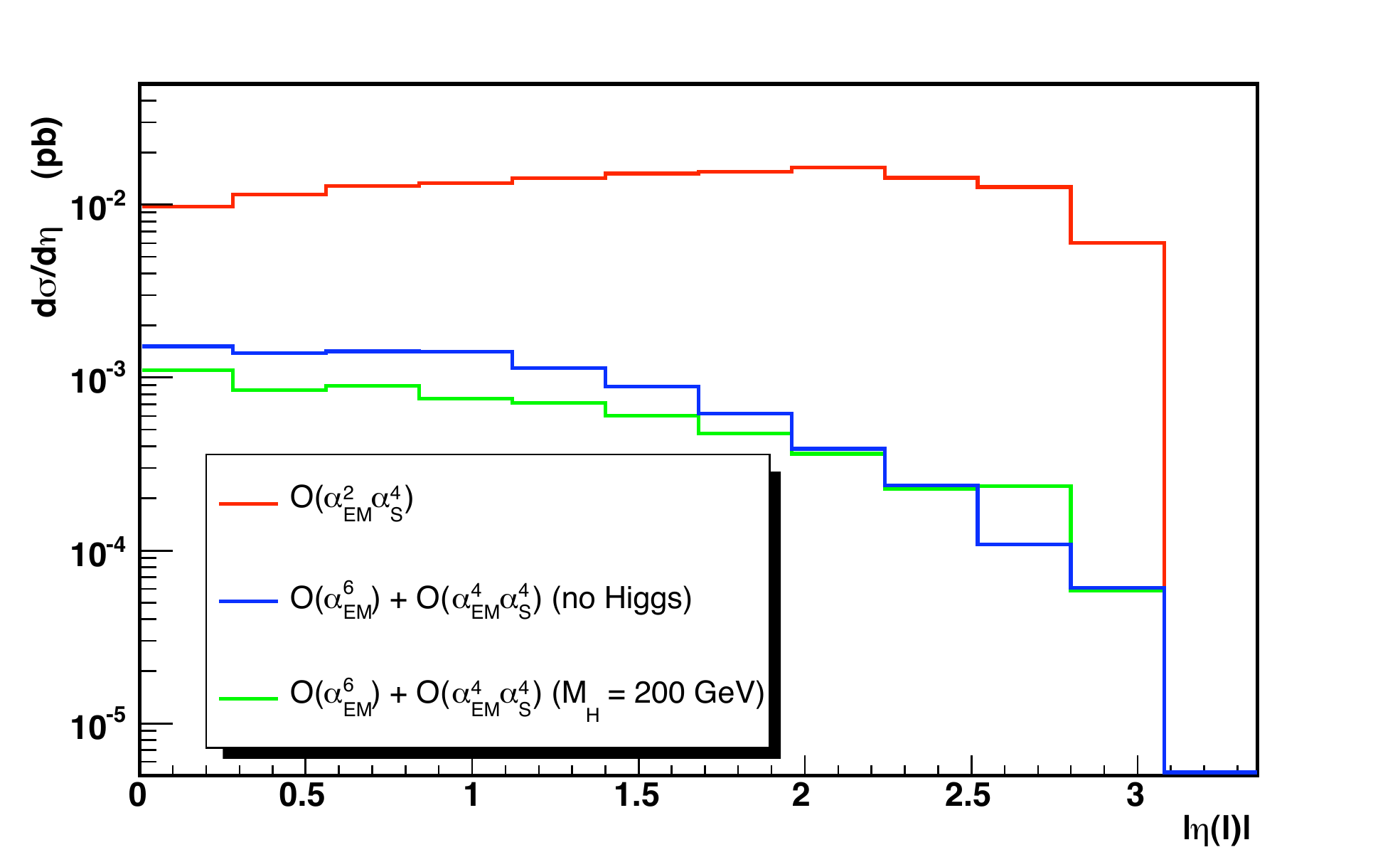}
\hspace*{-0.7cm}
\includegraphics*[width=8.3cm,height=6.2cm]{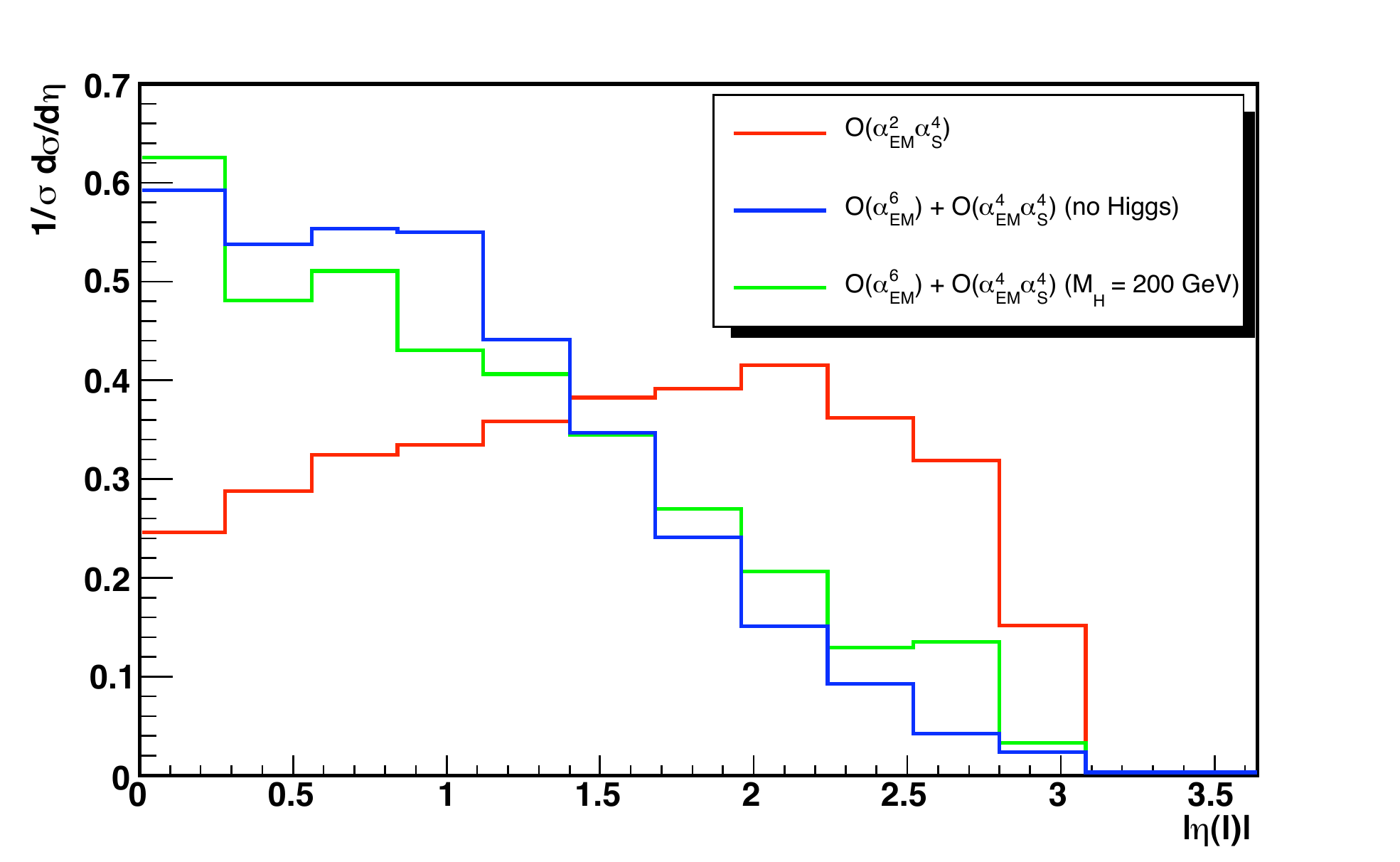}
\hspace*{-3cm}
}
\vspace{-0.4cm}
\subfigure[]{
\hspace*{-2.1cm}
\includegraphics*[width=8.3cm,height=6.2cm]{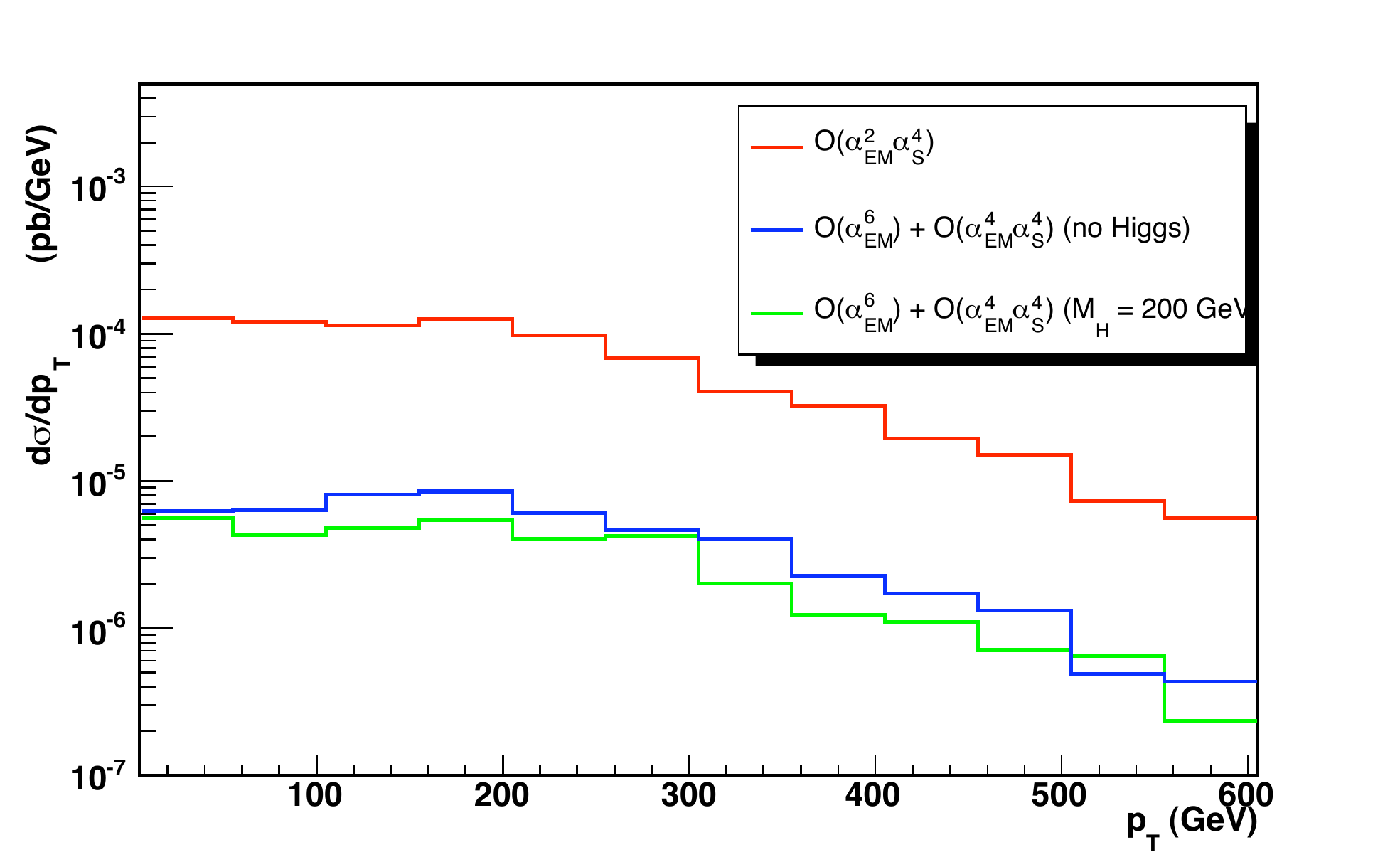}
\hspace*{-0.7cm}
\includegraphics*[width=8.3cm,height=6.2cm]{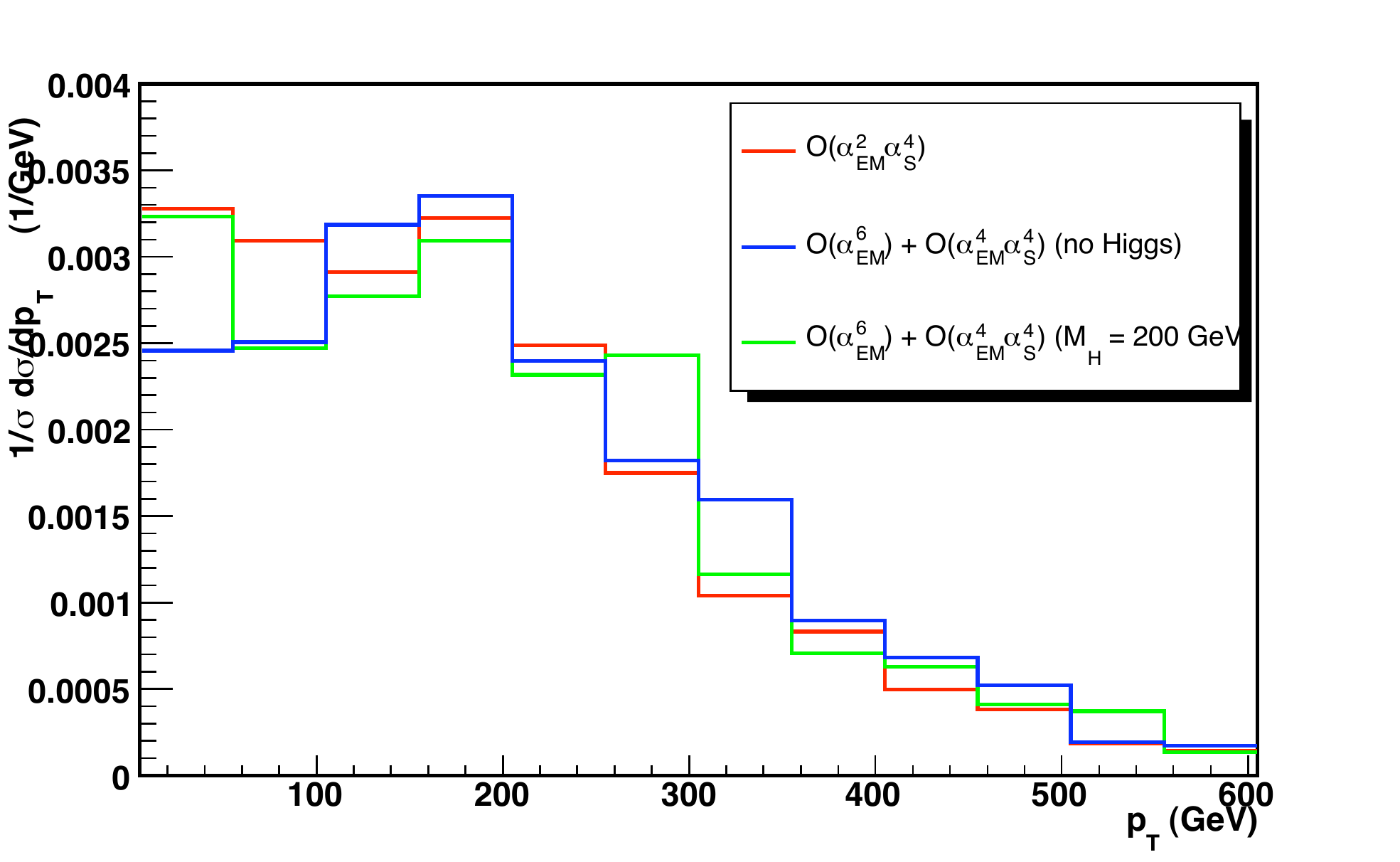}
\hspace*{-3cm}
}
\vspace{-0.4cm}
\subfigure[]{
\hspace*{-2.1cm}
\includegraphics*[width=8.3cm,height=6.2cm]{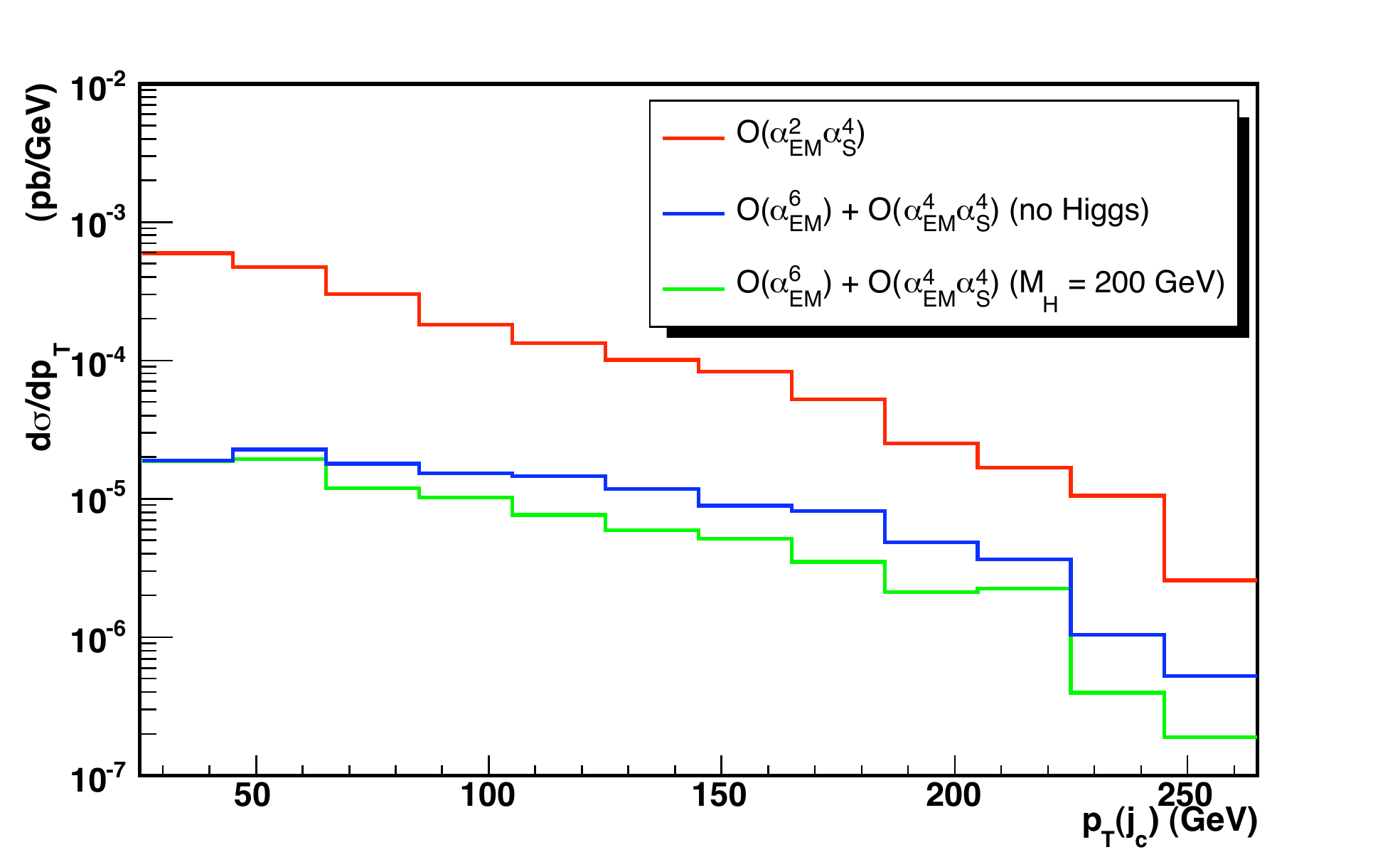}
\hspace*{-0.7cm}
\includegraphics*[width=8.3cm,height=6.2cm]{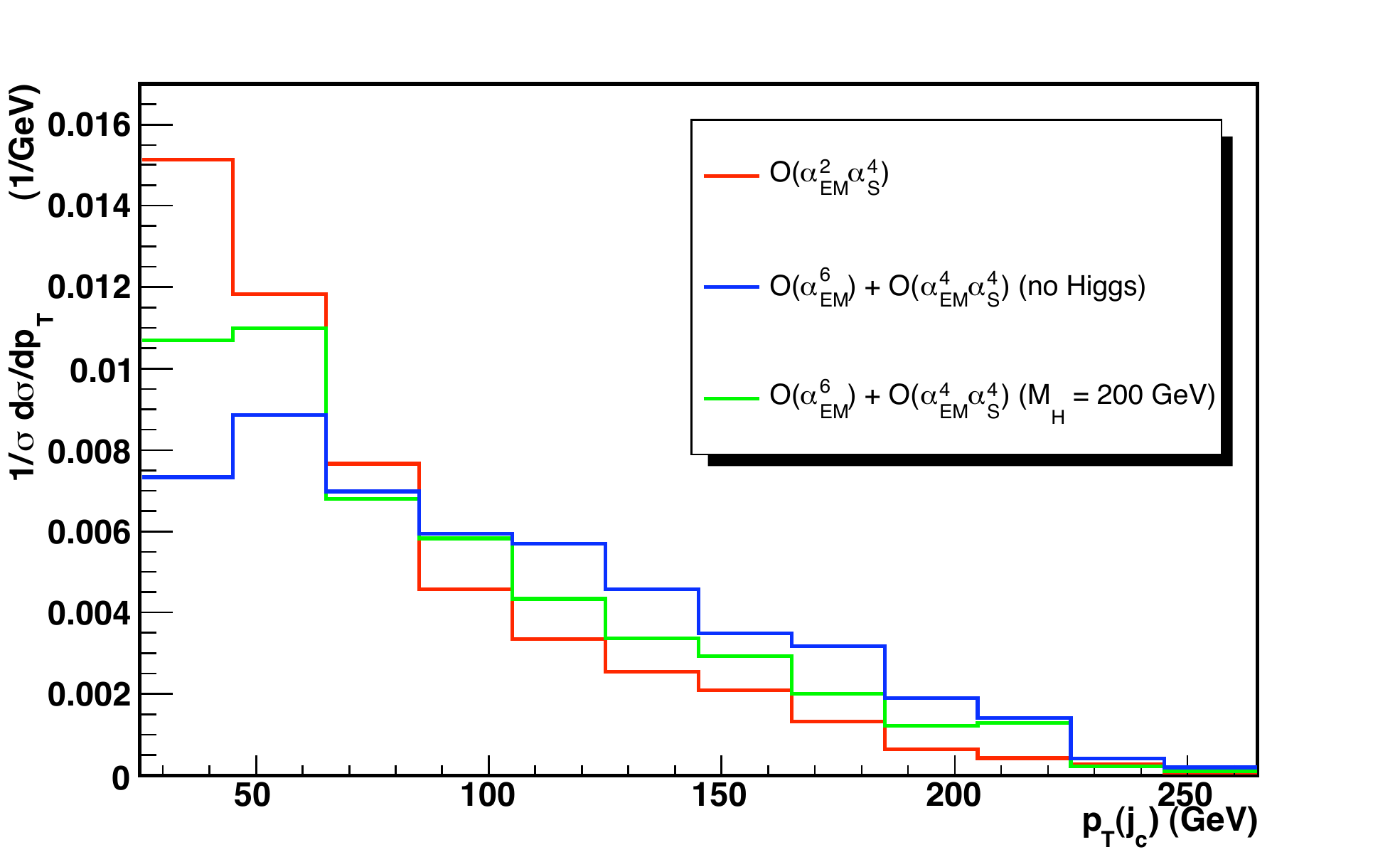}
\hspace*{-3cm}
}
\caption{\textit{From the top:} distribution of the pseudorapidity of the charged lepton, missing $p_T$ and
minimum $p_T$ of the two central jets with the cuts of
\tbnsc{tab:cuts_0}{tab:cuts_1} and of \tbn{tab:cuts_3}.
The plots on the right are
the same as those on the left, but normalized to one. The numbers refer to the
$\mu\nu + 4j$ channel only to the region
$M(j_cj_cl\nu) > 800 \mbox{GeV}$.
Interferences between the two perturbative orders are neglected.}
\label{fig:plots_step2_2}
\vspace{-0.5cm}
\end{figure}

The last step of the selection procedure is aimed at further enhancing the 
separation between the two analyzed scenarios while reducing $W+4j$.
Albeit the latter will be subtracted, it is nevertheless fundamental to achieve 
a good significance of the signal peak since the uncertainty of the background
will in the end determine the observability of the signal.
In Figure \ref{fig:plots_step2_1} the distributions of the absolute value of
the difference in pseudorapidity
between tag jets and of the invariant mass of the two tag jets is shown for the
different samples. On the right hand side we also present the normalized
distributions.  It is apparent that the $\ordQCDsq$ background has typically
tag jets with smaller invariant mass and separation in pseudorapidity.
In Figure \ref{fig:plots_step2_2} the distributions of the pseudorapidity of
the charged lepton, of the missing $p_T$ and of the
minimum $p_T$ of the two central jets are shown together with their
normalized counterparts. The distributions in Figure \ref{fig:plots_step2_2} 
clearly show that the vector bosons in the signal sample are
usually more central.
All the results presented in the two figures are obtained with the cuts
of \tbnsc{tab:cuts_0}{tab:cuts_1} and of \tbn{tab:cuts_3}.
Therefore, based on the results of Figure \ref{fig:plots_step2_1} and
\ref{fig:plots_step2_2}, we apply the cuts in \tbn{tab:cuts_4}.

Some representative cross sections at high invariant masses are reported in 
\tbn{tab:xsec_before_subtraction}. 
There are clear indications that the scattering cross section is sensitive
to effects of a strongly-coupled gauge dynamics provided the 
$W+4j$ background is conveniently subtracted.
Indeed, without subtraction, the separation with respect to
the Standard Model predictions  which has been achieved,
of order $\mathcal{O}(15\%)$, would lie within the accuracy of the 
calculation and, as a result, the underlying sensitivity would be completely 
spoiled.

Next, as described in Section \ref{sec:Outline_of_the_analysis}, we proceed with
the construction of the discriminator and of its probability density function.
In the following we define the background $B$ as the expected yield of the
$\ordQCDsq$ $W+4j$ processes and the signal $S$ as the expected number of
events from all $\ordEW$ and $\ordQCD$ processes. 
$B$ and $S$ are random variables representing the number of background and
signal events for a possible experimental outcome.
$\bar{B}$ and $\bar{S}$ are the corresponding average values which will be taken
equal to the predictions of our simulation.

\begin{table}[htb]
\begin{center}
\begin{tabular}{|c|}
\hline
\textbf{Selection cuts II} \\
\hline
$ M(j_fj_b) > 1000 \mbox{ GeV}$ \\
\hline
$|\Delta\eta(j_fj_b)| > 4.8$ \\
\hline
$\vert \eta(\ell^\pm) \vert < 2.0 $ \\
\hline
$\mbox{missing } p_T > 100 \mbox{ GeV}$ \\
\hline
$p_T(j_c) > 70 \mbox{ GeV}$ \\
\hline
\end{tabular}
\caption{Further selection cuts applied in the final analysis of all
samples in addition to the cuts in \tbnsc{tab:cuts_0}{tab:cuts_1}
and \tbn{tab:cuts_3}.}
\label{tab:cuts_4}
\end{center}
\end{table}

%\eject
%\newpage
% FIGURE 14

%\eject
%\newpage

%\eject
%\newpage

\begin{table}[htb]
\begin{center}
\begin{tabular}{|c|c|c|c|c|c|}
\cline{1-5}
\multirow{2}{*}{$M_{cut}$} & \multicolumn{2}{|c|}{no Higgs} & 
\multicolumn{2}{|c|}{$M_H = 200$ GeV} & \multicolumn{1}{c}{}\\
\cline{2-6}
 & $\sigma$ & events & $\sigma$ & events & ratio  \\
\hline
600 GeV & 6.074 (1.184) fb & 607 (118) & 5.414 (0.524) fb & 541 (52) & 1.12 (2.27) \\
800 GeV & 3.758 (0.779) fb & 376 ( 78) & 3.288 (0.309) fb & 329 (31) & 1.14 (2.52) \\
1.0 TeV & 2.255 (0.483) fb & 226 ( 48) & 1.941 (0.169) fb & 194 (17) & 1.16 (2.82) \\
1.2 TeV & 1.317 (0.263) fb & 132 ( 26) & 1.148 (0.094) fb & 115 ( 9) & 1.15 (2.89) \\
1.4 TeV & 0.683 (0.132) fb &  68 ( 13) & 0.601 (0.050) fb &  60 ( 5) & 1.13 (2.60) \\
\hline
\end{tabular}
\caption{Integrated $\ordEW+\ordQCD+\ordQCDsq$ cross section for 
$M(j_cj_cl\nu)> M_{cut}$ and number of expected events after one year at high 
luminosity ($\mathcal{L} = 100 \mbox{ fb}^{-1}$) with the set of cuts listed in 
\tbnsc{tab:cuts_0}{tab:cuts_1}
and \tbnsc{tab:cuts_3}{tab:cuts_4}.
In parentheses the results for the $\ordEW+\ordQCD$ are also shown.
Interferences between the different perturbative orders are neglected.
The numbers refer to the $\mu\nu + 4j$ channel only.}
\label{tab:xsec_before_subtraction}
\end{center}
\end{table}

For a reliable estimate of the exclusion limits, two different sources of 
uncertainty are taken into account. On one side, the number of 
expected events, both for $B$ and $S$, is affected by statistical fluctuations.
We assume they
are distributed according to the Poisson density function,
\begin{equation}
\label{eq:Poisson}
f(N,\bar{N}) = \frac{\bar{N}^N\,e^{{\small -\bar{N}}}}{N!} \,,
\end{equation}

where $\bar{N}$ is the number of expected events as a consequence of the
cross section $\sigma$ and of the given
luminosity $\mathcal{L}$. On the other hand, the predicted signal
cross section is 
affected by theoretical uncertainties, so the parameter $\bar{S}$ is itself 
subject to fluctuations. 
The p.d.f. is eventually calculated as a convolution of the two contributions.

We define the test statistics $D$ using the following prescription,
\begin{equation}
\label{eq:test_statistic_definition}
D = B + S - \bar{B}
\end{equation}

In this analysis, for $S$ we assume, in addition to the statistical
fluctuations, a theoretical error defined as a flat
distribution in
the window $\bar{S} \pm 30\%$ which,
in our opinion, is a 
reasonable choice to account for both pdf's and scale uncertainties for the
signal. The processes we are interested in require center of mass energies of
the order of the TeV and therefore involve rather large-$x$ quarks,
$x \approx 10^{-1}\div 10^{-2}$ at a typical scale $Q$ of about 100 GeV.
In this region the uncertainty due to the parton distribution functions is of
the order of 5\% \cite{Martin:2002aw,Martin:2003sk}.
As already stated, QCD corrections are in the range of 10\% and,
as a consequence theoretical uncertainties are expected to be well within this
order of magnitude.
Only  
statistical fluctuations have been taken into consideration
in the case of $B$. This is motivated by
the fact that the background is likely to be well measured experimentally 
from the region outside the signal peak, so that the theoretical error on
$W+4j$ is not 
expected to be an issue at the time when real data analysis will be performed.

\eject

\begin{figure}[htb]
% FIGURE 15
\vspace*{-1.9cm}
\centering
\subfigure[]{
\hspace*{-2.1cm}
\includegraphics*[width=8.3cm,height=6.5cm]{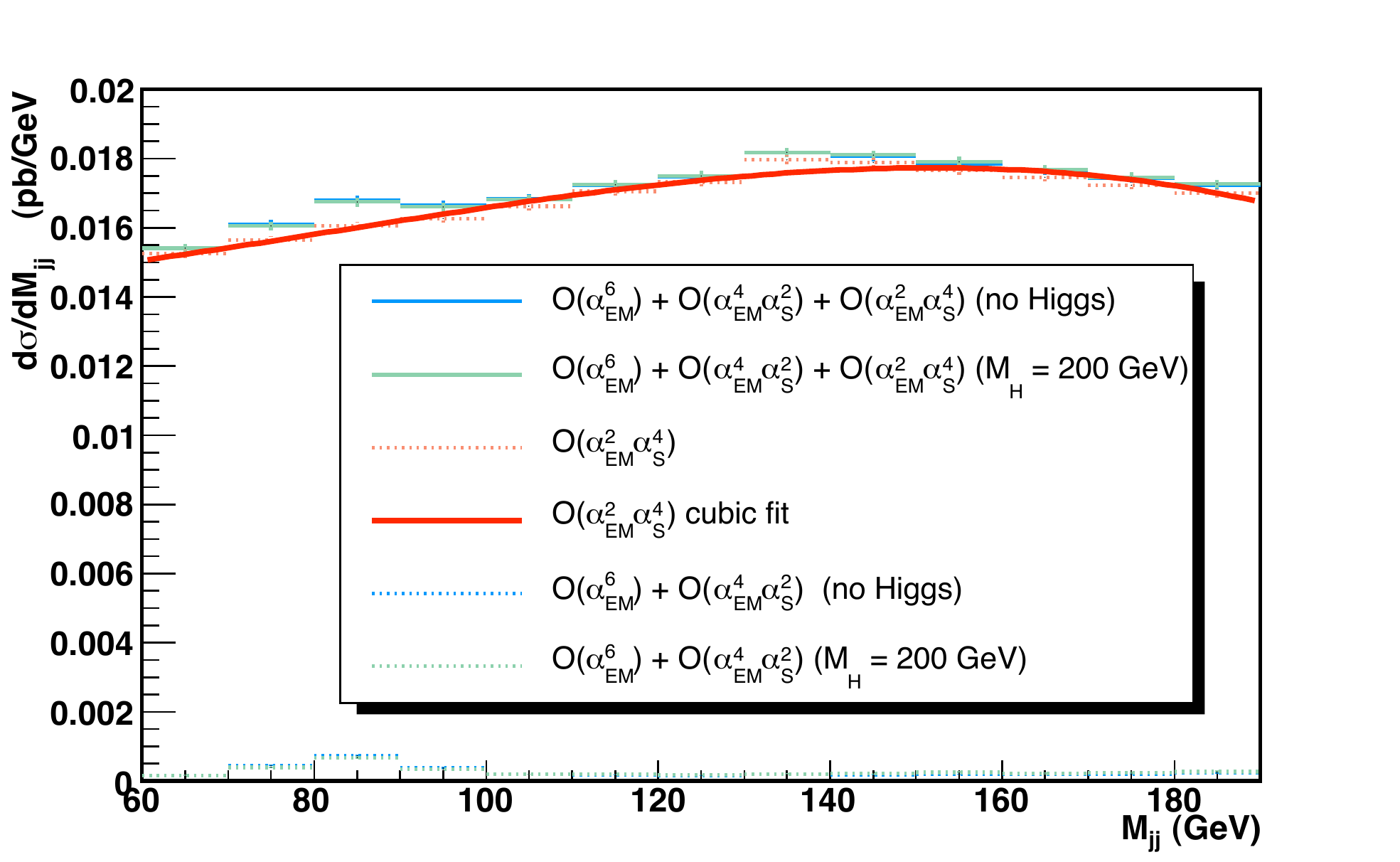}
\hspace*{-0.7cm}
\includegraphics*[width=8.3cm,height=6.5cm]{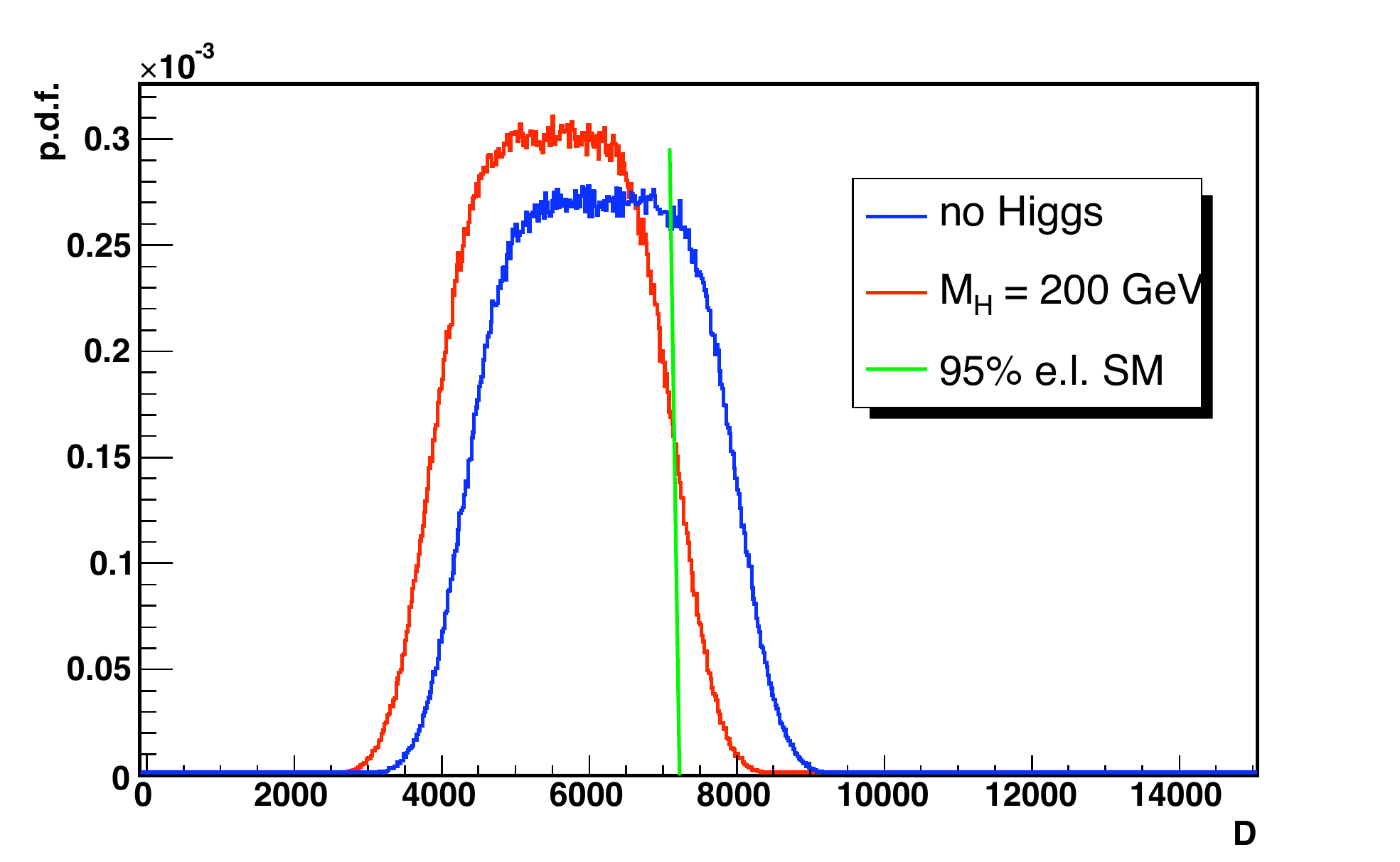}
\hspace*{-3cm}
}
\vspace{-0.4cm}
\subfigure[]{
\hspace*{-2.1cm}
\includegraphics*[width=8.3cm,height=6.5cm]{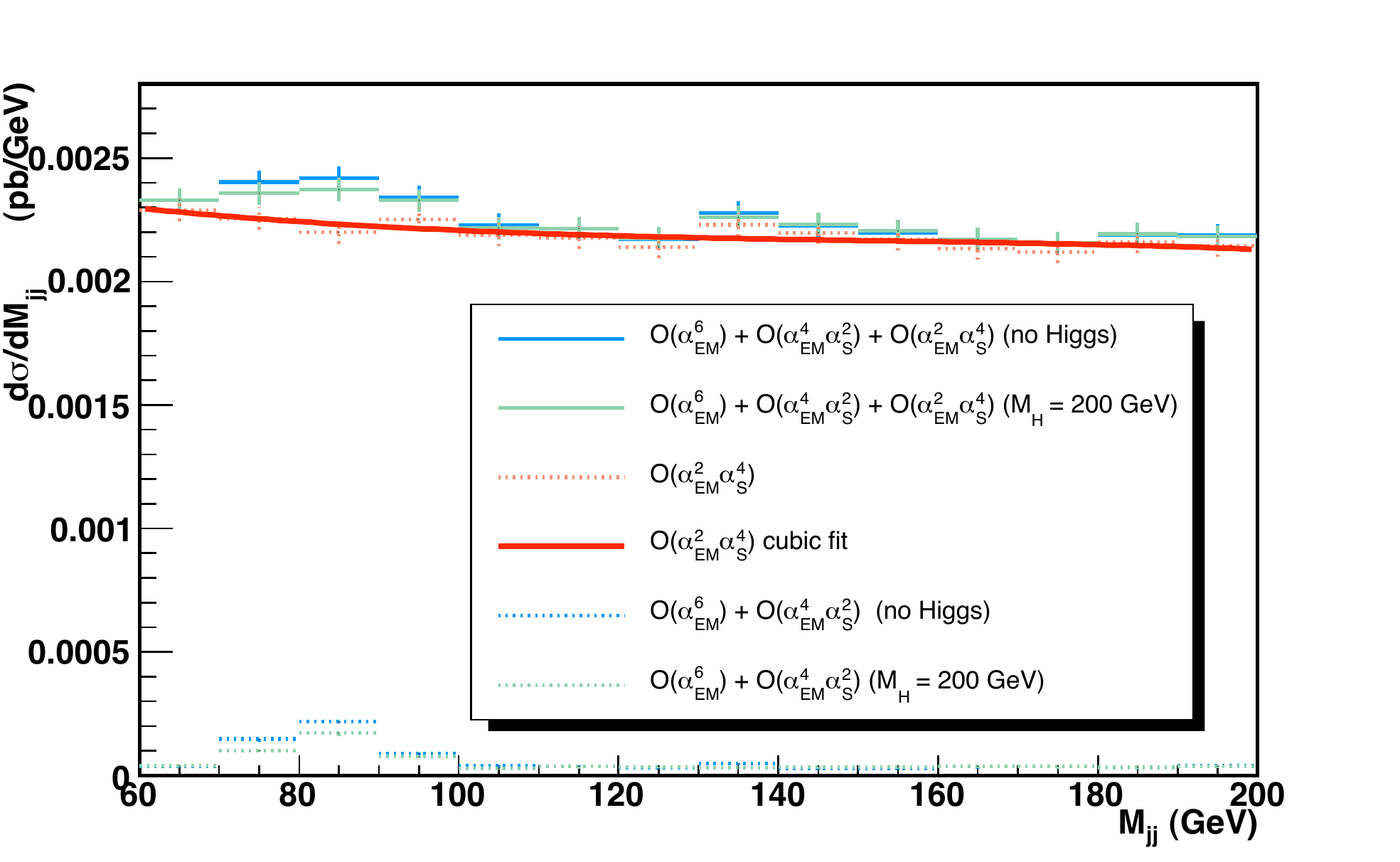}
\hspace*{-0.7cm}
\includegraphics*[width=8.3cm,height=6.5cm]{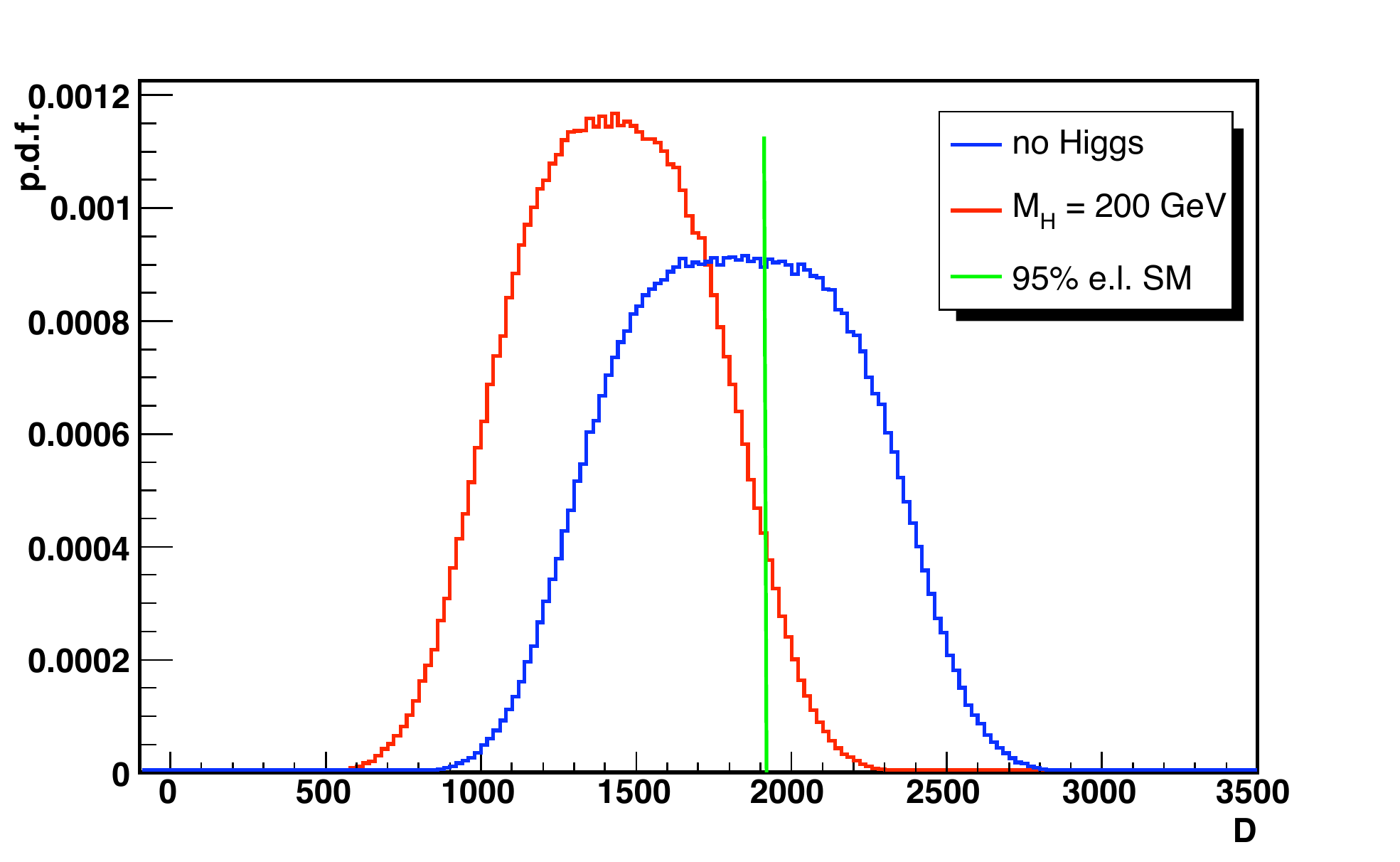}
\hspace*{-3cm}
}
\vspace{-0.4cm}
\subfigure[]{
\hspace*{-2.1cm}
\includegraphics*[width=8.3cm,height=6.5cm]{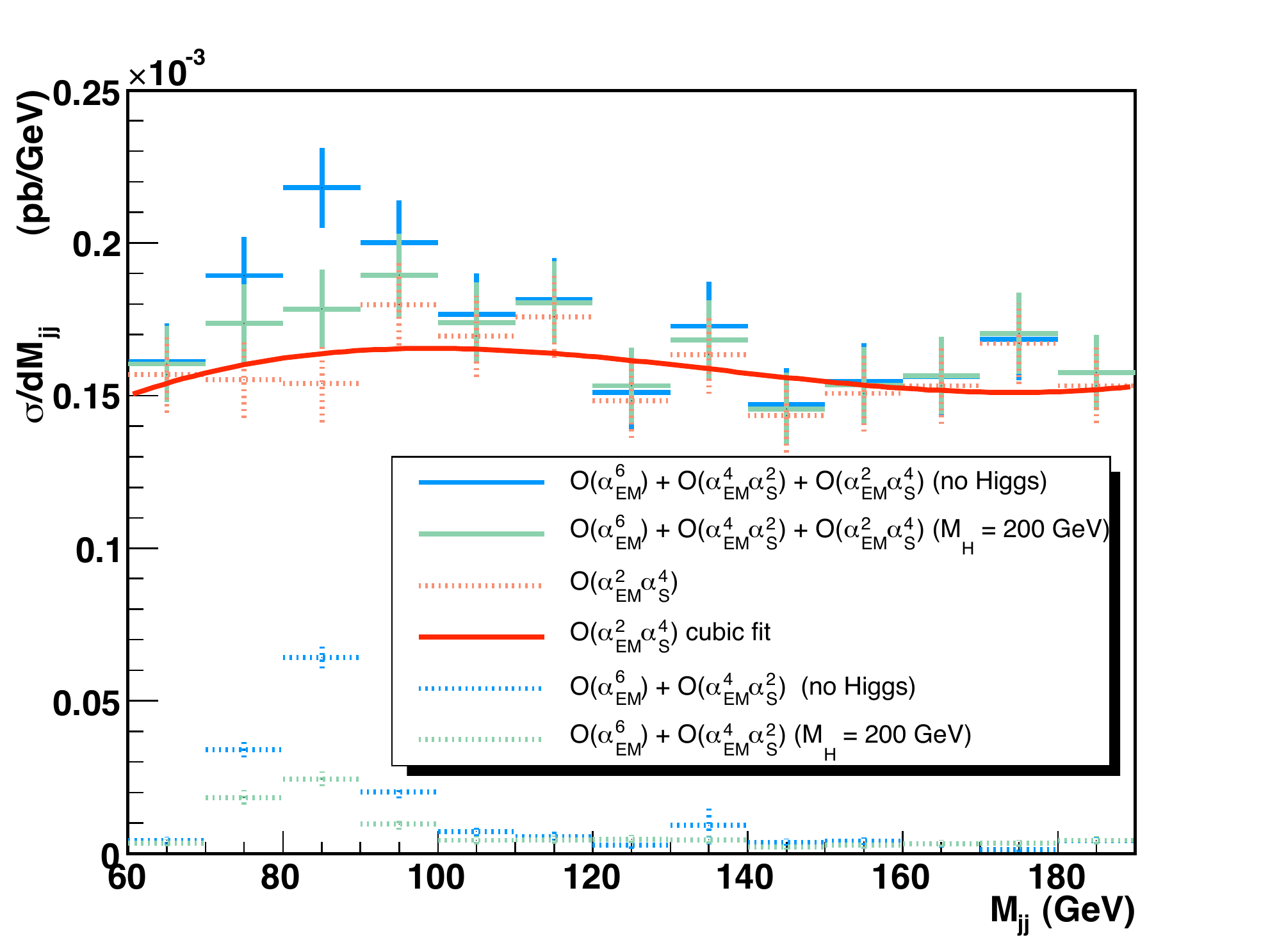}
\hspace*{-0.7cm}
\includegraphics*[width=8.3cm,height=6.5cm]{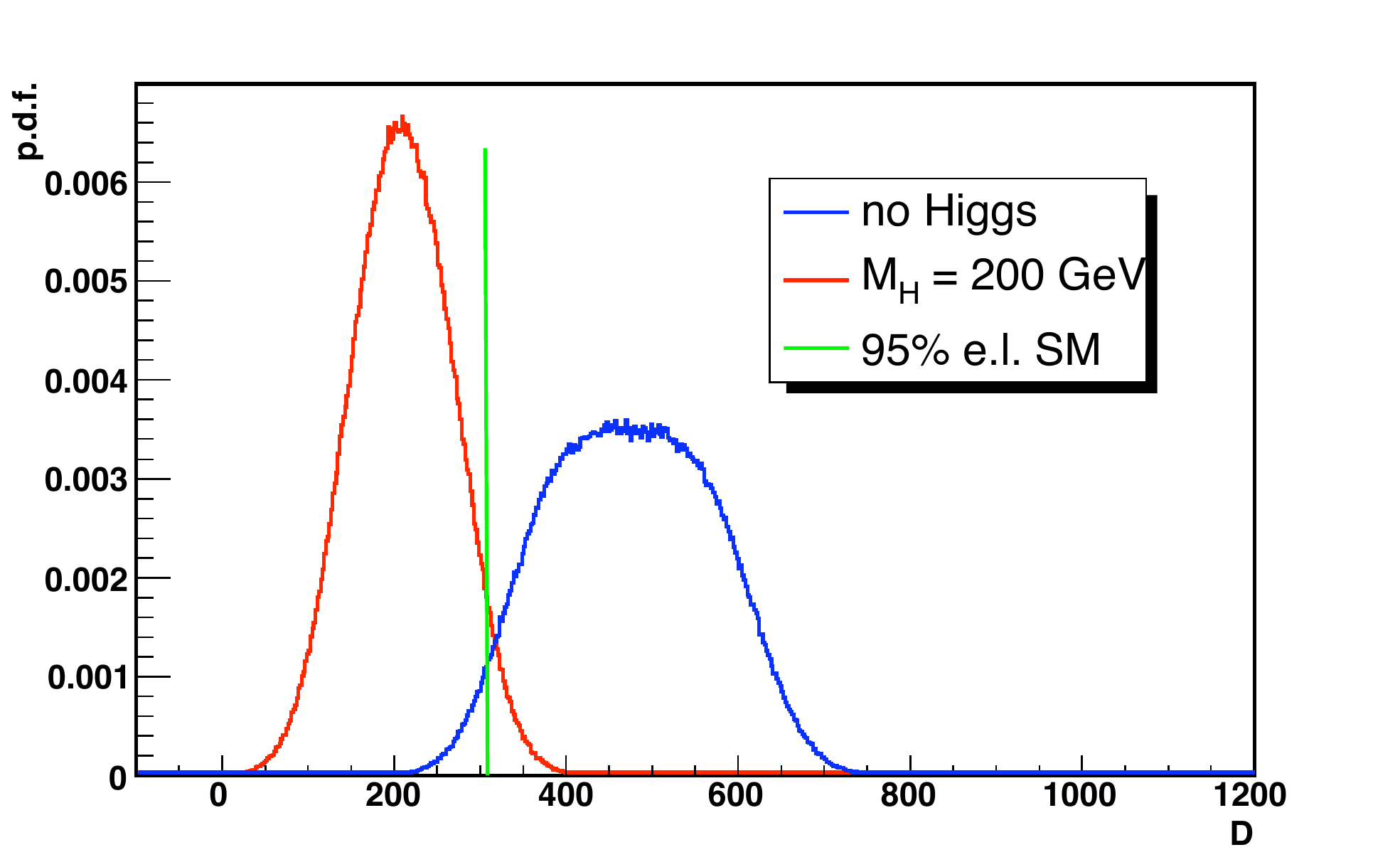}
\hspace*{-3cm}
}
\caption{\textit{On the left:} distribution of the invariant mass of the
two central jets:
(a) with the cuts of \tbnsc{tab:cuts_0}{tab:cuts_1},
(b) with the cuts of \tbnsc{tab:cuts_0}{tab:cuts_1} and \tbn{tab:cuts_3},
(c) with the full set of cuts \tbnsc{tab:cuts_0}{tab:cuts_1}
and \tbnsc{tab:cuts_3}{tab:cuts_4}.
\textit{On the right:} corresponding probability density function of the test statistics
(Eq. \ref{eq:test_statistic_definition}) for the two analyzed scenarios. The cross
sections refer to the muon channel only, while the right hand plots assume an integrated
luminosity of $\mathcal{L}=400 \mbox{ fb}^{-1}$.
}
\label{fig:plots_indicator}
\vspace{-0.7cm}
\end{figure}

\eject

%\newpage

%\eject
%\newpage

Two parameters which well exemplify the effectiveness of the adopted procedure 
are the significance $\bar{S}/\sqrt{\bar{B}}$ of the signal peak over
the non-resonant background and the probability of what we call,
for brevity and with a slight abuse of language, the BSM 
hypothesis at 95\% SM exclusion limit, namely the probability
that, in the absence of a Higgs boson, the result of an experimental outcome,
at fixed luminosity, has a chance of less than 5\% in the Standard Model. 
The first one is related to the possibility of detecting a clear vector boson
scattering signal 
independently of the model assumed. The second parameter is rather an indicator 
of the resolving power which can be achieved between the two considered 
prototypes of weakly-interacting and strongly-interacting scenarios.
Figure \ref{fig:plots_indicator} and \tbn{tab:significance&probabilities}
show how the sequence of selection cuts manages to
improve the separation and the significance of the signal. The red line in the
left hand plots is the result of a cubic fit of the $W+4j$ background in the
interval  $\mbox{60 GeV} < M_{j_cj_c} < \mbox{200 GeV}$. 
The two curves in the right hand plots are obtained by randomly generating
values for the $D$ parameter in Eq.(\ref{eq:test_statistic_definition}) as
discussed above.
In practice we first compute $\bar{S}$ and $\bar{B}$, where $\bar{S}$
is the number of events for $\ordEW + \ordQCD$ processes
in the mass window
70 GeV $< M_{j_cj_c} <$ 100 GeV corresponding to the plots on the left hand side
of Figure \ref{fig:plots_indicator} for a 
luminosity of $400 \mbox{ fb}^{-1}$.
$\bar{B}$ is the corresponding number of events for $\ordQCDsq$ processes,
obtained interpolating, as already mentioned,
the sidebands between 60 GeV and 200 GeV for the same luminosity and the same
mass interval.
Then we generate an  $\bar{S}^\prime$ from the probability density:
\be
P(\bar{S}^\prime) = \frac{1}{0.6 \; \bar{S}} \; \theta ( \bar{S}^\prime - 0.7\bar{S} )
\; \theta ( 1.3\;\bar{S} -\bar{S}^\prime)
\label{eq:ProbDens}
\ee
corresponding to the assumed theoretical uncertainty.
From $\bar{B}$ and $\bar{S}^\prime$ we generate $B$ and $S$ according to the
Poisson distributions $f(B,\bar{B})$ and $f(S,\bar{S}^\prime)$ respectively.
Finally from  $B$ and $S$ we construct $D$ according to
\eqn{eq:test_statistic_definition}. The normalized frequency of $D$ for the two
scenarios is reported in the right hand side of Figure \ref{fig:plots_indicator}.
The red curve refers to a Higgs of 200 GeV while the blue one
refers to the no--Higgs case.
The green vertical line in the right
hand plots marks the 95\% confidence limit for the SM predictions.

The right hand side of Figure \ref{fig:plots_indicator} and \ref{fig:plots_DR_dependence}
refer to
a luminosity of $\mathcal{L}=400 \mbox{ fb}^{-1}$. This corresponds 
to the sum of the muon and electron 
channels, $pp \rightarrow \ell\nu + 4j\,\, (\ell = \mu,e)$, and to   
$100 \mbox{ fb}^{-1}$ of data for each of the two LHC general-purpose 
experiments CMS and ATLAS.
The cross sections on the left hand side, on the contrary, refer to the muon channel only.

\begin{table}[htb]
\begin{center}
\begin{tabular}{|c|c|c|c|}
\hline
\rule[-2mm]{0mm}{7mm} & $\bar{S}/\sqrt{\bar{B}}$@no--Higgs & 
     $\bar{S}/\sqrt{\bar{B}}$@$M_H = 200$ GeV & $P_{BSM}@95\%CL$ \\
\hline
\tbnsc{tab:cuts_0}{tab:cuts_1}                                  & 14.07 (7.03) &  12.61 (6.30) & 22.14\% (16.54\%) \\
\tbnsc{tab:cuts_0}{tab:cuts_1}, \ref{tab:cuts_3}               & 11.20 (5.60) &  8.69 (4.34) &  43.94\% (29.17\%) \\
\tbnsc{tab:cuts_0}{tab:cuts_1}, \ref{tab:cuts_3} and \ref{tab:cuts_4} & 10.72 (5.36) &  4.75 (2.37) &  96.78\% (78.11\%) \\
\hline
\end{tabular}
\caption{Significances and BSM probabilities with a
luminosity $\mathcal{L} = 400 \mbox{ fb}^{-1}$ and the set of cuts listed in 
\tbnsc{tab:cuts_0}{tab:cuts_1}
and \tbnsc{tab:cuts_3}{tab:cuts_4} are progressively applied. 
In parentheses the results for a luminosity of $\mathcal{L}=100 \mbox{ fb}^{-1}$ are also shown.
}
\label{tab:significance&probabilities}
\end{center}
\end{table}

\eject

The matching significances and BSM probabilities are shown in
\tbn{tab:significance&probabilities}. In parentheses
the results for a luminosity of $\mathcal{L}=100 \mbox{ fb}^{-1}$ are also presented: 
With $\mathcal{L}=400 \mbox{ fb}^{-1}$, with the full set of cuts, 
\tbnsc{tab:cuts_0}{tab:cuts_1} and \tbnsc{tab:cuts_3}{tab:cuts_4}, 
we find that $209 \pm 59$ signal events are expected in the light Higgs SM
scenario and $474 \pm 96$ in the no Higgs case. 
Under the given 
conditions, a number of signal events of approximately 300 units or more would 
represent an evidence for new physics beyond a light-intermediate SM Higgs at 
95\% confidence level for SM. Conversely, the probability for a BSM-like 
experimental outcome to lie below this threshold is 4\%.

%%%%%%%%%%%%%%%%%%%%%%%%%%%%%%%%%%%%%%%%%%%%%%%%%%%%%%%%%%%%%%%%%%%%%%%%%
\section{Dependence on the jet cone size}
\label{sec:discuss}

Jet resolution is an important experimental issue.
At detector level, many analyses make use of the cone algorithm for jet 
reconstruction which requires hadronic jets to be well separated from each 
other in the pseudorapidity ($\eta$) \--- azimuthal angle ($\phi$) plane within 
a cone of radius $\Delta R$. The cone algorithm establishes a natural 
correspondence between jets and the underlying partons.
However, due to the large center-of-mass energy 
available at LHC, some fraction of EW bosons will be highly boosted in the
transverse direction and,
consequently, the jets from their decays will tend to have a small $\Delta R$ separation.
Thus, an approach based on the cone separation of jets would eventually result
in a severe  
loss of potentially interesting events: the larger the radius of the cone, the
larger the number of discarded events. This could seriously damage 
the possibility to evidentiate effects of new physics, since
the EWSB mechanism is expected to manifest itself in events with 
very energetic, high-$p_T$ vector bosons. 

On the other hand, alternative jet finding algorithms have been proposed
\cite{JetFinding} 
which may prove useful in connection with this kind of studies as they lead to
encouraging results in identifying hadronic decays of heavy bosons via a cut on 
the sub-jet separation scale. 

At the partonic level we are considering, there is no room for addressing the 
issue of jet resolution in much more detail. The only thing we can do is to 
investigate what is the impact of requiring a minimum $\Delta R$ separation 
among coloured particles.
Of course, the final word is left to a realistic study at the hadronic level, 
including the full chain of detector simulation and reconstruction.

\begin{figure}[htb]
\vspace*{-2.cm}
\centering
\subfigure[]{
\hspace*{-2.1cm}
\includegraphics*[width=8.3cm,height=6.5cm]{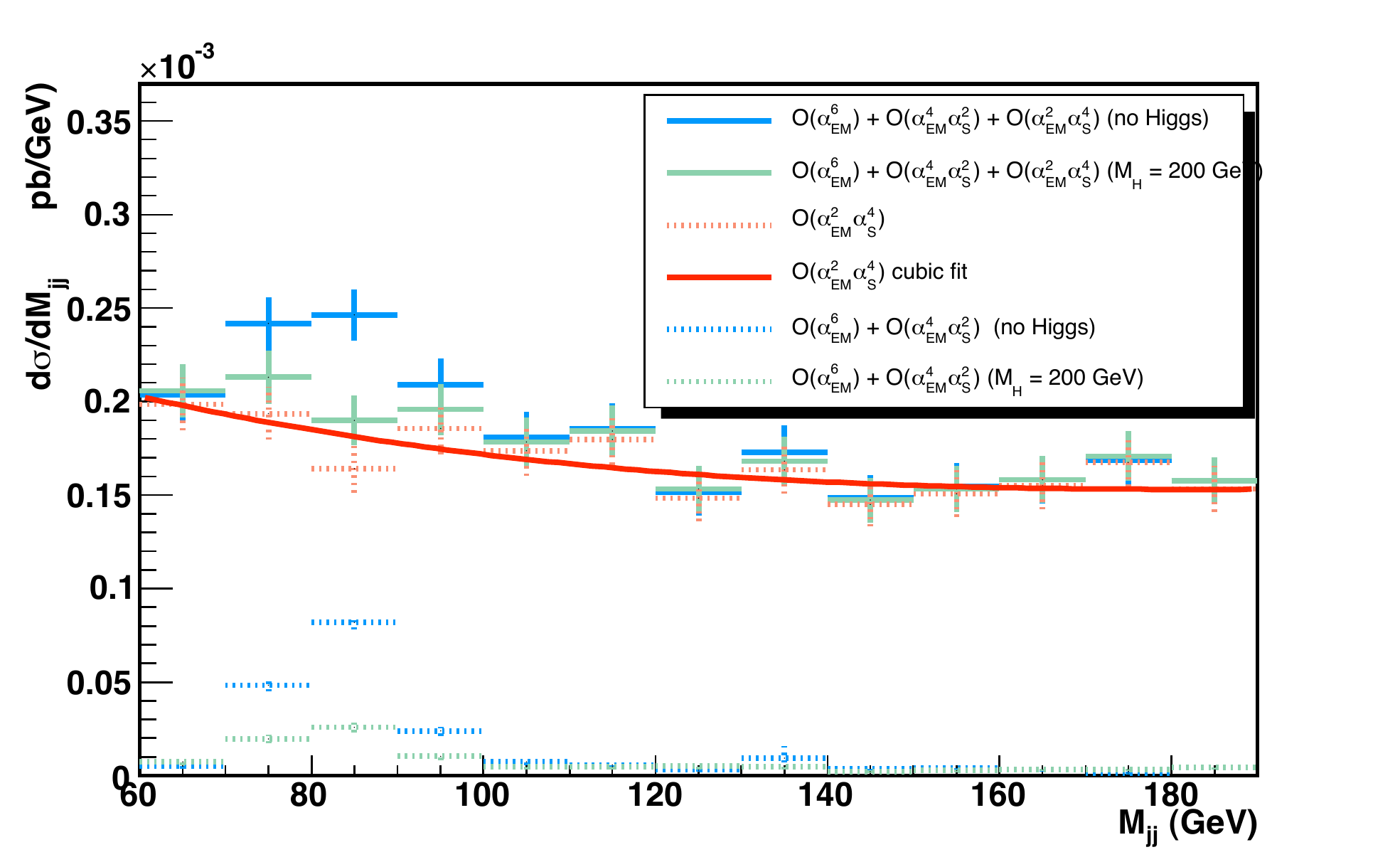}
\hspace*{-0.7cm}
\includegraphics*[width=8.3cm,height=6.5cm]{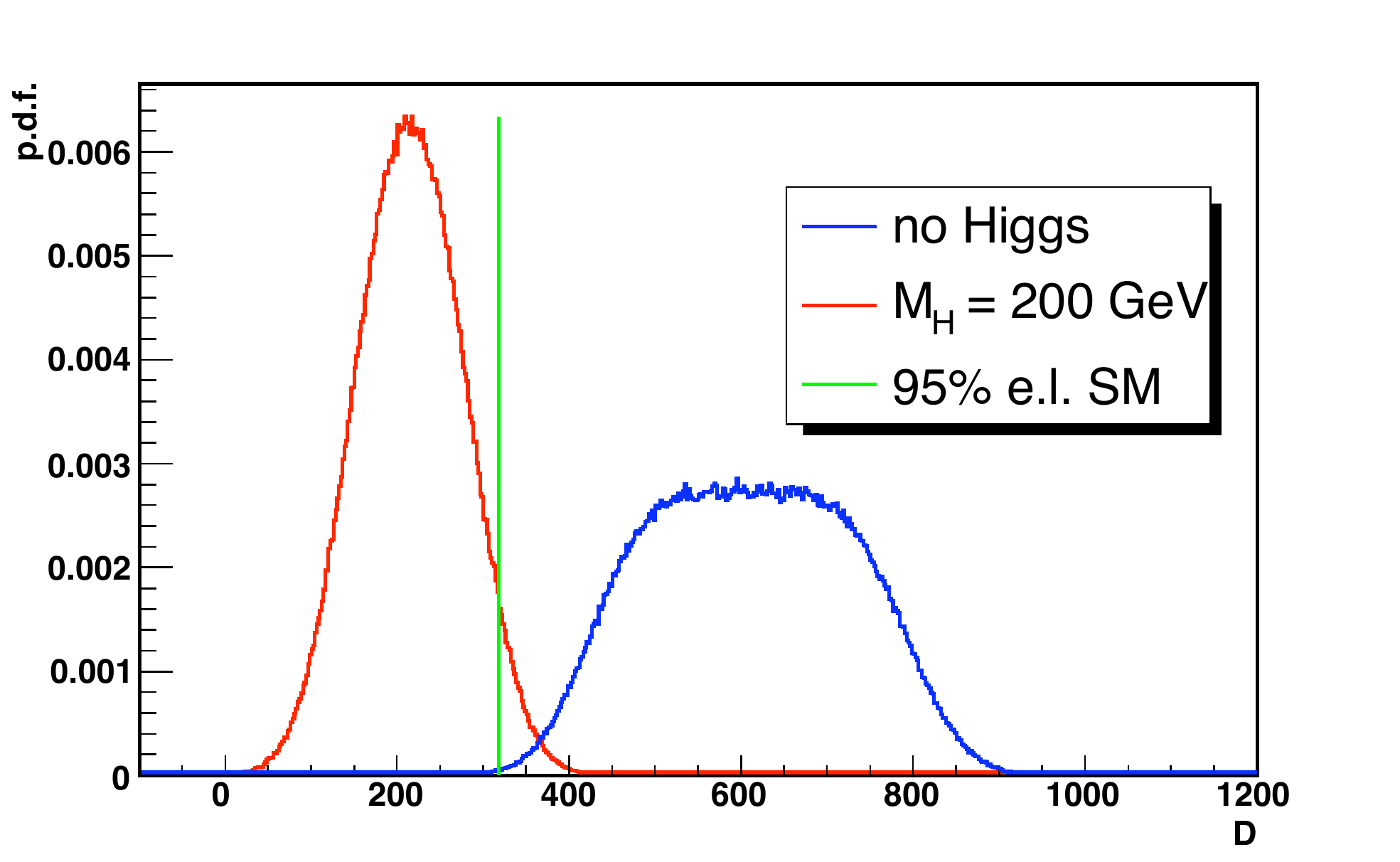}
\hspace*{-3cm}
\vspace*{-0.6cm}
}
\subfigure[]{
\hspace*{-2.1cm}
\includegraphics*[width=8.3cm,height=6.5cm]{plot_Mjcjc_MVV600_step3.pdf}
\hspace*{-0.7cm}
\includegraphics*[width=8.3cm,height=6.5cm]{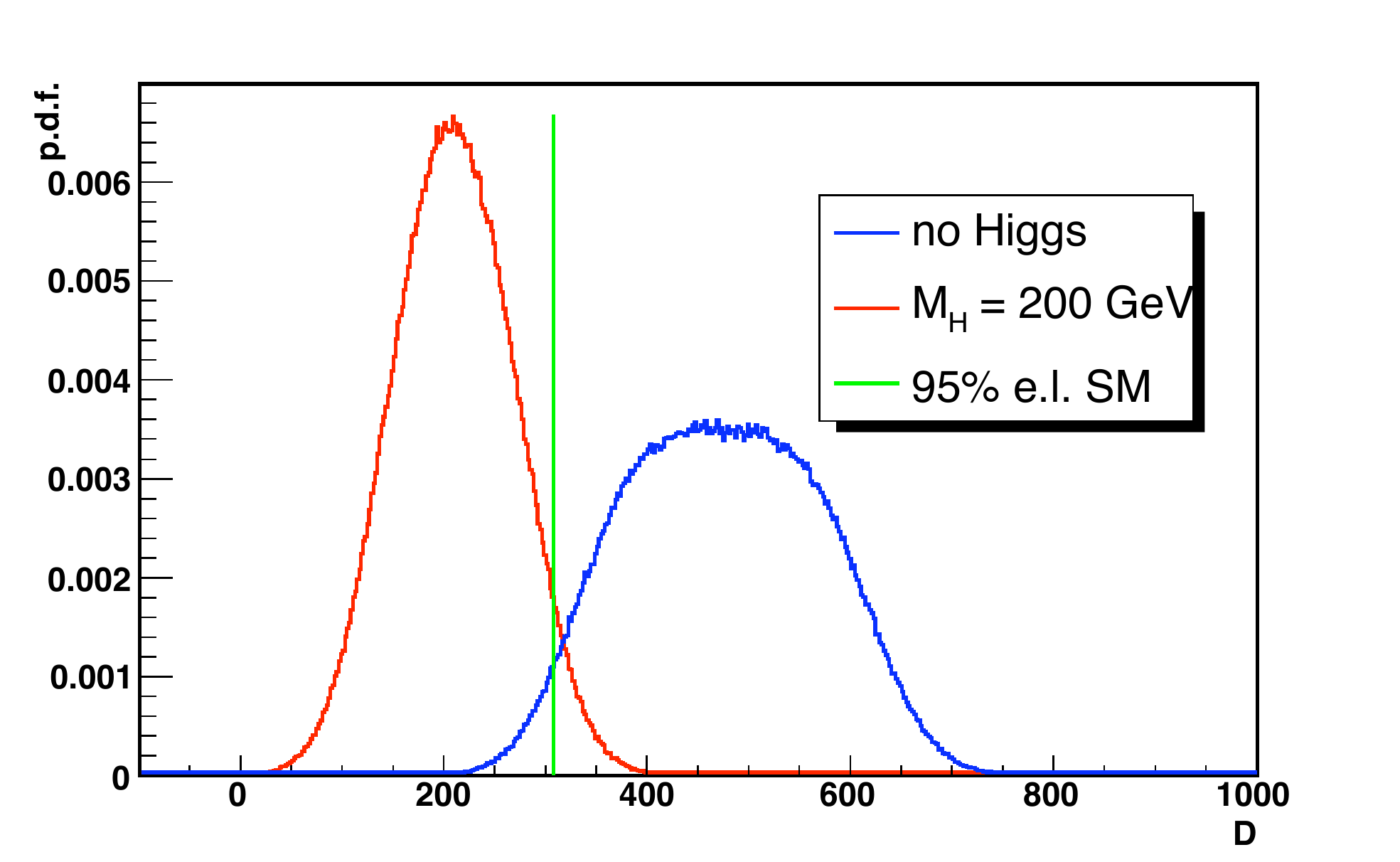}
\hspace*{-3cm}
\vspace*{-0.4cm}
}
\subfigure[]{
\hspace*{-2.1cm}
\includegraphics*[width=8.3cm,height=6.5cm]{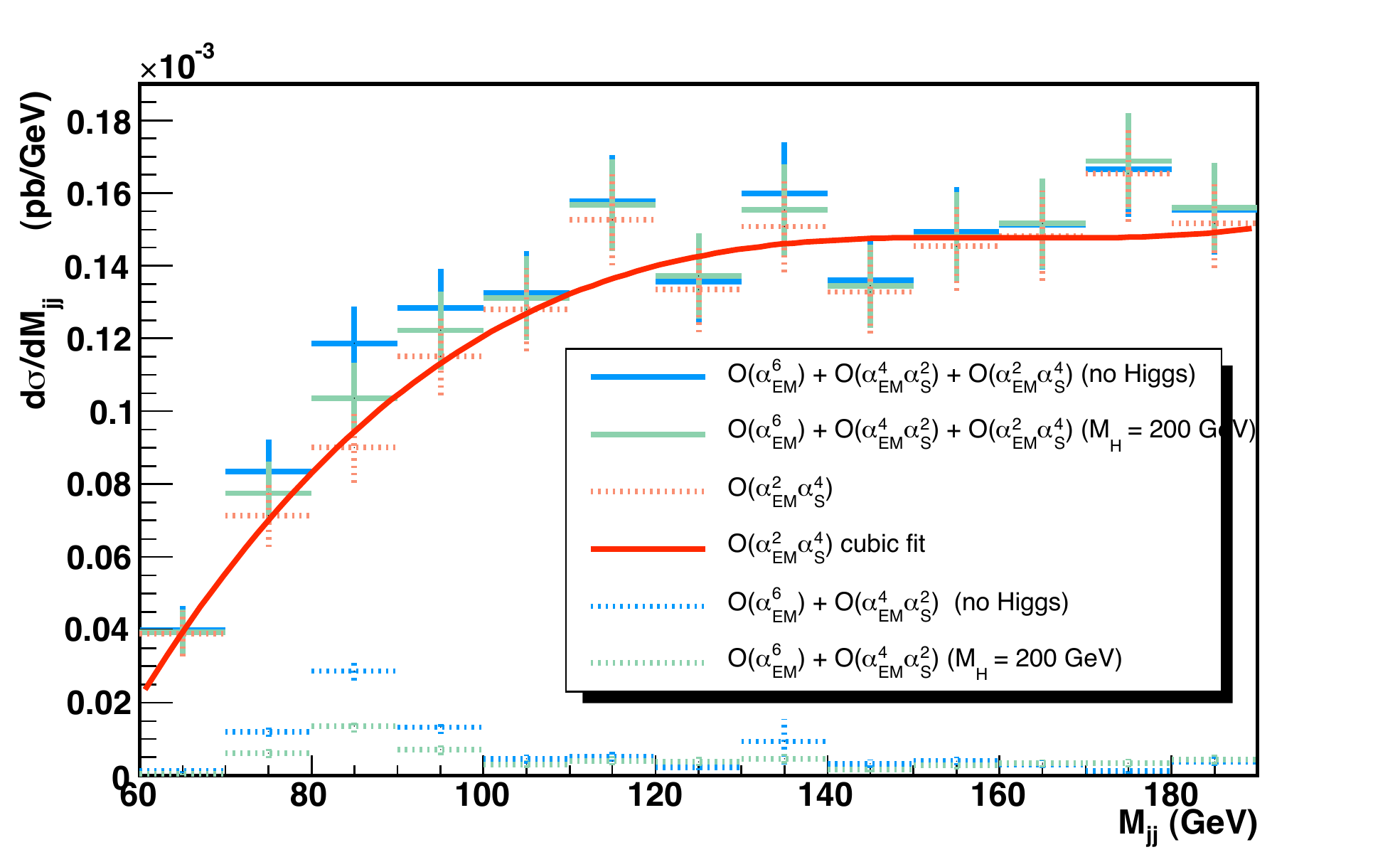}
\hspace*{-0.7cm}
\includegraphics*[width=8.3cm,height=6.5cm]{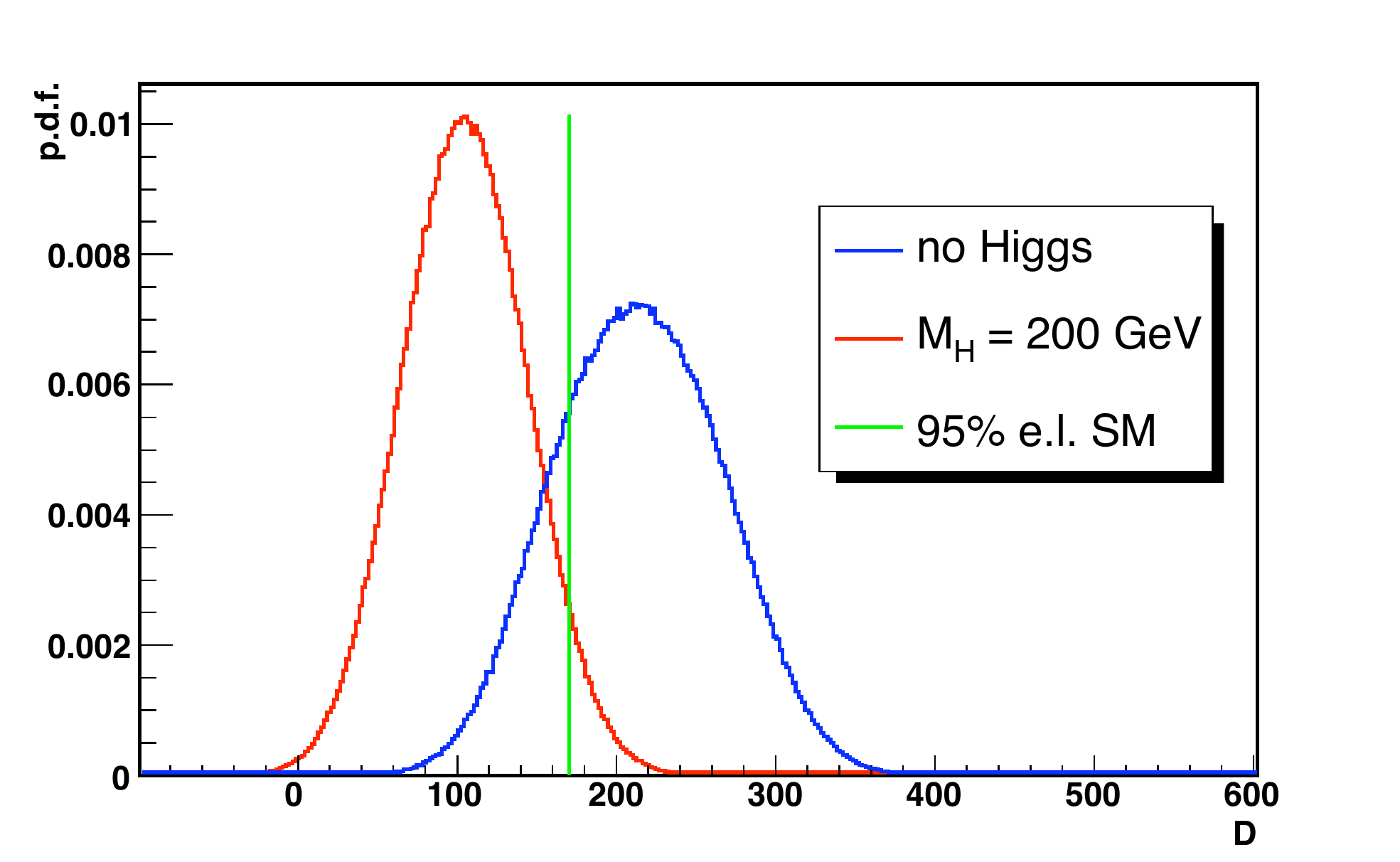}
\hspace*{-3cm}
}
\caption{Impact of $\Delta R(jj)$ as a criterion for jet separation.
\textit{On the left:} distribution of the invariant mass of the two central jets:
(a) without any cut on $\Delta R$ among coloured particles,
(b) with $\Delta R > 0.3$, (c) with $\Delta R > 0.5$. Apart from the
$\Delta R$ cut, in all cases
the full set of cuts in \tbnsc{tab:cuts_0}{tab:cuts_1}
and \tbnsc{tab:cuts_3}{tab:cuts_4} is applied. In particular a minimum invariant
mass $M(jj) > 60$
GeV has been imposed.
\textit{On the right:} corresponding probability density function of the test
statistics
(Eq. \ref{eq:test_statistic_definition}) for the two analyzed scenarios.
The cross
sections refer to the muon channel only, while the right hand plots assume
an integrated
luminosity of $\mathcal{L}=400 \mbox{ fb}^{-1}$. }
\label{fig:plots_DR_dependence}
\vspace*{-2.cm}
\end{figure}

\begin{figure}[htb]
\begin{center}
\includegraphics[width=0.80\textwidth]{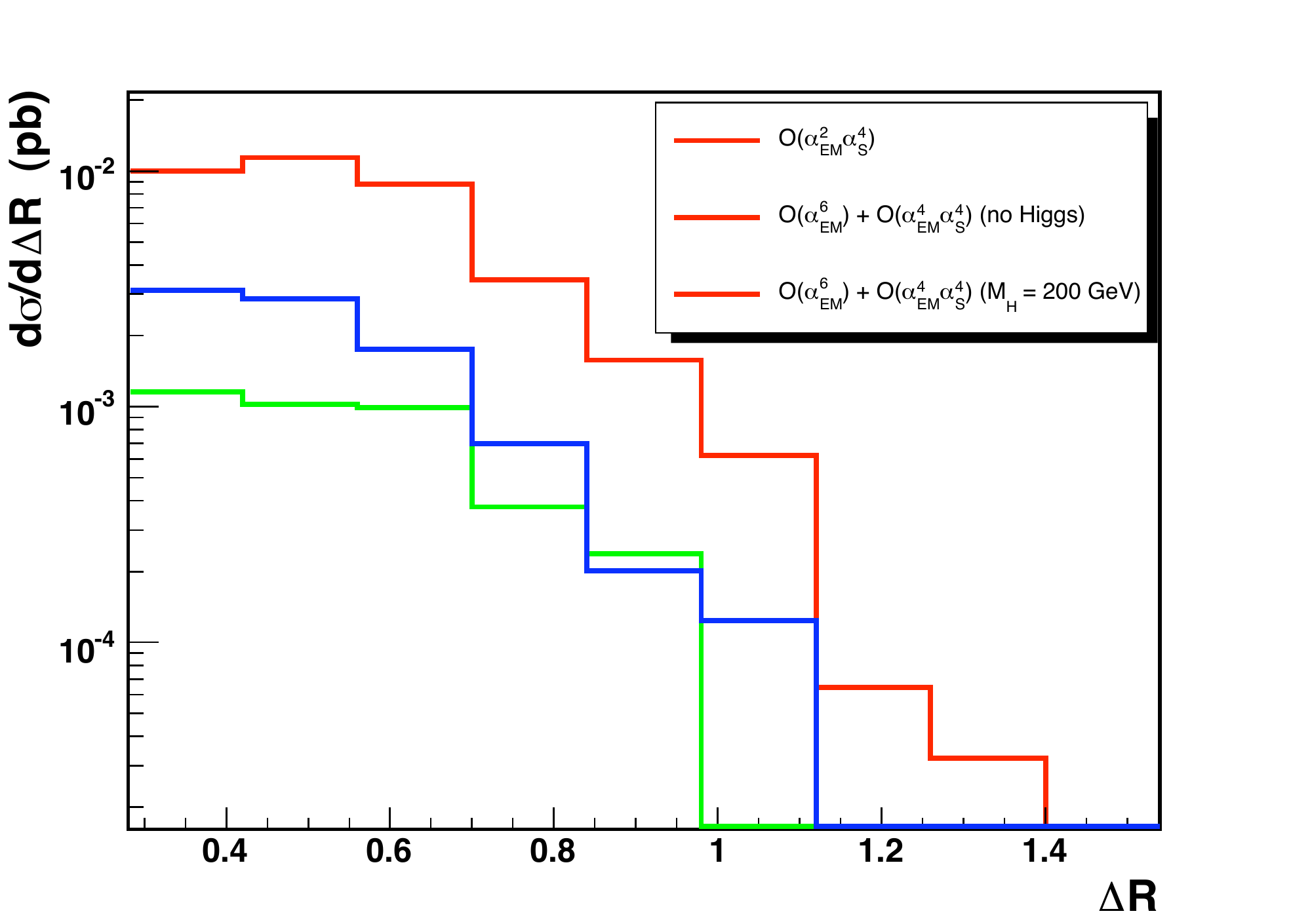}
\end{center}
\caption{Distribution of the minimum $\Delta R$ among all jets with the cuts of
\tbnsc{tab:cuts_0}{tab:cuts_1} and of \tbnsc{tab:cuts_3}{tab:cuts_4}
The plot refers to the region
$M(j_cj_cl\nu) > 600 \mbox{GeV}$.
}
\label{fig:DR_step3}
\end{figure}

Many experimental analyses adopt a cone size larger than 
$\Delta R(jj) = 0.3$ . It is therefore extremely important to monitor how the 
final results change with the minimum $\Delta R$ imposed for any pair of 
coloured particles. It should be however mentioned that in all our results we
impose a minimum invariant mass of 60 GeV for each pair of coloured particles
which partially accounts for jet separation.
As shown in Figure \ref{fig:plots_DR_dependence}, the two
scenarios tend to align their predictions at larger $\Delta R$ thresholds.
This effect is related to the fact that, without a Higgs, the outgoing vector 
bosons are more central and have a larger $p_T$ than in presence
of a Higgs boson. As a consequence the probability that the two jets
from a boson decay eventually merge into a single jet is higher
in the first case.
The significances and BSM probabilities corresponding to 
Figure \ref{fig:plots_DR_dependence} are shown in
\tbn{tab:significance&probabilitiesDR}. 
In Figure \ref{fig:DR_step3} we show the distribution
of the  minimum $\Delta R$ separation among jets.
The results show a rapid decrease of the discriminating power when thresholds
above $\Delta R(jj) = 0.5$ are imposed.

\eject

\begin{table}[htb]
\begin{center}
\begin{tabular}{|c|c|c|c|}
\hline
\rule[-2mm]{0mm}{7mm} & $\bar{S}/\sqrt{\bar{B}}$@no--Higgs & 
     $\bar{S}/\sqrt{\bar{B}}$@$M_H = 200$ GeV & $P_{BSM}@95\%CL$ \\
\hline
No  $\Delta R$    & 13.05 (6.51) &  4.63 (2.31) & 99.92\% (93.28\%) \\
$\Delta R > 0.3$  & 10.72 (5.36) &  4.75 (2.37) &  96.78\% (78.11\%) \\ 
$\Delta R > 0.5$  &  6.43 (3.24) &  3.15 (1.56) &  79.91\% (48.59\%) \\
\hline
\end{tabular}
\caption{Significances and BSM probabilities with a
luminosity $\mathcal{L} = 400 \mbox{ fb}^{-1}$ and the set of cuts listed in 
\tbnsc{tab:cuts_0}{tab:cuts_1}
and \tbnsc{tab:cuts_3}{tab:cuts_4} for three values of $\Delta R$ separation.
In parentheses the results for a luminosity of $\mathcal{L}=100 \mbox{ fb}^{-1}$ are also shown.
}
\label{tab:significance&probabilitiesDR}
\end{center}
\end{table}

%%%%%%%%%%%%%%%%%%%%%%%%%%%%%%%%%%%%%%%%%%%%%%%%%%%%%%%%%%%%%%%%%%%%%%%%%
\section{Conclusions}
We have examined at parton level
the process $pp\rightarrow \ell\nu + 4j$ including all irreducible backgrounds
contributing to this six parton final state. We have considered two scenarios:
a light Higgs Standard Model framework with $M_H = 200$ GeV
and an infinite mass Higgs scenario in order to determine whether the two models
can be distinguished at
the LHC using boson--boson scattering. The largest background is $W+4j$ which can
be subtracted looking at the distribution of the invariant mass of the two most
central jets in the region outside the weak boson mass window.
We have estimated the probability, in the no Higgs scenario, of finding a result 
outside the 95\% confidence limit in the Standard Model. 
This probability turns out to be about 97\% for an integrated luminosity of
$\mathcal{L}=400 \mbox{ fb}^{-1}$ and a mass of the reconstructed pair of vector
bosons larger than 600 GeV. 
With the full set of cuts, 
\tbnsc{tab:cuts_0}{tab:cuts_1} and \tbnsc{tab:cuts_3}{tab:cuts_4}, 
$209 \pm 59$ signal events are expected in the light Higgs SM
scenario and $474 \pm 96$ in the no Higgs case. 
Jet resolution plays a crucial role in the present analysis as in all processes
in which high transverse momentum vector bosons or top particles are present
and decay hadronically.

\section *{Acknowledgments}
A.B. wishes to thank the Department of Theoretical Physics of Torino University for support.

This work has been supported by MIUR under contract 2006020509\_004 and by the
European Community's Marie-Curie Research 
Training Network under contract MRTN-CT-2006-035505 `Tools and Precision
Calculations for Physics Discoveries at Colliders' 

%\hfill*
%\eject
%\newpage
\vspace{2cm}

%%%%%%%%%%%%%%%%%%%%%%%%%%%%%%%%%%%%%%%%%%%%%%%%%%%%%%%%%%%%%%%%%%%%%%%%%

\end{document}